\documentclass[aip,apl,reprint,dvipdfmx]{revtex4-1}

\usepackage{graphicx}
\usepackage{amsmath}
\usepackage{dcolumn}
\usepackage[bookmarks=false,colorlinks=true,linkcolor=blue,citecolor=blue,urlcolor=blue]{hyperref}
\usepackage{xr}
\usepackage{here}
\usepackage{bm}
\usepackage{framed}

\begin{document}

\title{Microscopic mechanism of ultrashort-pulse laser ablation of metals: a molecular dynamics study incorporating electronic entropy effects}

\author{Yuta Tanaka}
\email[\scriptsize{Present Address: ENEOS Corporation, 8 Chidoricho, Naka-ku, Yokohama 231-0815, Japan; }]{tanaka.yuta.735@eneos.com}
\affiliation{$^{1}$Department of Physics, The University of Tokyo, 7-3-1 Hongo, Bunkyo-ku, Tokyo 113-0033, Japan}
\author{Shinji Tsuneyuki}
\affiliation{$^{1}$Department of Physics, The University of Tokyo, 7-3-1 Hongo, Bunkyo-ku, Tokyo 113-0033, Japan}

\date{\today}

\begin{abstract}

The microscopic mechanism of metal ablation induced by ultrashort laser pulse irradiation is investigated.
A two-temperature model scheme combined with molecular dynamics (TTM-MD) is developed to incorporate electronic entropy effects into the simulation of metal ablation while satisfying the energy conservation law.
Simulation with the TTM-MD scheme reveals that ultrashort laser pulse irradiation near the ablation threshold causes high-energy ion emission and sub-nanometer depth ablation, as observed experimentally, due to the electronic entropy effect.
It is also shown that the electronic entropy effect is also significant in spallation.

\end{abstract}


\maketitle

\section{introduction}

Laser ablation is widely employed in industry as a method for laser processing (cutting, drilling), pulsed laser deposition,~\cite{Watanabe_1998,Yoshitake_2000} and nanoparticle  production.~\cite{Fojtik_1993,Neddersen_1993}
The physical mechanism of laser ablation, especially with ultrashort-pulse lasers (fs laser), has attracted attention in science and industry~\cite{Kobayashi_2020,Kobayashi_2021} because it involves remarkable phenomena that cannot be observed with long-pulse lasers, such as almost no thermal damage,~\cite{Chichkov_1996, Shaheen_2013} depth of less than 1 nm,~\cite{Hashida_1999,Hashida_2002,Miyasaka_2012} and emission of high-energy ions.~\cite{Miyasaka_2012,Hashida_2010,Dachraoui_2006,Dachraoui_2006_2}
This ablation, which cannot be explained under the assumption of thermal equilibrium, is referred to as non-thermal ablation, whose effects have been reported to be dominant near the ablation threshold fluence.~\cite{Hashida_2002,Miyasaka_2012,Hashida_2010,Chichkov_1996,Shaheen_2013,Momma_1996}
Although tremendous efforts using both experimental and theoretical approaches have been devoted to elucidating the physical mechanism of the non-thermal ablation of metals, discrepancies exist  between experiments and previous theoretical simulations.

Molecular dynamics (MD) simulation is a powerful computational tool to elucidate the microscopic mechanisms of metal ablation.
To date, MD simulation of laser ablation  in the low-laser-fluence region has been reported for several metals [aluminum (Al),~\cite{Wu_2013} silver (Ag),~\cite{Ji_2017} copper  (Cu),~\cite{Foumani_2018,Schafer_2002} gold (Au),~\cite{Zhiglei_2003} nickel (Ni),~\cite{Zhiglei_2003} and platinum (Pt)~\cite{Rouleau_2014}].
These calculation results proposed the following explanation for the ablation mechanisms of metals caused by irradiation with ultrashort laser pulse with low laser fluence.
With irradiation by ultrashort laser pulse near the ablation threshold fluence, the laser-deposited energy raises the surface temperature so that the surface starts to expand and begins to melt.
Subsequently, tensile stress occurs near the surface region, and as a result,  a molten surface layer is spalled, whose thickness is more than $10\,\text{nm}$.~\cite{Zhigilei_2009,Ji_2017,Foumani_2018,Schafer_2002,Gan_2009,Zhiglei_2003}
This ablation process is called spallation, and has been observed in experiments.~\cite{Linde_1997,Sokolowski_1998,Sokolowski2_1998,Libde_2000,Rouleau_2014}
As the  laser fluence increases, the thickness of the spalled layer decreases, and eventually small clusters and atoms are emitted from the overheated surface.~\cite{Wu_2013}
This ablation process is called phase explosion, and the main cause of this process is considered to be the thermodynamic instability of the overheated surface.~\cite{Miotello_1999,Miotello_2001}

Previous MD simulation studies have argued that isolated atoms are not emitted with irradiation by ultrashort laser pulse near the ablation threshold, and that the phenomenon that occurs near the ablation threshold is spallation.
This means that these explanations  have a fatal problem in describing the metal ablation induced by ultrashort laser pulses near the ablation threshold fluence, since the emission of high-energy ions and sub-nanometer depth ablation have been experimentally observed  in this fluence region.~\cite{Hashida_1999,Hashida_2002,Miyasaka_2012,Hashida_2010,Dachraoui_2006,Dachraoui_2006_2}

It is considered that this discrepancy between MD simulations and experiments comes from a lack of  physical mechanisms in previous MD simulations, where the force acting on atoms is assumed to not be changed even in a highly excited laser-irradiated system. 
Based on this consideration, some physical mechanisms have been proposed to explain the process of the non-thermal ablation of metals.~\cite{Tao_2014, Li_2015,Norman_2012,Norman_2013,Stegailov_2015,Stegailov_2016,Ilnitsky_2016,Miyasaka_2012,Dachraoui_2006,Dachraoui_2006_2}
One of the most famous ones is the Coulomb explosion (CE), which has been experimentally verified in the case of a semiconductor~\cite{Zhao_2013} an insulator,~\cite{Stoian_2000,Stoian_2000_2} and a molecular system.~\cite{Sato_2008}
CE describes the physical mechanism of non-thermal ablation as follows.
Under intense laser irradiation, electrons are emitted from a laser-irradiated surface due to the photoelectric effect and/or the thermionic emission process so that strong Coulomb interaction occurs between positively charged ions at the ionized surface.
Hence,  when the Coulomb interaction is strong enough to overcome the bonding forces between these ions, they are emitted from the surface.
If CE plays a dominant role in the laser ablation process, the peak velocity of the emitted ions is scaled by the valence of the emitted ions, which has been observed by time-of-flight experiments in a semiconductor,~\cite{Zhao_2013} insulator,~\cite{Stoian_2000} and a molecular system.~\cite{Sato_2008}
These observations have been regarded as conclusive evidence of CE in these materials.
On the other hand, the peak velocity of the emitted Cu ions is not scaled by the valence of the ions.~\cite{Zhao_2013}
In addition, other experimental result~\cite{Li_2011} showed that the electric field  near the surface created by the laser irradiation is shielded within the time duration of the probe pulse  ($200\,\text{fs}$),  and this fast electrostatic shielding  is expected to be natural because the inverse of the plasma frequency is very fast ($< 1\,\text{fs}$) in bulk Cu.
This confirms the consideration based on the continuum model (CM) calculation~\cite{Lin_2012} that the electric field near surface due to the electron emission is shielded by high-mobility electrons in the bulk metal before the CE can occur.
Hence, the validity of the CE in metals is questionable.
Besides CE, other possible origins of the non-thermal ablation of metals have been proposed,  for example,  the kinetic energy of free electrons and changes in the charge distribution.~\cite{Norman_2012,Norman_2013,Stegailov_2015,Stegailov_2016,Ilnitsky_2016}
However,  the validity of these explanations is still under debate.

Recently, we have shown by finite-temperature density functional theory (FTDFT) calculations  that the laser-irradiated bulk metal becomes unstable due to the electronic entropy effect.~\cite{Tanaka_2018} 
This result suggests that the non-thermal ablation of metals is induced by the electronic entropy effect, and this explanation for the non-thermal ablation of metals is called the electronic entropy-driven (EED) mechanism.~\cite{Tanaka_2018} 
Based on the EED mechanism, we have developed a CM where the well-known two-temperature model (TTM)~\cite{Anisimov_1973} and the electronic entropy effect are incorporated, and succeeded in  quantitatively describing the experimental ablation depth~\cite{Colombier_2005, Nielsen_2010}  in the low-laser-fluence region.

To further discuss  the validity of the EED mechanism and  investigate the effect of electronic entropy on the non-thermal ablation of metals, MD simulation is preferred over CM simulation since  it can directly describe atom emission and sub-nanometer scale ablation, which are characteristic of non-thermal ablation.~\cite{Hashida_1999,Hashida_2002,Miyasaka_2012,Hashida_2010,Dachraoui_2006,Dachraoui_2006_2}  
Previously, to elucidate the microscopic mechanism of laser-irradiated metals, a two-temperature model combined with molecular dynamics (TTM-MD) scheme has been employed.~\cite{Murphy_2015,Murphy_2016,Daraszewicz_2013}
However, the previous TTM-MD scheme is not appropriate for a system in which the electronic entropy  makes a large contribution, and to our knowledge,  there have been no TTM-MD schemes that satisfy the law of conservation of energy in such a system.
Therefore, to carry out reliable TTM-MD simulation of the non-thermal ablation of metals, where the electronic entropy effect is proposed to be large,~\cite{Tanaka_2018}  it is necessary to develop a new TTM-MD scheme.

The purpose of this study was twofold.
The first was to develop a  TTM-MD scheme that satisfies the law of conservation of energy even in a system where electronic entropy effects make a dominant contribution.
The other was bridging the discrepancy between experiment and previous theoretical simulations regarding the explanation of the non-thermal ablation of metals by elucidating the effect of electronic entropy in these phenomena.

The outline of this paper is as follows.
In Sec.~\ref{sec:method},  the developed TTM-MD scheme and computational details  are explained. 
Owing to this development, the TTM-MD simulation can be performed while satisfying the law of conservation of energy even in a system where electronic entropy effects are large.
In Sec.~\ref{sec:results},  it is firstly shown that  the law of conservation of energy is satisfied with reasonable accuracy in the developed TTM-MD simulation.
Subsequently, calculation results for the ultrashort-pulse laser ablation of a Cu film using the developed TTM-MD simulation are exhibited.
Here,  the microscopic mechanisms of the metal ablation and the effect of the electronic entropy  are investigated.
To confirm the validity of the TTM-MD simulation and the EED mechanism,  the ablation depth in the TTM-MD simulation is compared with previous calculations~\cite{Tanaka_2018}  and experimental data.~\cite{Colombier_2005,Nielsen_2010}
Finally, a brief conclusion is provided in  Sec.~\ref{sec:conc}.


\section{Calculation Methods}
\label{sec:method}
\subsection{Two-temperature model (TTM)}
\label{sec:TTM}

\begin{figure}[b]
  \begin{center}
    \begin{tabular}{cc}

      \begin{minipage}{1.0\hsize}
        \begin{center}
          \includegraphics[clip, width=7cm]{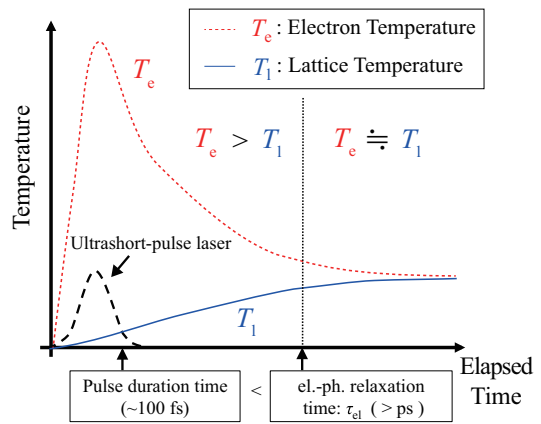}
       
        \end{center}
      \end{minipage} \\
	
   \end{tabular}
    \caption{Schematic image of the main concept of the TTM and  the time development of $T_e$ and $T_l$.
                 } 
    \label{fig: timescale}
  \end{center}
\end{figure}

Fig.~\ref{fig: timescale} represents a schematic image of the two-temperature model (TTM),~\cite{Anisimov_1973}  which has been widely used to describe laser-irradiated systems.~\cite{Daraszewicz_2013,Ernstorfer_2009,Giret_2011,Norman_2012,Norman_2013,Recoules_2006,Inogamov_2012,Wang_2018,Murphy_2015,Murphy_2016}
Ultrashort-pulse laser irradiation of a metal surface changes the electron subsystem (ES) from the ground state into excited states by the absorption of single or multiple photons.
The ES is thermalized to the Fermi-Dirac distribution with the electron temperature $T_e$ via the electron-electron (el.-el.) interaction, of which the scattering time $\tau_{ee}$ is approximately $10 \mathchar`-100\,\text{fs}$ in metals.~\cite{Mueller_2013, Brown_2016_2}

In this time scale, the ES and the lattice subsystem (LS) do not reach equilibrium with each other, so $T_e$ is higher than the  lattice  temperature $T_l$.
Ordinarily, the maximum $T_e$ reaches values more than 10 times higher  than the final equilibrium temperature ($T_e \approx T_l$) since the heat capacity of an electron is very much smaller than that of the lattice.
$T_l$ begins to increase by energy transfer from the ES via electron-phonon (el.-ph.) scattering, for which the relaxation time $\tau_{el}$ is larger than several picoseconds.~\cite{Schoenlein_1987,Elsayed-Ali_1987,Elsayed-Ali_1991,Hohlfeld_2000}
Therefore, under the assumption of the instantaneous and local thermalization in the ES and the LS,
the ultrashort-laser-irradiated metals can be described by $T_e > T_l$  before $\tau_{el}$.
This explanation is the main concept of the TTM.
Based on the TTM, many previous studies~\cite{Daraszewicz_2013,Ernstorfer_2009,Giret_2011,Norman_2012,Norman_2013,Recoules_2006,Inogamov_2012,Wang_2018} have been successful in description of the experimental data.


\subsection{Two-temperature model combined with molecular dynamics (TTM-MD) scheme}
\label{sec:TTM-MD}
\begin{figure}[tbp]
  \begin{center}
    \begin{tabular}{cc}

      \begin{minipage}{1.0\hsize}
        \begin{center}
          \includegraphics[clip, width=8cm]{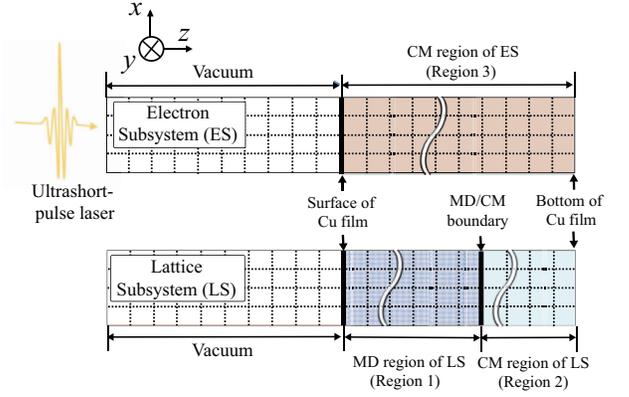}
          \hspace{0.6cm}
       
        \end{center}
      \end{minipage} \\
	
   \end{tabular}
    \caption{Schematic image of TTM-MD scheme used to simulate the laser-irradiated Cu film. 
             The laser comes from the left side of the figure.
             The local electron temperature $T_e^n$ and the local lattice temperature $T_l^n$  are defined for the  $n$-th 3D cell (in dotted region). 
              Periodic boundary conditions are used in the $x$, $y$ directions (parallel to the surface). 
             The free boundary condition is used at the bottom of the MD region of LS (Region 1).
             In the MD region of LS, the atomic dynamics are calculated  using MD simulation.
             On the other hand, to reduce calculation cost, the time development of $T_l^n$ in the CM region of LS (Region 2) is calculated  using the CM.
             The time development of all $T_e^n$ is calculated  using the CM.
              }
    \label{fig:TTM-MD}
  \end{center}
\end{figure}

Here, we explain a newly developed calculation scheme for simulating the atom dynamics of metal ablation caused by irradiation with an ultrashort-pulse laser.
In the scheme, the MD scheme is hybridized with the TTM scheme to express the non-equilibrium state between the ES  and the LS.
To decrease the computational cost of large-scale atomistic simulations, the CM is partly employed in LS as well as in ES.
Fig.~\ref{fig:TTM-MD} represents a schematic image of this calculation scheme.
Hereafter in this paper, this calculation scheme is called the TTM-MD scheme.

In the TTM-MD scheme, electronic effects,  such as a highly excited ES near the surface, electronic thermal diffusion, electron-phonon scattering, and energy absorption due to the electronic entropy effect, are incorporated into the MD simulation through the TTM.
In the TTM-MD simulation, atom dynamics are calculated based on the MD scheme, and at the same time, other time developments such as that of $T_e$ are calculated by employing the TTM.
For reduction of computational cost, the CM is also used to calculate the time development of $T_l$ deep inside the Cu film (Region 2 in Fig.~\ref{fig:TTM-MD}).
The dynamics of each atom in this region is not as dominant in the atom dynamics of laser ablation, so only the time development of $T_e$ and $T_l$ are calculated in this region.  
This region is called the CM region of LS and plays an important role in the thermal dissipation of the energy deposited by laser irradiation.
On the other hand, the region near the surface in which atoms exist is called the MD region of LS (Region 1 in Fig.~\ref{fig:TTM-MD}).
With the volume change due to expansion or ablation,  the position of the surface and the CM region change during simulation.
Periodic boundary conditions are used in the $x$-axis and $y$-axis directions in Fig.~\ref{fig:TTM-MD}.
The free boundary condition is employed  between  the CM and the MD regions of LS.
 
The local electron temperature $T_e^n$  and the local lattice temperature $T_l^n$ are defined in three-dimensional (3D) cells, where $n$ is the index of the 3D cells.
A region surrounded by dotted lines in  Fig.~\ref{fig:TTM-MD} represents one of the 3D cells.
Although $T_l^n$ is referred to as the local ``lattice'' temperature, we do not imply that a crystalline structure is assumed in the 3D cells.
In other words, $T_l^n$ represents not only the lattice temperature but also the temperature of the atoms.
Besides, it is noted that $T_l^n$ of the MD region of LS represents the instantaneous temperature of atoms.  

The time development of $T_e^n$ is calculated by solving the following nonlinear differential equation:
\begin{align}
 C_e^n \frac{dT_e^n}{dt} = {\nabla } \cdot & (\kappa _e^n {\nabla }T_e^n)  - G^n (T_e^n - T_l^n)  \notag  \\
                                        & - {\sum}_i^{N^n} \bm{v}_i \frac{{\partial}}{{\partial}\bm{r}_i} \left(  S^nT_e^n \right)  + I^n,  \label{eq:electron} 
\end{align}
where $\bm{r}_i$ and  $\bm{v}_i$ are the position and the velocity of atom $i$  in the $n$-th cell, respectively.
$C_e^n$ is the electronic heat capacity, 
$\kappa _e^n$ is the electronic thermal conductivity, 
$G^n$ is the electron-phonon heat transfer constant, 
$N^n$ is the number of atoms, 
$S^n$ is the electronic entropy,
and $I^n$ is the energy deposited by laser irradiation at each $n$-th 3D cell.
These quantities are calculated at each 3D cell by the following equations: 
\begin{subequations}
\begin{align}
 C_e^n &=  \frac{N^n}{N_0} C_e (T_e^n),  \label{eq:Cen} \\
 G^n &=  \frac{N^n}{N_0} G,  \label{eq:Cen} \\
 \kappa _e^n &= \frac{N^n}{N_0} \kappa _e (T_e^n, T_l^n), \label{eq:kappan} \\ 
  I^n &= \frac{N^n}{N_0} I (T_e^n,T_l^n), \label{eq:depo} \\ 
 N_0 &=\rho _0 V_c.  \label{eq:numn}
\end{align}
\end{subequations}
Here, $\rho _0$ is the bulk density in the equilibrium states and $V_c$ is the volume of each 3D cell. 
$C_e (T_e^n)$, $G$, $\kappa_e (T_e^n,T_l^n)$, and $I(T_e^n,T_l^n)$ represent each physical property per unit volume.
The values for these properties are the same as those used in the previous study,~\cite{Tanaka_2018} whose details are explained in the Supplemental Material.
The third term on the right-hand side of Eq.~(\ref{eq:electron})  represents the absorption of energy by the electronic entropy.
The derivation of Eq.~(\ref{eq:electron})  based on the law of conservation of energy is explained in Sec.~\ref{sec:convergedquantity}.

$T_l^n$ in the CM and the MD regions of LS is calculated by solving  Eqs.~(\ref{eq:lattice}) and~(\ref{eq:instantaneoustemp}), respectively:
\begin{subequations}
\begin{align}
 C_l^n \frac{\text{d}T_l^n}{\text{d}t} &=  G^n(T_e^n - T_l^n),     \label{eq:lattice} \\
 T_l^n &= \frac{1}{3k_BN^n} \sum_i^{N^n} \left(\bm{v}_i - \bm{v}^{n}_c\right)^2. \label{eq:instantaneoustemp}
\end{align}
\end{subequations}
Here, $k_B$ is the Boltzmann constant, $\bm{v}_c^n$ is the average velocity of atoms (center-of-mass velocity) in the $n$-th 3D cell, while $C_l^n=  \frac{N^n}{N_0} C_l $ is the lattice heat capacity in the $n$-th 3D cell. 
$C_l $ is the lattice heat capacity per unit volume, and details are also explained in Supplemental Material.

The atomic dynamics in the MD region of LS are calculated by solving the following equations:
\begin{subequations}
\begin{align}
 \frac{d \bm{r}_i}{dt}&=\bm{v}_i, \label{eq:motionx1} \\
m \frac{d \bm{v}_i}{dt}&=-\frac{{\partial}F^n}{{\partial}\bm{r}_i} - m {\xi}^n \bm{v}_i. \label{eq:motionv1}
 \end{align}
\end{subequations}
Here, $m$ is the mass of an atom and  ${\xi}^n$ is a coefficient that represents the force deriving from the electron-phonon interaction.
$F^n$ of Eq.~(\ref{eq:motionv1}) is the free energy of the ES in the $n$-th 3D cell, and the definition is given in the following equation:
\begin{eqnarray}
F^n = E^n -S^n T_e^n. \label{eq:def_free}
\end{eqnarray}
Here, $E^n$ represents the internal energy.
In this study, $F^n$ and $E^n$  are calculated using the $T_e$-dependent inter-atomic potential (IAP), which is based on the embedded atom method (EAM) potential.
The functional form and parameter values for the $T_e$-dependent IAP of Cu were proposed in a previous study.~\cite{Tanaka_2021}
The previous study reported that this $T_e$-dependent IAP can reproduce the FTDFT results of $T_e$-dependent physical properties, such as the volume dependence of $F_n$ and  $E_n$, and the phonon dispersion.
Moreover,  MD simulations using the  $T_e$-dependent IAP  quantitatively reproduce the results of MD simulation using FTDFT; for example, the time development of the elastic properties of nano-scale slabs and an ablation threshold $T_e$.


\subsection{Law of conservation of energy }
\label{sec:convergedquantity}

In this study, the developed TTM-MD simulations were performed to investigate the microscopic mechanism of metal ablation induced by irradiation with an ultrashort-pulse laser.
Although the $T_e$-dependent IAP is used in some previous TTM-MD simulations,~\cite{Daraszewicz_2013,Murphy_2015,Murphy_2016} the law of conservation of energy is considered not to be satisfied, the reason for which  is explained below. 
In these simulations, because the laser has too small a fluence to cause ablation, the effect of electronic entropy may not be very large and deviation from the law of conservation of energy might be negligible. 
On the other hand, the electronic entropy effect is proposed to be dominant in metal ablation induced by irradiation with an ultrashort-pulse laser.~\cite{Tanaka_2018}
In this case we must carefully take the electron entropy effect into account to realize energy conservation in the TTM-MD simulation.
In the following Sec.\ref{sec:convergedquantity}, it is shown that the law of conservation of energy is satisfied in the developed TTM-MD scheme, theoretically.

First,  to simplify the situation, the laser-deposited energy and the energy flow among the 3D cells are neglected.
In other words, only energy exchange between the ES and the LS in the 3D cells is considered.
In this situation,  the conserved energy of  the 3D cells is the internal energy: $E^n+\sum_i^{N^n}\frac{1}{2}m\bm{v}_i^2$.

The time derivative of the conserve energy can be calculated easily as
\begin{align} 
  \frac{d}{dt}  & \left( E +  \sum_i^{N} \frac{1}{2}  m  \bm{v}_i ^2  \right) \notag & \\
  & =    \frac{d T_e }{d t}   \frac{\partial E}{\partial T_e}  + \sum_i^{N}  \frac{d \bm{r}_{i}}{dt} \frac{\partial E}{{\partial } \bm{r}_{i}}  +  \sum_i^{N}  \bm{v}_{i} \left( m \frac{d\bm{v}_i }{dt} \right)  \notag\\
       &= \frac{d T_e }{d t}   \frac{\partial  E}{\partial T_e}   + \sum_i^{N}  \bm{v}_{i} \frac{\partial E}{{\partial }\bm{r}_{i}}  -  \sum_i^{N}  \bm{v}_{i} \left[  \frac{\partial (E-ST_e )}{\partial \bm{r}_i} + m\xi  \bm{v}_i \right] \notag \\
       & = C_e  \frac{d T_e }{d t}  + \sum_i^{N}  \bm{v}_{i} \frac{\partial }{{\partial} \bm{r}_{i}} \left(  ST_e \right) - \sum_i^{N} m  \xi  \bm{v}_{i}^2 . \label{eq:econv1}
                 \end{align}
Here, to simplify notation, the 3D cell index $n$ is omitted.
In the second equality, Eqs.~(\ref{eq:motionx1}), (\ref{eq:motionv1}), and ({\ref{eq:def_free}}) are used.
In the third equality, the definition of the electronic heat capacity $C_e(T_e) = {\partial E(T_e)}/{\partial T_e}$ is used.
Since the time derivative of the conserved quantity is 0, the following equation can be derived:
\begin{eqnarray}
   C_e  \frac{d T_e }{d t}   = -  \sum_i^{N}   \bm{v}_{i} \frac{\partial}{{\partial}\bm{r}_{i}} \left(  ST_e \right)    + \sum_i^{N} m \xi \bm{v}^2_{i}.  \label{eq:econv2}
\end{eqnarray}
The first term on the right-hand side of this equation represents the absorbed energy due to the electronic entropy and the second term is the exchange energy due to the electron-phonon interaction.
It is a fundamental assumption in TTM that the electron-phonon interaction is represented by a single linear coupling term of the form of $G^n (T_e^n - T_l ^n)$.~\cite{Hohlfeld_2000}
Previously, based on this assumption, the value of $G^n$ for Cu has been investigated by experiment~\cite{Elsayed-Ali_1987} and theoretical calculations,~\cite{Lin_2008,Migdal_2016,Migdal_2016,Petrov_2013,Migdal_2015} for which  the details are explained in the Supplemental Material.
Therefore, for energy conservation with respect to the electron-phonon interaction between ES and LS, the following equation must be satisfied:
\begin{eqnarray}
  \sum_i^{N^n} m  \xi ^n \bm{v}_{i}^2  + G^n(T_e^n - T_l^n) = 0.  \label{eq:econv4}
\end{eqnarray}

Subsequently, we added the effect of the electronic thermal diffusion energy $D^n_{\text{tot}}$ and laser-deposited energy of the $n$-th 3D cell $I^n_{\text{tot}}$  to this scenario.
The former effect can be expressed as
\begin{eqnarray}
      D^n_{\text{tot}}   =  - \int {\nabla }\cdot(\kappa _e^n{\nabla }T_e^n) dt,  \label{eq:Dtot}
\end{eqnarray}
and the latter effect can be written as
\begin{eqnarray}
        I^n_{\text{tot}}   = \int I^n  dt. \label{eq:Itot} 
\end{eqnarray}
In this situation, the conserved energy in each 3D cell is  $E^n+\sum_i^{N_n}\frac{1}{2}mv_i^2 + D^n_{\text{tot}}  - I^n_{\text{tot}} $.
Therefore, the time derivative of the conserved energy can be written as
\begin{align}
  \frac{d}{dt} & \left( E^n +  \sum_i^{N^n} \frac{1}{2}  m \bm{v}_i^2 + D^n_{\text{tot}}  - I^n_{\text{tot}}   \right) \notag  \\
  & =  C_e^n \frac{d T_e^n }{d t}    + \sum_i^{N^n}  \bm{v}_{i} \frac{\partial}{{\partial} \bm{r}_{i}} \left(  S^nT_e^n \right) - \sum_i^{N^n} m  \xi ^n \bm{v}_{i}^2  \notag \\
      & \hspace{10mm} -  {\nabla }\cdot(\kappa _e^n {\nabla }T_e^n) - I^n \notag  \\
   & = C_e^n \frac{d T_e^n }{d t}   + \sum_i^{N^n}  \bm{v}_{i} \frac{\partial }{{\partial} \bm{r}_{i}}\left(  S^nT_e^n \right)  + G^n(T_e^n - T_l^n)  \notag \\
   & \hspace{10mm} -  {\nabla }\cdot(\kappa _e^n {\nabla }T_e^n) - I^n. \label{eq:econv3}
                 \end{align}
Here,  Eqs.~(\ref{eq:econv1}), (\ref{eq:Dtot}), and (\ref{eq:Itot}) are used in the first equality, and Eq.~(\ref{eq:econv4}) is used in the second equality.
Since the time derivative of the conserved quantity is 0,  Eq.~(\ref{eq:electron}) can be derived using Eq.~(\ref{eq:econv3}).
Consequently, Eq.~(\ref{eq:electron}) is derived based on the law of conservation of energy.

\begin{figure}[tp]
  \begin{center}
    \begin{tabular}{c}

      \begin{minipage}{1.0\hsize}
        \begin{center}
          \includegraphics[clip, width=8cm]{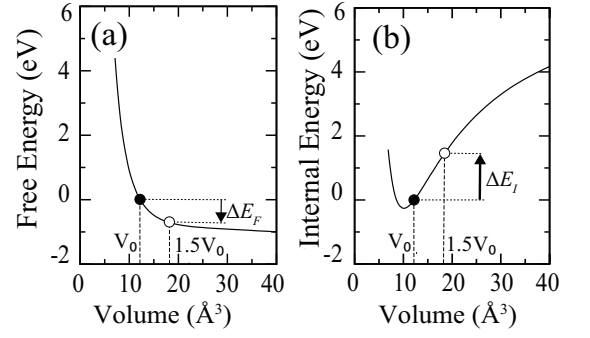}
       
        \end{center}
      \end{minipage} \\
	
   \end{tabular}
    \caption{Schematic image of the energy absorbed by  the electronic  entropy.
    Volume dependence of (a) the free energy and (b) the internal energy at $T_e = 25,000\,\text{K}$,  which are calculated using FTDFT.
    Calculation conditions are the same as those used in the previous study.~\cite{Tanaka_2018} 
    The filled and blank circles are the energies at the equilibrium volume ($V_0$) and $1.5\,V_0$, respectively.
    $\Delta E_F$ represents the difference between the free energy  at $V_0$ and that at $1.5\,V_0$.
    Also, $\Delta E_I$ represents the difference between the internal energy  at $V_0$ and that at $1.5\,V_0$.
    When the volume changes from $V_0$ to $1.5\,V_0$, these energies should be absorbed from the ES as the electronic entropy effect to accelerate atoms and to raise the internal energy surface. 
              }
    \label{fig:image_pre}
  \end{center}
\end{figure}

In previous studies,~\cite{Daraszewicz_2013,Murphy_2015,Murphy_2016} forces acting on the atoms  were calculated by the spatial derivative of the free energy calculated using the $T_e$-dependent IAP, and the energy exchange due to the electron-phonon interaction was considered.
However, the absorbed energy due to the electronic entropy effect was ignored.
This means that, in conventional simulations, the time development of $T_e$ is calculated by the following equation:
 \begin{eqnarray}
  C_e^n \frac{dT_e^n}{dt} = {\nabla }\cdot(\kappa _e^n {\nabla }T_e^n) - G^n(T_e^n - T_l^n)  + I^n.  \label{eq:conv_Te}
 \end{eqnarray}
Hence, energy that is used  to accelerate atoms and to raise the internal energy surface is supplied from the virtual electron thermal bath  [Fig.~\ref{fig:image_pre}], because the third term on the right-hand side of Eq.~(\ref{eq:electron}) is ignored in the conventional TTM-MD scheme [Eq.~(\ref{eq:conv_Te})].
In this study, we developed the TTM-MD scheme by adding the $- {\sum}_i^{N^n} \bm{v}_i  \frac{{\partial}}{{\partial}\bm{r}_i}  \left(  S^nT_e^n \right)$ term to the equation of the conventional TTM-MD scheme, which enabled us to perform simulations that satisfy the law of conservation of energy even in a system where the electronic entropy effects are large.

    \subsection{Calculation Conditions}
\label{sec:calculation_detail}

Here, we explain the calculation conditions. 
The lateral dimensions of a laser-irradiated Cu film are $3.615\,\text{nm}\times3.615\,\text{nm}$, which is  ten times  the lattice constant of the conventional unit cell of the face-centered cubic (fcc) structure of Cu.
The initial MD and CM regions of LS are about $361.5\,\text{nm}$ and $638.5\,$nm, respectively.
Hence, the thickness of the computational Cu film is  $1\, \mu$m.
The total number of atoms in the computational cell is about $4.0\times 10^5$.
The surface of the film is a (001)  free surface of the fcc structure.
The laser pulse shape is assumed to be Gaussian.
The pulse duration times of an ultrashort-pulse laser and a ps-pulse laser are $100\,\text{fs}$ and  $200\,\text{ps}$, respectively.
The size of the 3D cells is  $1.205\,\text{nm} \times 1.205\,\text{nm} \times 1.205\,\text{nm}$.
Therefore, the space step $\Delta x_{\text{CM}}$ is  $1.205\,\text{nm}$.

Eqs.~(\ref{eq:electron}) and~(\ref{eq:lattice}) are solved by a finite difference method (FDM).
To solve Eqs.~(\ref{eq:motionx1}) and (\ref{eq:motionv1}), the velocity Verlet algorithm is used.
A value of the time step  $ {\Delta}t$  is  $10\,\text{as}$.
This value is much shorter than the time step for ordinary  MD simulations.
To reduce the calculation cost, the time step for MD calculation ${\Delta t}_{\text{MD}}$ is set to ${\Delta t}_{\text{MD}} = n_{\text{MD}} {\Delta} t$, where $ n_{\text{MD}}$ is an integer number.
In Sec.~\ref{sec:Test_TTM-MD}, we determine a suitable time step $\Delta t_{\text{MD}}$ so that the law of conservation of energy is satisfied with little error.

In our simulations, before irradiation of a laser pulse on the Cu film,  the computational cell was relaxed  using the Nos$\acute{\text{e}}$-Hoover thermostat~\cite{Hoover_1985} at $300\,\text{K}$ for $800\,\text{ps}$, where $\Delta t_{\text{MD}}=5\,\text{fs}$ was used. 
Details of the calculation flow of the TTM-MD  are explained in the Supplemental Material.

\section{Results and Discussion}
\label{sec:results}

\subsection{Conservation of energy}
\label{sec:Test_TTM-MD}

\begin{figure}[b]
  \begin{center}
    \begin{tabular}{c}

      \begin{minipage}{1.0\hsize}
        \begin{center}
          \includegraphics[clip, width=8cm]{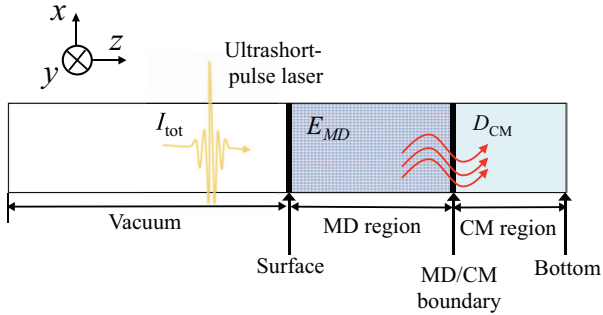}
          \hspace{1.6cm}
       
        \end{center}
      \end{minipage} \\
	
   \end{tabular}
    \caption{Schematic image of the conserved energy of the MD region ($E_{\text{cons}}$) whose definition is $E_{\text{cons}} = E_\text{MD}  + D_{\text{CM}} -I_{\text{tot}}$.
    Here, $E_\text{MD}$,  $D_{\text{CM}}$, and  $I_{\text{tot}}$ represent the internal energy of the MD region, the energy thermally diffusing to the CM region, and the energy deposited on the Cu film by the laser, respectively.
              }
    \label{fig:converge_image}
  \end{center}
\end{figure}

\begin{figure}[tbp]
  \begin{center}
    \begin{tabular}{cc}

      \begin{minipage}{1.0\hsize}
        \begin{center}
          \includegraphics[clip, width=6cm]{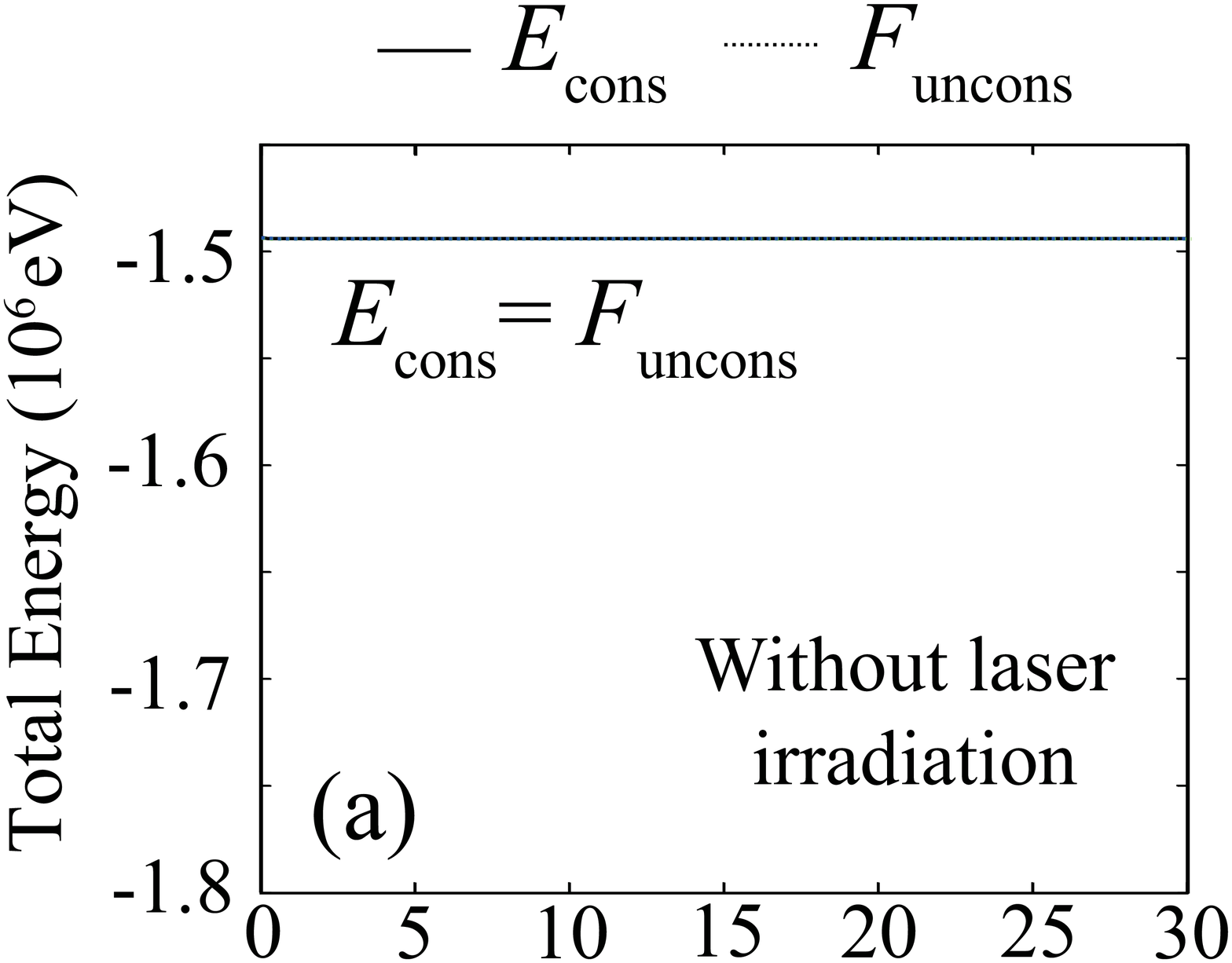}
        \hspace{0.8cm}
       
        \end{center}
      \end{minipage} \\
	
      \begin{minipage}{1.0\hsize}
        \begin{center}
          \includegraphics[clip, width=6cm]{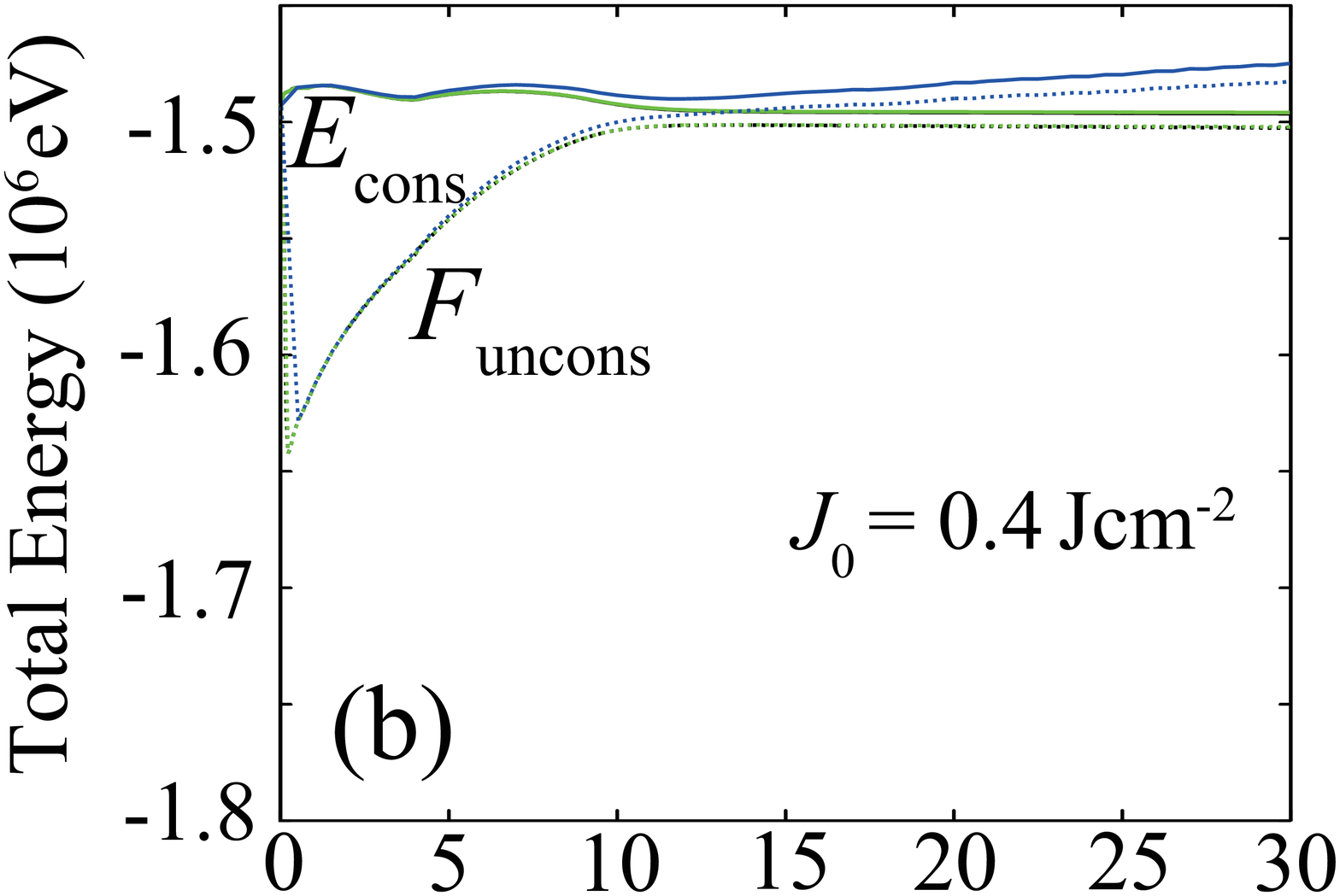}
         \hspace{0.8cm}
       
        \end{center}
      \end{minipage} \\
      
      \begin{minipage}{1.0\hsize}
        \begin{center}
          \includegraphics[clip, width=6cm]{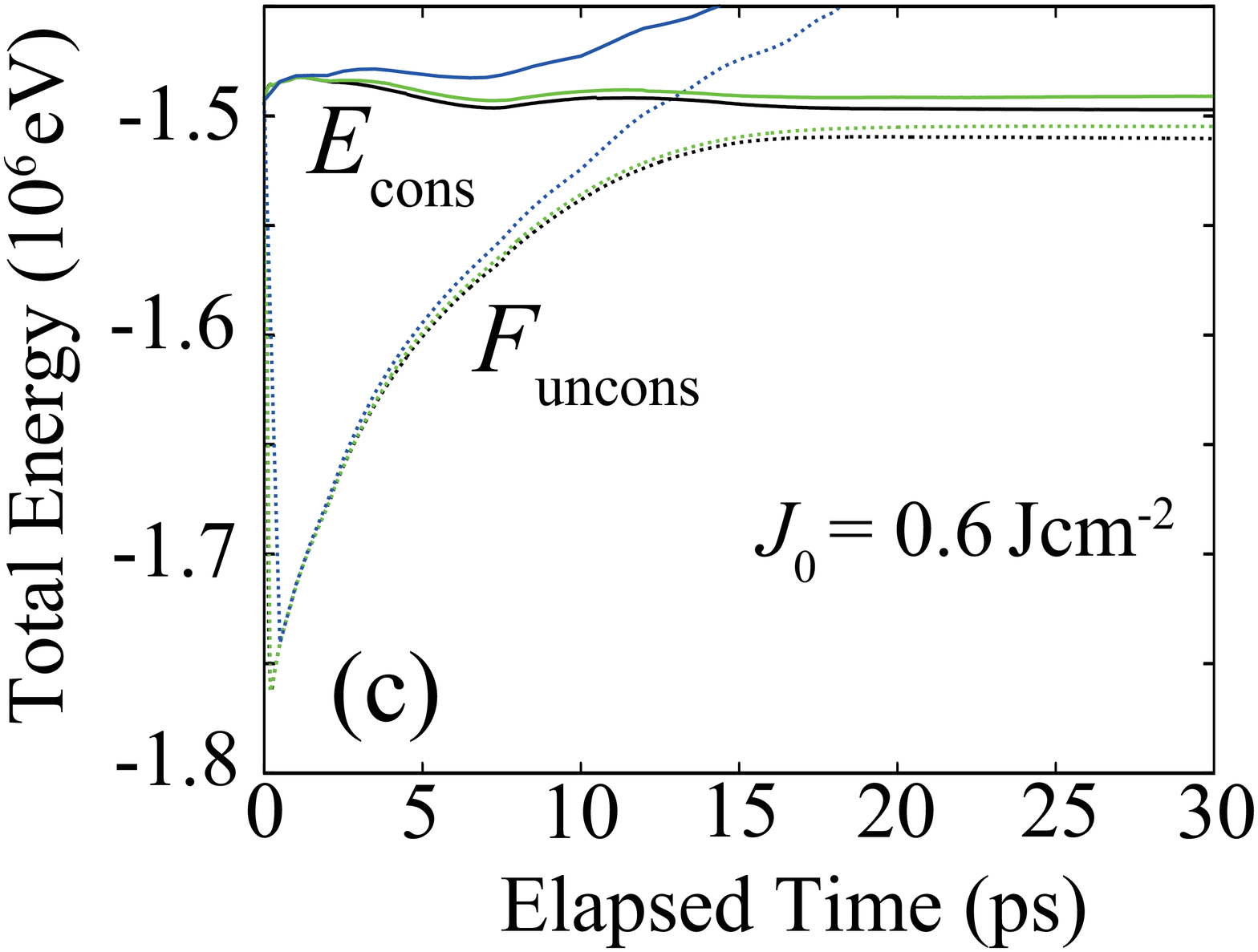}
         \hspace{0.8cm}
       
        \end{center}
      \end{minipage} \\

   \end{tabular}
    \caption{Time development of $E_{\text{cons}}$ (solid line)  and $F_{\text{uncons}}$  (dotted line).
    (a) Calculation results of TTM-MD simulations in which a laser is not applied.
    Calculation results for the Cu film irradiated by an ultrashort-pulse laser with (b) $J_0=0.4\,\text{J}\,\text{cm}^{-2}$, where ablation is not caused, and with (c) $J_0=0.6\,\text{J}\,\text{cm}^{-2}$, where ablation is caused. 
Black, green, and blue lines represent calculation results of $t_{\text{MD}} = 0.5, 1.0$, and $5.0\,\text{fs}$, respectively.
The total number of atoms in the computational cell is approximately $4.0\times10^{5}$.
              }
    \label{fig:e_converge}
  \end{center}
\end{figure}

Here, it is shown that the developed TTM-MD scheme satisfies the law of conservation of energy with a small error.
In addition, we show calculation results of the $\Delta t_{\text{MD}}$ dependence of the conserved energy, which  is investigated to choose an appropriate $t_{\text{MD}}$ for the following simulations.

Fig.~\ref{fig:converge_image} represents a schematic image of the conservation of energy  for the MD region ($E_{\text{cons}}$), which is calculated to investigate whether the developed TTM-MD simulation satisfies the law of conservation of energy.
$E_{\text{cons}}$ can be written as follows:
 \begin{eqnarray}
 E_{\text{cons}} = E_\text{MD} + D_{\text{CM}} -I_{\text{tot}}. \label{eq:E_cons}
 \end{eqnarray}
Here, $E_\text{MD}$, $D_{\text{CM}}$, and  $I_{\text{tot}}$ represent  the internal energy of the MD region, the energy thermally diffusing to the CM region, and the  energy deposited on the Cu film by the laser, respectively.  
The internal energy of the MD region is expressed as $ \sum_n^{\text{MD\,cells}} ( E^n+\sum_i^{N^n}\frac{1}{2}m\bm{v}_i^2 )$, where the first summation is taken over all MD cells.
For comparison, we calculate the free energy of the MD region ($ F_{\text{uncons}}$), which is regarded as the conserved energy in the conventional TTM-MD scheme.
The definition of $ F_{\text{uncons}}$ is as follows:
 \begin{eqnarray}
   F_{\text{uncons}} =  E_{\text{cons}} - \sum_n^{\text{MD\,cells}} S^n T_e^n. \label{eq:S_cons}
 \end{eqnarray}

Fig.~\ref{fig:e_converge} represents the $\Delta t_{\text{MD}}$ dependence of  $E_{\text{cons}}$ (solid line)  and $F_{\text{uncons}}$  (dotted line)  in TTM-MD simulations.
Black, green, and blue lines represent calculation results using $t_{\text{MD}} = 0.5, 1.0$, and $5.0\,\text{fs}$, respectively.
Fig.~\ref{fig:e_converge}(a) represents results for the TTM-MD simulation for $30\,\text{ps}$ without laser irradiation.
Figs.~\ref{fig:e_converge}(b) and (c) represent the results of TTM-MD simulations where the Cu film is irradiated by the ultrashort-pulse laser of  (b) $J_0=0.4\,\text{J}\,\text{cm}^{-2}$ and by the laser of (c) $J_0=0.6\,\text{J}\,\text{cm}^{-2}$.
Here, $J_0$ represents the laser fluence.
Ablation does not occur in (b) the former case; on the other hand, ablation occurs in (c) the latter case.
In these two simulations, $T_e$ near the surface increases to approximately  $20,000\,\text{K}$.

Fig.~\ref{fig:e_converge} shows that when sufficiently small $\Delta t_{\text{MD}}$ is used,  our simulations satisfy the law of conservation of energy with error of several $10\,\text{meV}\,\text{atom}^{-1}$, and that $F_{\text{uncons}} $ is not conserved.
Since Figs.~\ref{fig:e_converge}(b) and (c) show that $E_{\text{cons}}$ returns back to the initial value at $t>20\,\text{ps}$, where low electron temperatures ($T_e \simeq 1,000\,\text{K}$) are realized, these errors of $E_{\text{cons}}$ exist only at high $T_e$.
The previous study~\cite{Tanaka_2018} showed that the electronic heat capacity ($C_e$) calculated using the $T_e$-dependent IAP  is slightly overestimated compared to that calculated by FTDFT.
In this study, $E^n$ is calculated using the $T_e$-dependent IAP; on the other hand, the time development of $T_e^n$ is calculated according to Eq.~(\ref{eq:econv1}), where $C_e$ is calculated by FTDFT. 
Therefore, $E^n$ calculated by the $T_e$-dependent IAP is expected to be over-estimated, which can be shown in the calculation results of the laser-irradiated system [Figs.~\ref{fig:e_converge}(b) and (c)].
Hence,  to decrease the error for the law of conservation of energy, it is necessary to develop $T_e$-dependent IAP that can reproduce the electronic specific heat of FTDFT with higher accuracy.

Fig.~\ref{fig:e_converge}(a) shows that $E_{\text{cons}}$  is conserved with little or no error $\Delta t_{\text{MD}}$.
On the other hand, Figs.~\ref{fig:e_converge}(b) and (c) represent that small $\Delta t_{\text{MD}}$ is needed to conserve $E_{\text{cons}}$ when the ultrashort-pulse laser is applied.
The reason that energy conservation is not satisfied in  the long time step $\Delta t_{\text{MD}} = 5.0\,\text{fs}$ can be attributed to high-velocity atoms accelerated by laser irradiation.

In all TTM-MD simulations shown in the following, appropriate $t_{\text{MD}}$ are used after verifying whether $E_{\text{cons}}$ is conserved in each simulation.
We carry out the TTM-MD simulations using $\Delta t_{\text{MD}}=1.0\,\text{fs}$ when the irradiation laser fluence is $J_0<0.9\,\text{J}\,\text{cm}^{-2}$,
and  $\Delta t_{\text{MD}}=0.5\,\text{fs}$ is used in the TTM-MD simulations when the irradiation laser fluence is $J_0\ge0.9\,\text{J}\,\text{cm}^{-2}$.

\subsection{Microscopic mechanisms of metal ablation}

\begin{figure*}[t]
  \begin{center}
    \begin{tabular}{c}

      \begin{minipage}{1.0\hsize}
        \begin{center}
          \includegraphics[clip, width=12cm]{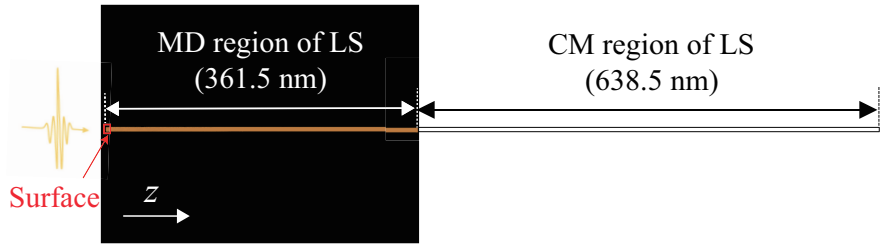}
       
        \end{center}
      \end{minipage} \\
	
   \end{tabular}
    \caption{Schematic image of the computational cell for the TTM-MD simulations.
    A laser pulse comes from the left side of the figure.
    The $z$ direction represents the depth direction of the Cu film.
    Periodic boundary conditions are employed in the directions parallel to the surface.
    The CM region  is connected to the MD region on the right  (see Fig.~\ref{fig:TTM-MD}).
    In addition to this figure, the following snapshots of atomic configurations are visualized using the Open Visualization Tool~\cite{OVITO} (OVITO).
                  }

    \label{fig:compcell}
  \end{center}
\end{figure*}

In this section,  results and analyses of the TTM-MD simulations of  the ultrashort-pulse laser ablation are described.
The computational cell for the TTM-MD simulations is shown in Fig.~\ref{fig:compcell}.
The laser comes from the left side of the Cu film, which consists of MD and CM regions (see Figs.~\ref{fig:TTM-MD} and \ref{fig:compcell}).
Fig.~\ref{fig:compcell} and  snapshots of the atomic configurations are visualized  using Open Visualization Tool~\cite{OVITO} (OVITO). 
The calculation results shown in this section were obtained under conditions in which  the fluence of the applied laser changed from $0.54$ to $1.00\,\text{J}\,\text{cm}^{-2}$ while its pulse width was fixed at $100\,\text{fs}$.

\subsubsection{Ablation near the ablation threshold: emission of atoms}
\label{sec:results_indep}

 
 \begin{figure}[tbp]
  \begin{center}
    \begin{tabular}{cc}

      \begin{minipage}{0.5\hsize}
        \begin{center}
          \includegraphics[clip, width=4cm]{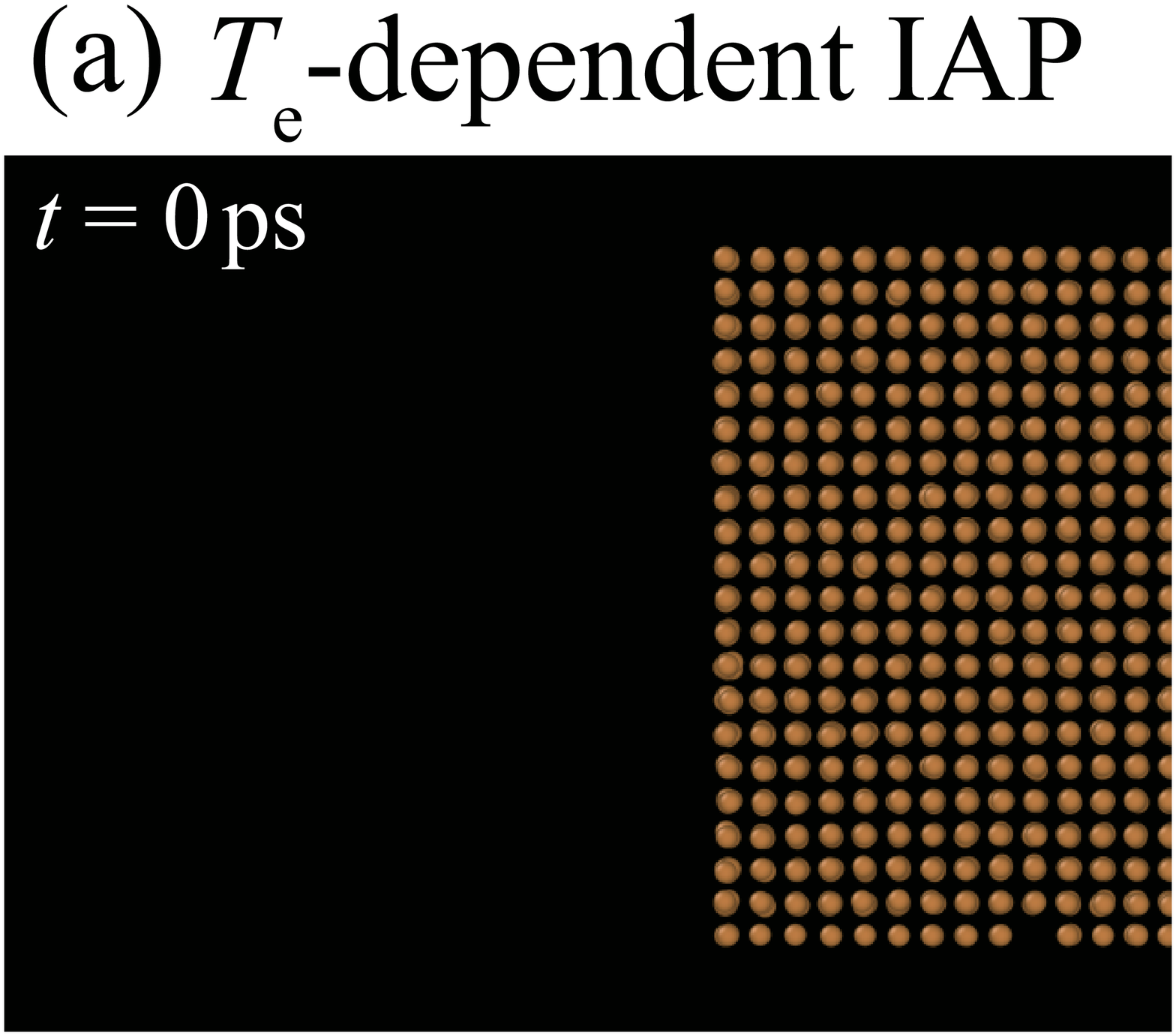}
        \end{center}
      \end{minipage}
      \begin{minipage}{0.5\hsize}
        \begin{center}
          \includegraphics[clip, width=4cm]{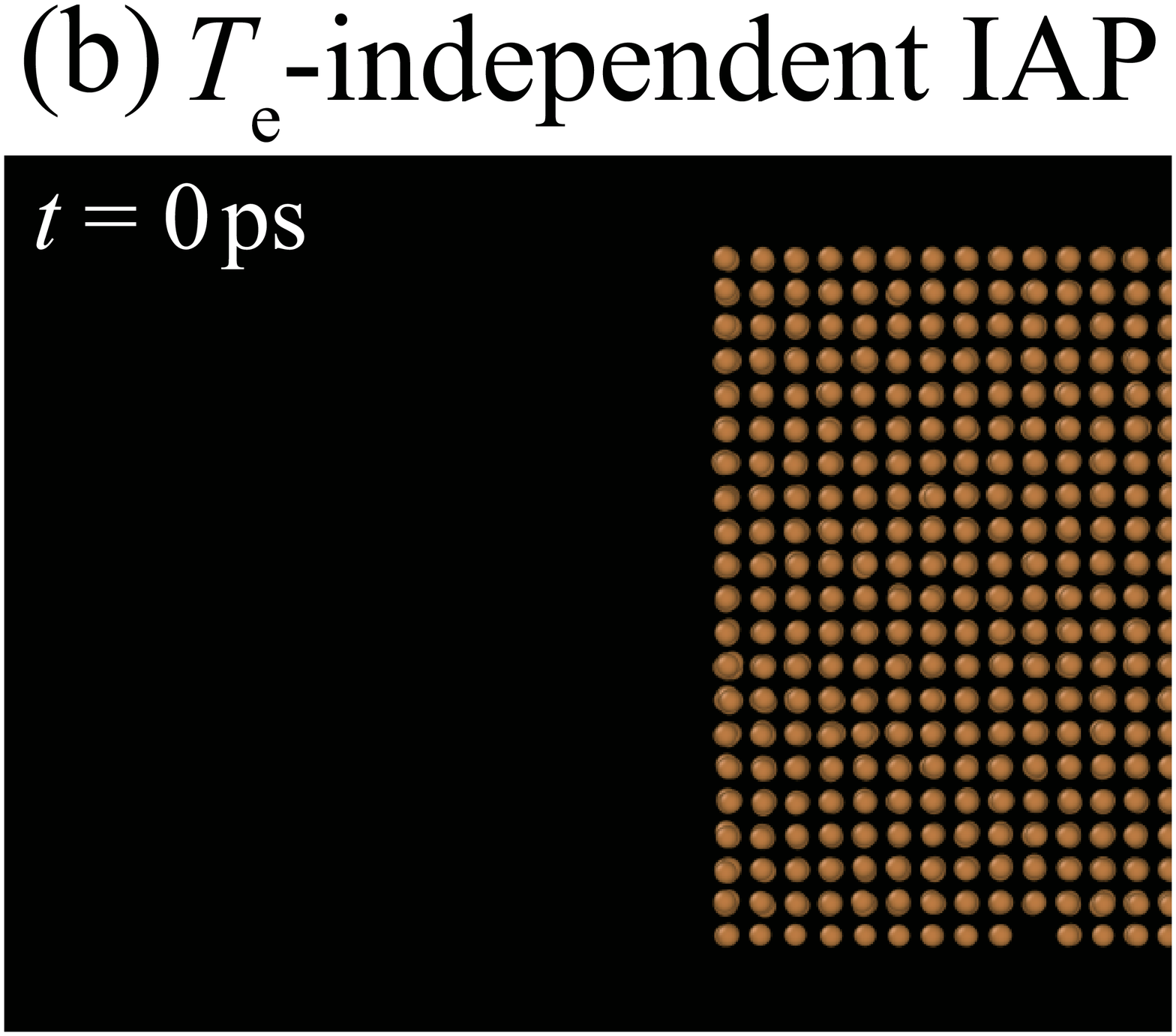}
        \end{center}  
      \end{minipage}
      \\
      \begin{minipage}{0.5\hsize}
        \begin{center}
         \includegraphics[clip, width=4cm]{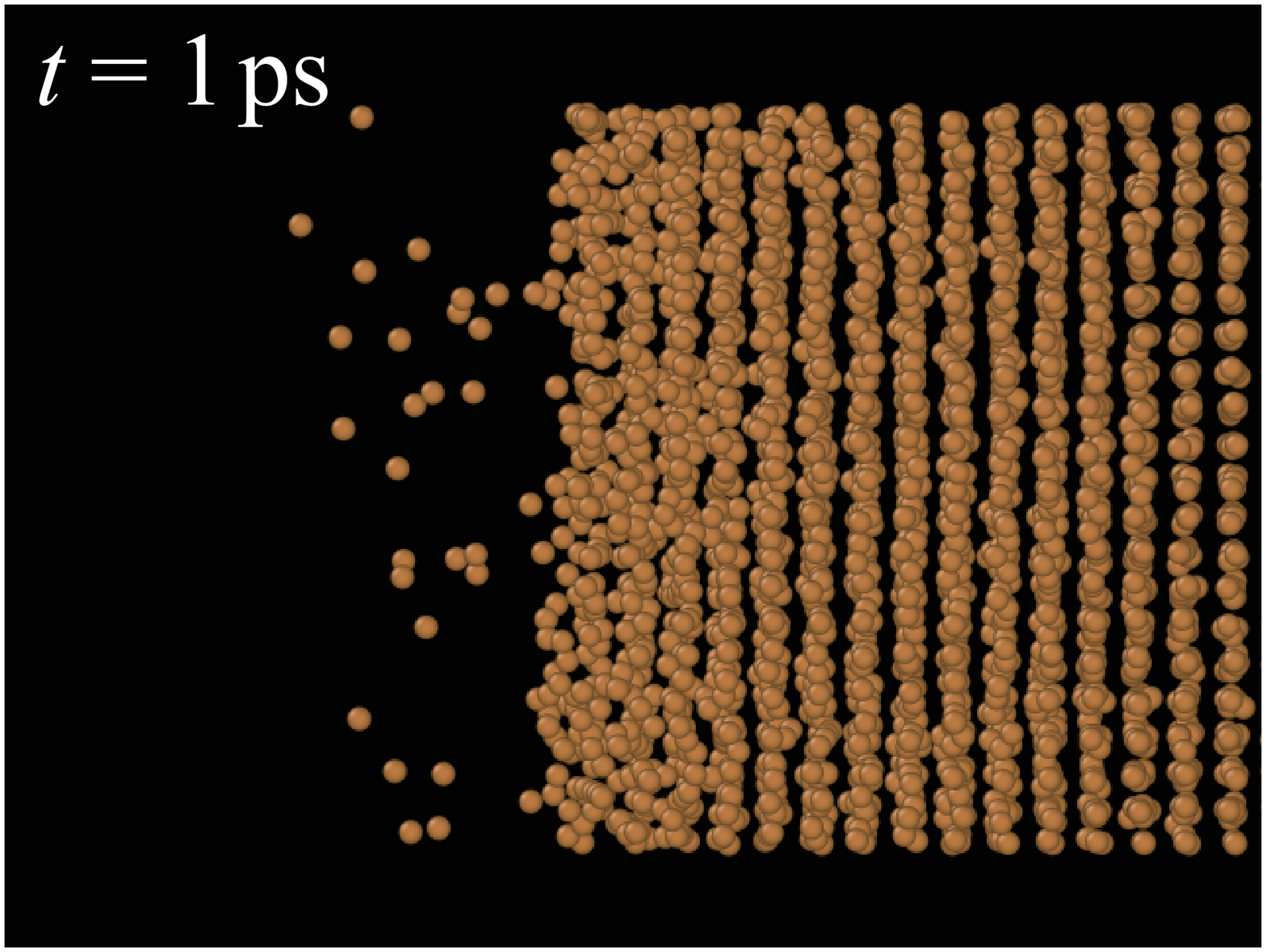}
        \end{center}
      \end{minipage}
      \begin{minipage}{0.5\hsize}
        \begin{center}
          \includegraphics[clip, width=4cm]{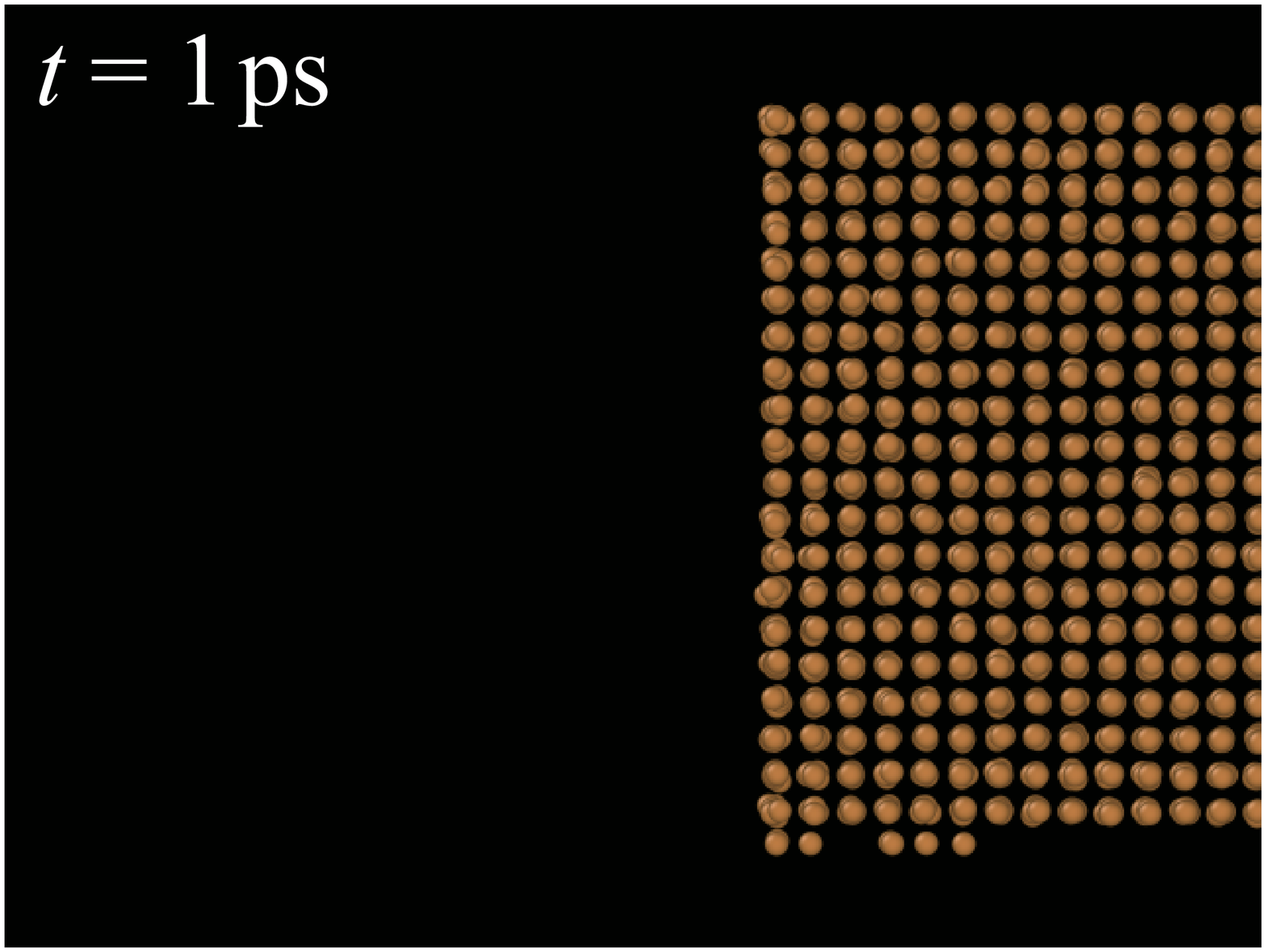}
        \end{center}
      \end{minipage}
     \\
      \begin{minipage}{0.5\hsize}
        \begin{center}
          \includegraphics[clip, width=4cm]{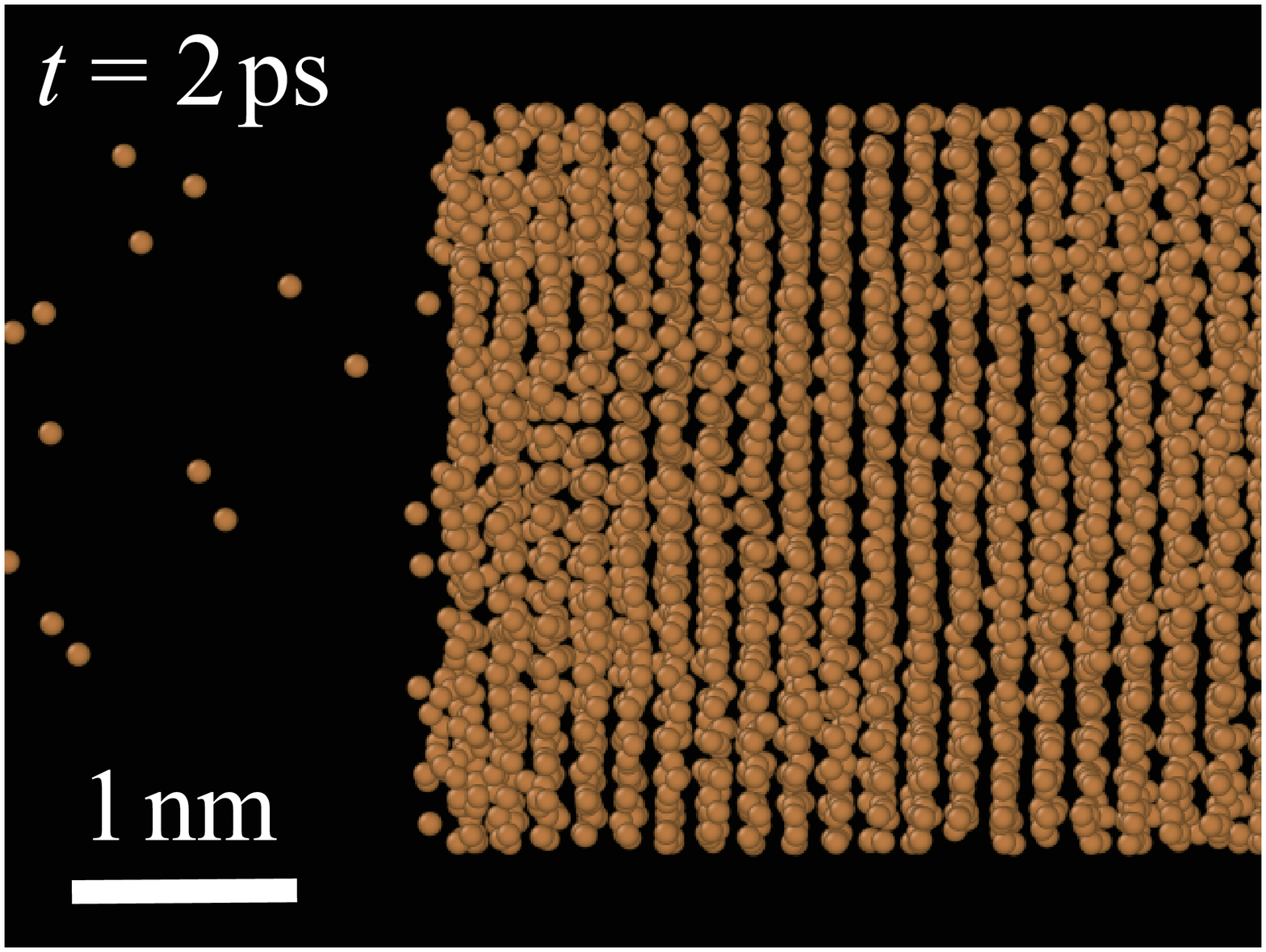}
        \end{center}
      \end{minipage}
      \begin{minipage}{0.5\hsize}
        \begin{center}
          \includegraphics[clip, width=4cm]{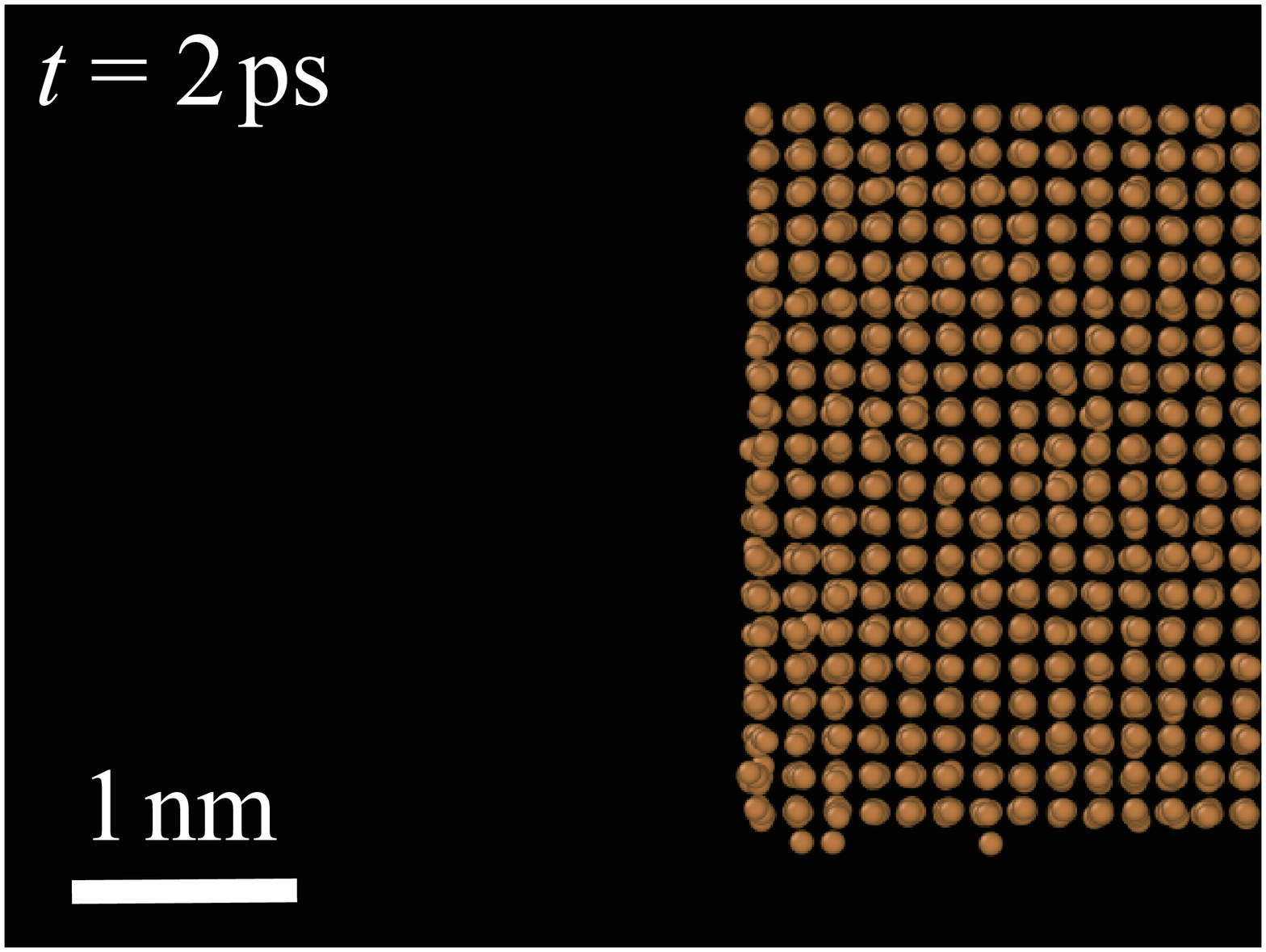}
        \end{center}  
      \end{minipage}

   \end{tabular}
      \caption{Snapshots of atomic configurations near the surface after $0$-$2\,\text{ps}$ irradiation with a  $100\,\text{fs}$ pulse laser of $J_0=0.57\,\text{J}\,\text{cm}^{-2}$.  These simulations are carried out using (a) the $T_e$-dependent IAP and (b) the $T_e$-independent IAP.
             }
    \label{fig:057snaps}
  \end{center}

\end{figure}

 \begin{figure}[tbp]
  \begin{center}
    \begin{tabular}{cc}

      \begin{minipage}{0.5\hsize}
        \begin{center}
          \includegraphics[clip, width=4.5cm]{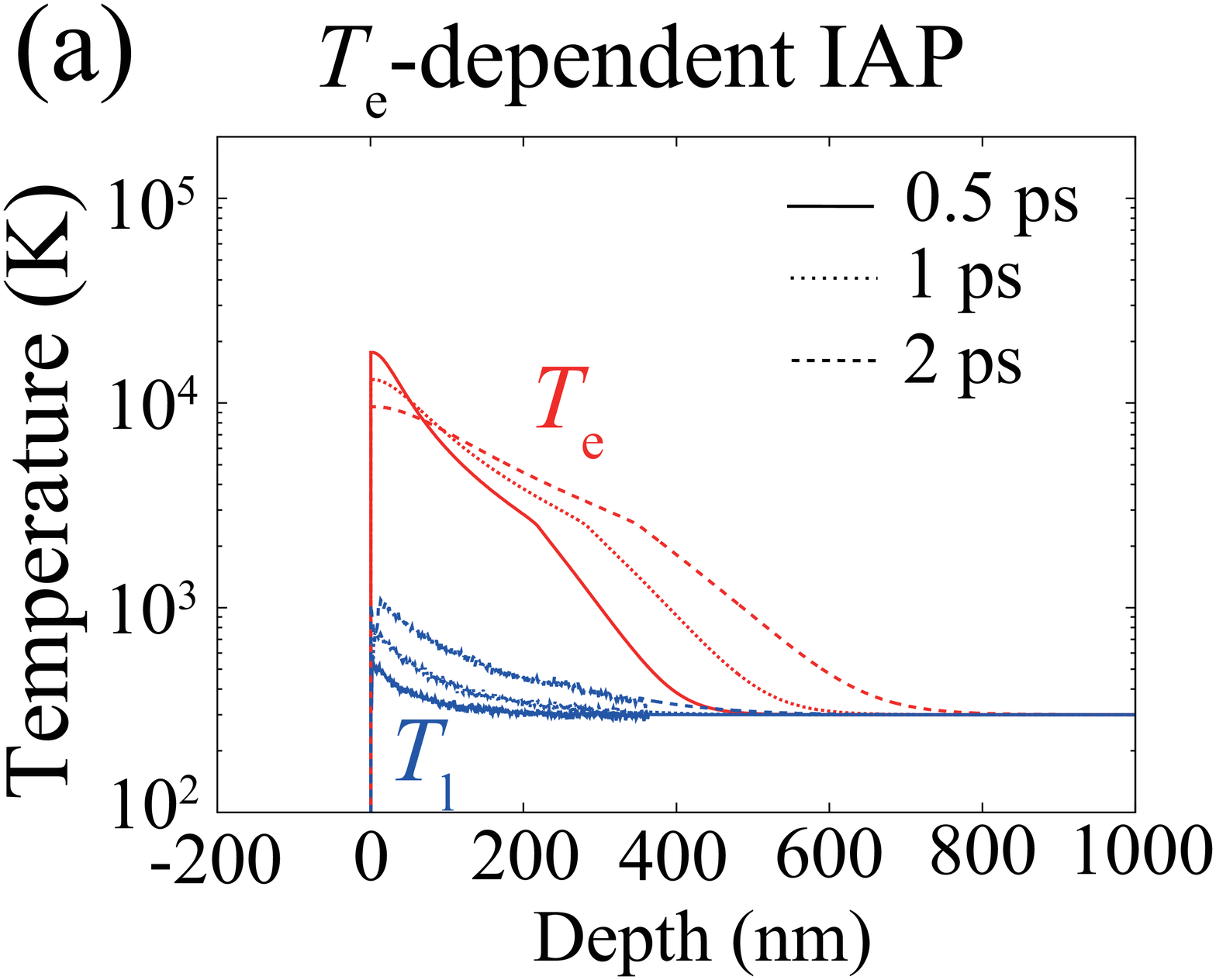}
        \end{center}
      \end{minipage}
      
      \begin{minipage}{0.5\hsize}
        \begin{center}
          \includegraphics[clip, width=4.5cm]{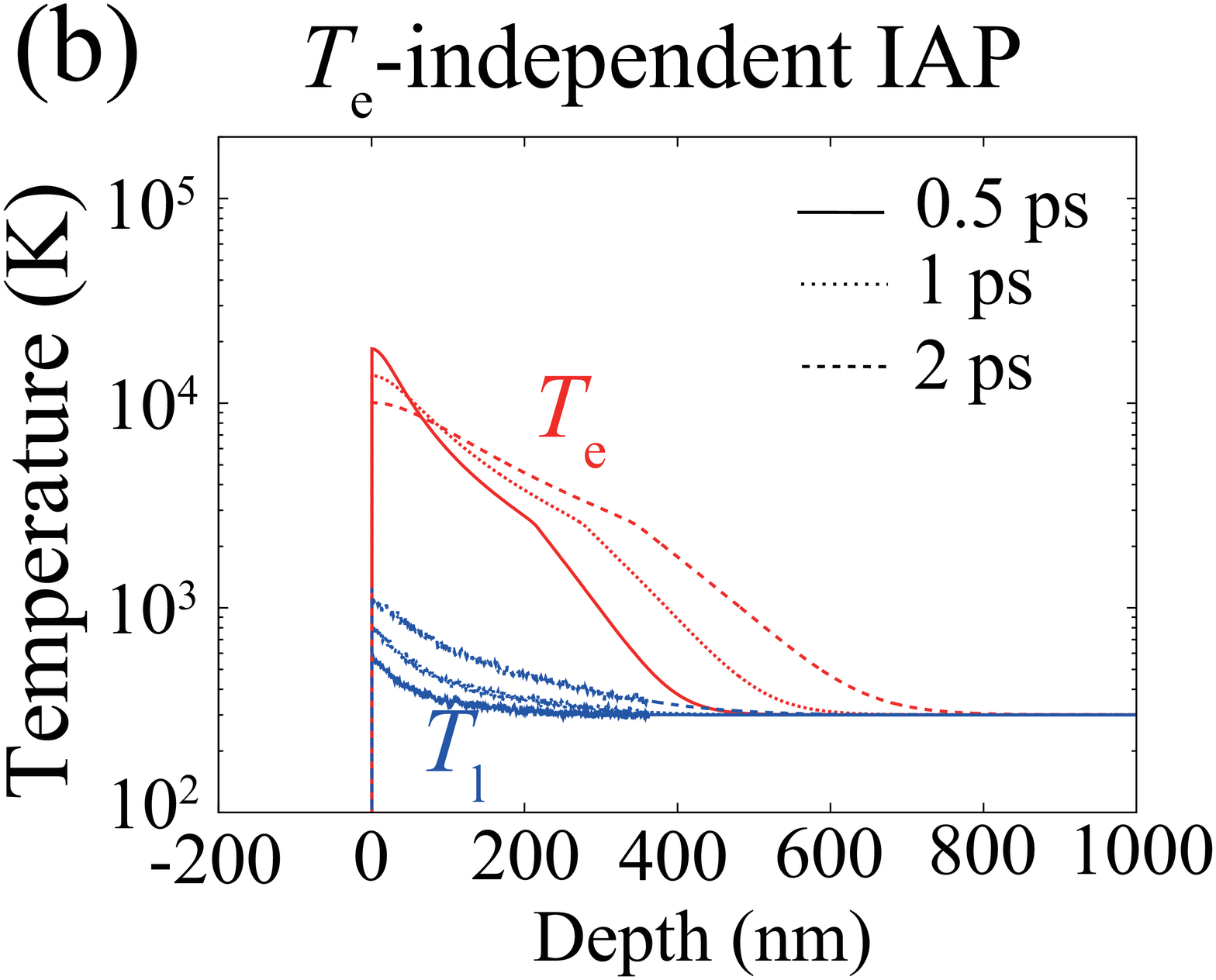}
        \end{center}  
      \end{minipage}
     
   \end{tabular}
      \caption{$T_e$ and $T_l$  space distributions. 
      The irradiated laser fluence is $J_0=0.57\,\text{J}\,\text{cm}^{-2}$.
      These figures represent the simulation results using (a) the $T_e$-dependent IAP and  (b) the $T_e$-independent IAP.
      Solid, dashed, and dotted lines represent the results at $t=0.5$, $1.0$, and $2.0\,\text{ps}$, respectively.
                  }
    \label{fig:057temp}
  \end{center}
\end{figure}

In the TTM-MD simulations using the $T_e$-dependent IAP,~\cite{Tanaka_2021} emission of an atom is observed when the Cu film is irradiated by a  laser pulse with $J_0=0.55\,\text{J}\,\text{cm}^{-2}$.
Whereas, irradiation of the Cu film by a laser with $J_0=0.54\,\text{J}\,\text{cm}^{-2}$ does not cause atom emission.
From these results, the ablation threshold fluence is estimated to be $J_0=0.55\,\text{J}\,\text{cm}^{-2}$, which is about the same as our previous CM simulation results ($J_0=0.47\,\text{J}\,\text{cm}^{-2}$).~\cite{Tanaka_2018}
The kinetic energy of the emitted atom is estimated to be $46.5\,\text{eV}$.
This excessively high atom energy is consistent with the experimental value (about $30\,\text{eV}$~\cite{Hashida_2010}), which is the most probable energy for the Cu$^+$ emitted on irradiation by a laser with the ablation threshold fluence.

\begin{figure}[tbp]
  \begin{center}
    \begin{tabular}{cc}

      \begin{minipage}{1.0\hsize}
        \begin{center}
          \includegraphics[clip, width=5cm]{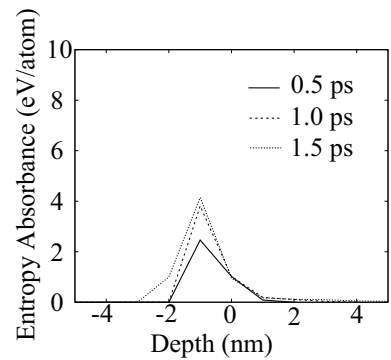}
       
        \end{center}
      \end{minipage} \\
	
   \end{tabular}
    \caption{The energy absorbed by the electronic entropy effect at each depth.
    The irradiated laser fluence is $J_0=0.57\,\text{J}\,\text{cm}^{-2}$.
    Solid, dashed, and dotted lines represent results at $t=0.5$,  $1.0$, and  $1.5\,\text{ps}$, respectively.
    Zero on the $x$-axis represents the initial surface position.
              }
    \label{fig:057entropy}
  \end{center}
\end{figure}

 \begin{figure}[t]
  \begin{center}
    \begin{tabular}{cc}

      \begin{minipage}{0.5\hsize}
        \begin{center}
          \includegraphics[clip, width=4.5cm]{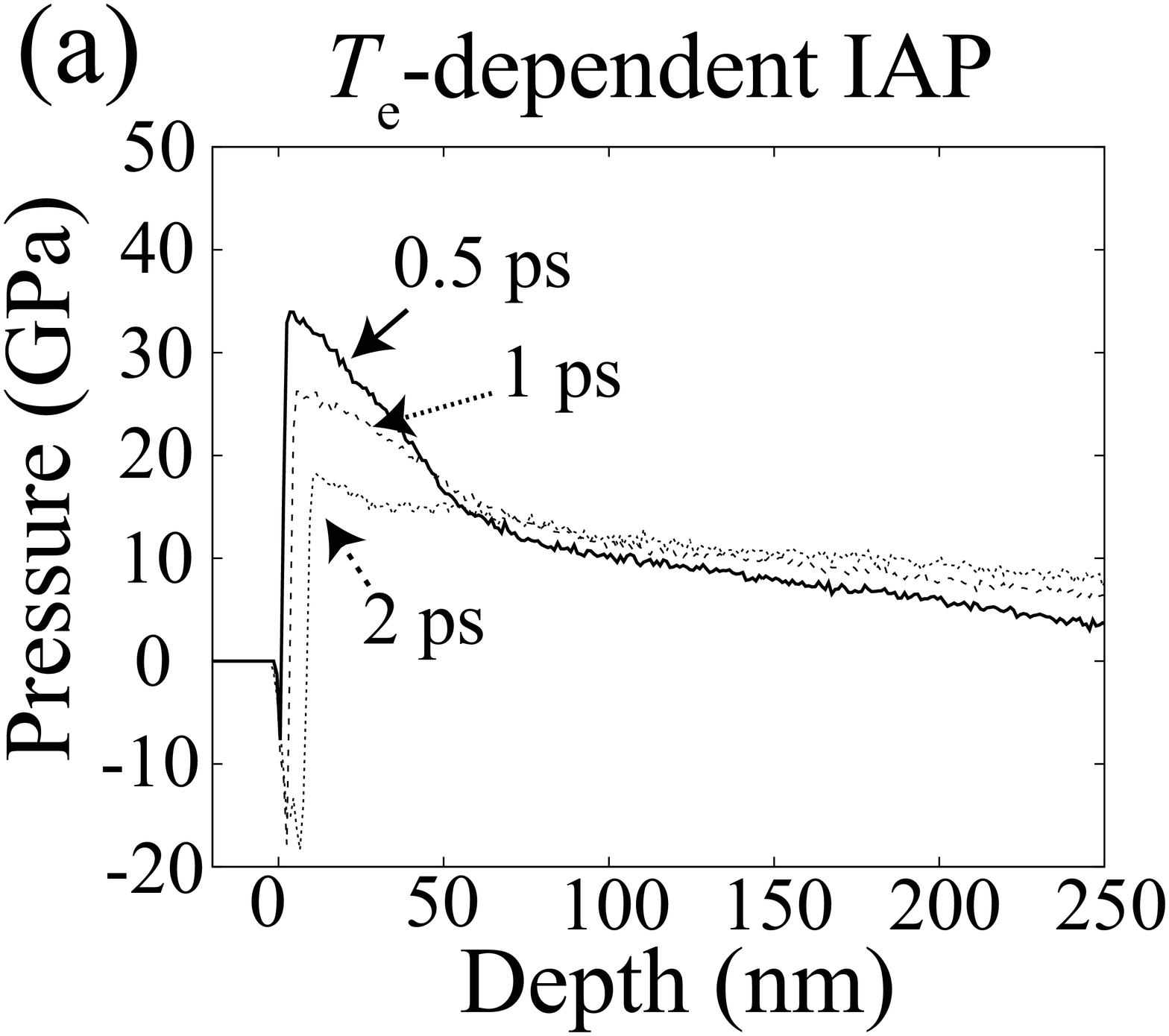}
        \end{center}
      \end{minipage}
      
      \begin{minipage}{0.5\hsize}
        \begin{center}
          \includegraphics[clip, width=4.5cm]{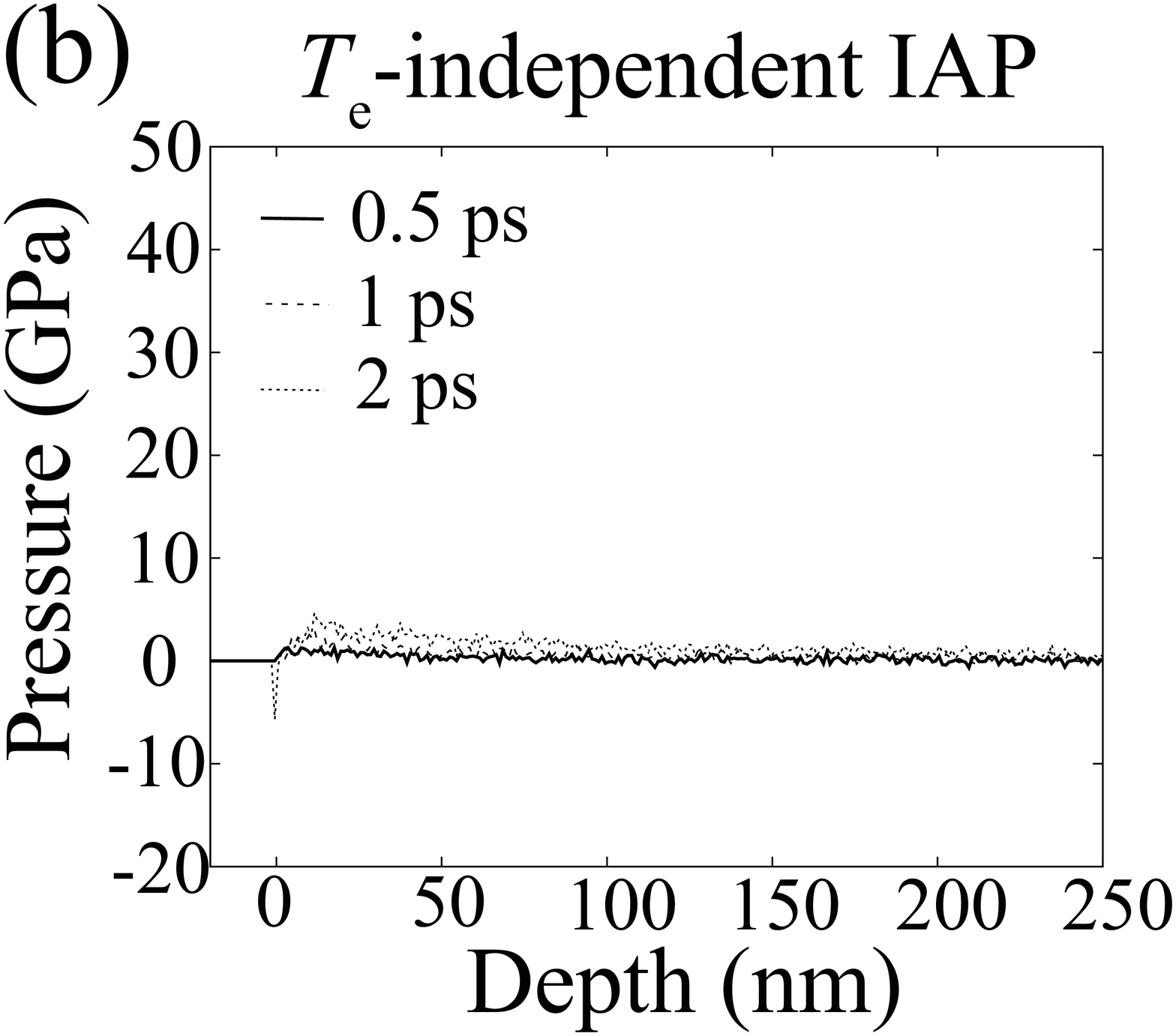}
        \end{center}  
      \end{minipage}
     
   \end{tabular}
      \caption{The spatial distribution of the local pressure along the $z$ direction of simulations using (a) the $T_e$-dependent IAP and  (b) the $T_e$-independent IAP.
      Bold, dashed, and dotted lines represent the results at $t=0.5$, $1.0$, and $2.0\,\text{ps}$, respectively.
            }
    \label{fig:057pres}
  \end{center}
\end{figure}

As shown in Fig.~\ref{fig:057snaps}(a), several atoms are emitted from the surface irradiated by a laser with $J_0=0.57\,\text{J}\,\text{cm}^{-2}$.
These atomic configurations are simulated using the $T_e$-dependent IAP,  in which the electronic entropy effect is incorporated.
TTM-MD simulations with $T_e$-independent IAPs were also performed using $300\,\text{K}$ potential parameters at all temperatures.
Snapshots of the atomic configurations obtained by this simulation are shown in Fig.~\ref{fig:057snaps}(b).
As can be seen from this figure, laser irradiation at $J_0=0.57\,\text{J}\,\text{cm}^{-2}$, which causes atom emission when simulated with the $T_e$-dependent IAP, does not cause atom emission when simulated with the $T_e$-independent IAP.

Fig.~\ref{fig:057temp} shows the spatial distribution of $T_e$ and $T_l$.
As can be seen from Fig.~\ref{fig:057temp}, at $ t = 0$, $1$, and $2\,\text{ps}$ there is little difference between the simulation results using the $T_e$-dependent IAP and that using the $T_e$-independent IAP.
Therefore, the atom emission  cannot be explained by the thermalized kinetic energy of the atoms.
Since previous studies~\cite{Tanaka_2018, Tanaka_2021} show that the internal energy ($E$) becomes more attractive at high $T_e$, it is considered that the origin of the atom emission comes from  the electronic entropy ($-ST_e$) effect, which is reported to induce large repulsion forces at high $T_e$.

The contribution of the electronic entropy effect can be investigated more directly by calculating the energy absorbed due to the electronic entropy effect.
Fig.~\ref{fig:057entropy} represents the energy absorbed due to the electronic entropy term of Eq.~(\ref{eq:def_free}) at each depth.
The solid, dashed, and dotted lines in this figure represent results at  $t=0.5, 1.0,$ and  $1.5\,\text{ps}$, respectively.
From  this figure, it can been seen that the electronic entropy effect is large near the surface.

Furthermore, the electronic entropy effect regarding the pressure is investigated since the laser-induced pressure is considered to be important for the occurrence of spallation and phase explosion.~\cite{Wu_2013}
Fig.~\ref{fig:057pres} represents the distribution of  the local pressure along the $z$ direction ($p_z$).
According to the simple deviation by Basinski $et\,al$.~\cite{Basinski_1971} based on the virial theorem, the local pressure $p^n$ in the $n$-th 3D cell also can be calculated from the following equation:
\begin{align}
p^n & = \frac{1}{3V^n} \notag \\
    \times  &  \left[ < \sum_i^{N^n} m(v_{i})^2 > + < \frac{1}{2}  \sum_{i \ne j} ^{N^n} \sum_j^{N^{\text{tot}}} r_{ij} \cdot f_{ij} > \right].
\end{align}
Here,  $V^n$ and  $N^{\text{tot}}$ are the volume of the $n$-th 3D cell and the total number of atoms, respectively.
$r_{ij}$ and  $ f_{ij}$ are the distance and force between the $i$-th and the $j$-th atoms, respectively.
The bracket means the time average.
In our calculation, the value  of $p^n$ is averaged within $100\,\text{fs}$.
We focus only on the local pressure along the $z$ direction, which is the most important for the ablation dynamics.

Fig.~\ref{fig:057pres}(b) shows that a  pressure of less than $5\,\text{GPa}$  is created in the simulation using the $T_e$-independent IAP.
On the other hand, Fig.~\ref{fig:057pres}(a) shows that a large pressure ($\sim 35\,\text{GPa}$) is created  near the surface in the simulation using  the $T_e$-dependent IAP, and more than  $5\,\text{GPa}$ pressure is created even deep inside the region ($> 200\,\text{nm}$).
The compressive pressure wave created by atom emission and that created by expansion near the surface  would not reach the deep interior region because the velocity of sound of bulk Cu is $4.76\,\text{nm}\,\text{ps}^{-1}$.~\cite{Linde_2003}
Therefore, this high compressive pressure in the deep interior region is considered to arise from the repulsion force between atoms due to the electronic entropy effect. 

As shown in Fig.~\ref{fig:057pres}(a), large negative pressure (tensile stress), which has the potential to induce spallation, is created near the surface ($\sim 5\,\text{nm}$) in the simulation using the $T_e$-dependent IAP.
Although a negative pressure is created, laser irradiation with $J_0=0.57\,\text{J}\,\text{cm}^{-2}$ does not cause spallation even after the compressive pressure wave reaches the MD/CM boundary ($361.5\,\text{nm}$).
Therefore, we thought that a higher negative pressure and a higher temperature were necessary to cause spallation, and we found that ion emission occurred at a lower laser irradiation fluence than the fluence that caused spallation.
From the number of emitted atoms, the ablation depth is estimated to be $0.65\,\text{nm}$ for laser-irradiation with $J_0=0.57\,\text{J}\,\text{cm}^{-2}$.
Not only the ablation at $J_0=0.55\,\text{J}\,\text{cm}^{-2}$ but also this result has the potential to explain the non-thermal ablation of metals where sub-nanometer ablation was observed.~\cite{Hashida_1999,Hashida_2002,Miyasaka_2012}

\subsubsection{Ablation with a laser fluence slightly larger than the ablation threshold: spallation}
\label{sec:results_indep}

 
 \begin{figure}[b]
  \begin{center}
    \begin{tabular}{cc}

      \begin{minipage}{0.5\hsize}
        \begin{center}
          \includegraphics[clip, width=4cm]{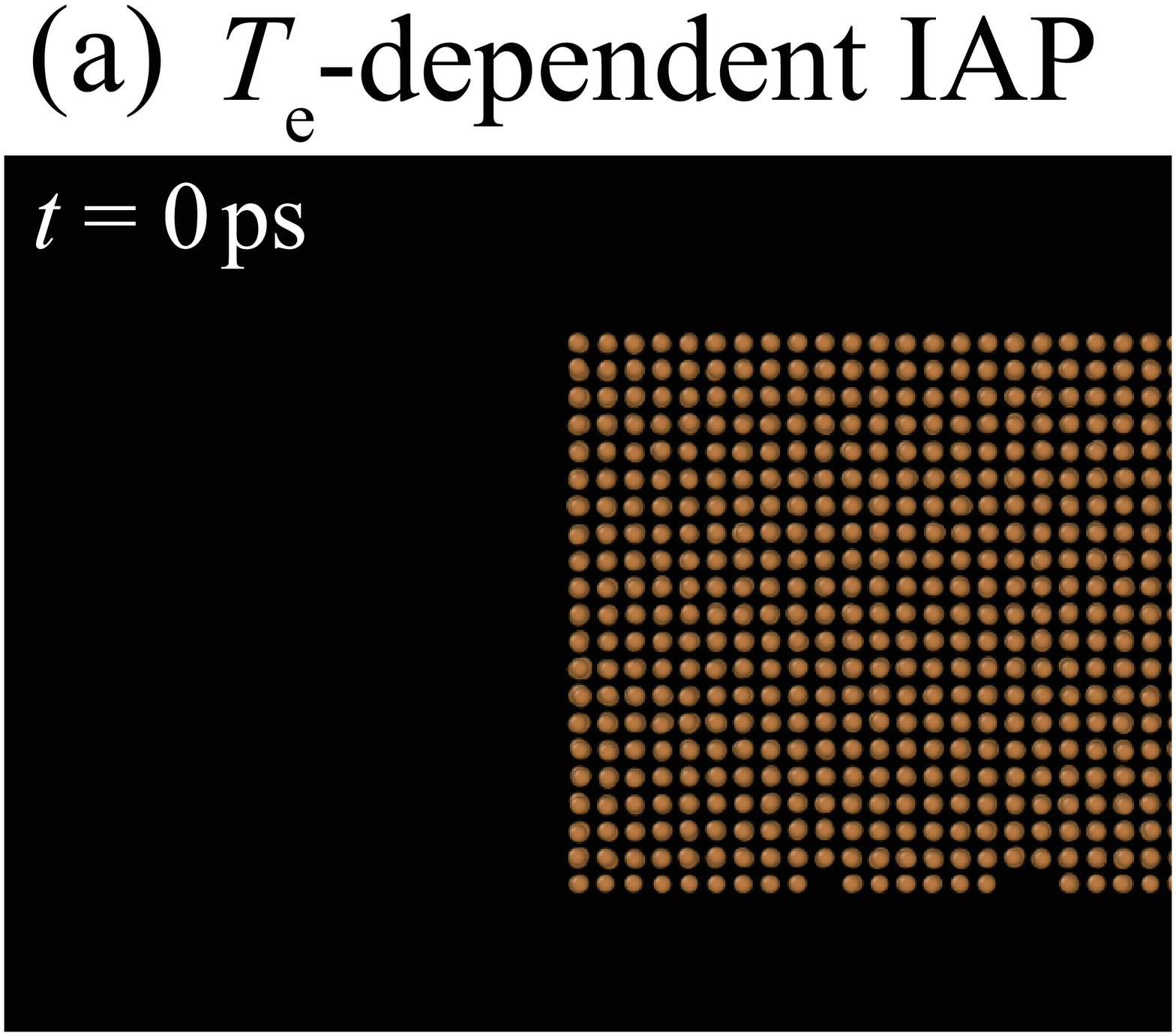}
        \end{center}
      \end{minipage}
      \begin{minipage}{0.5\hsize}
        \begin{center}
          \includegraphics[clip, width=4cm]{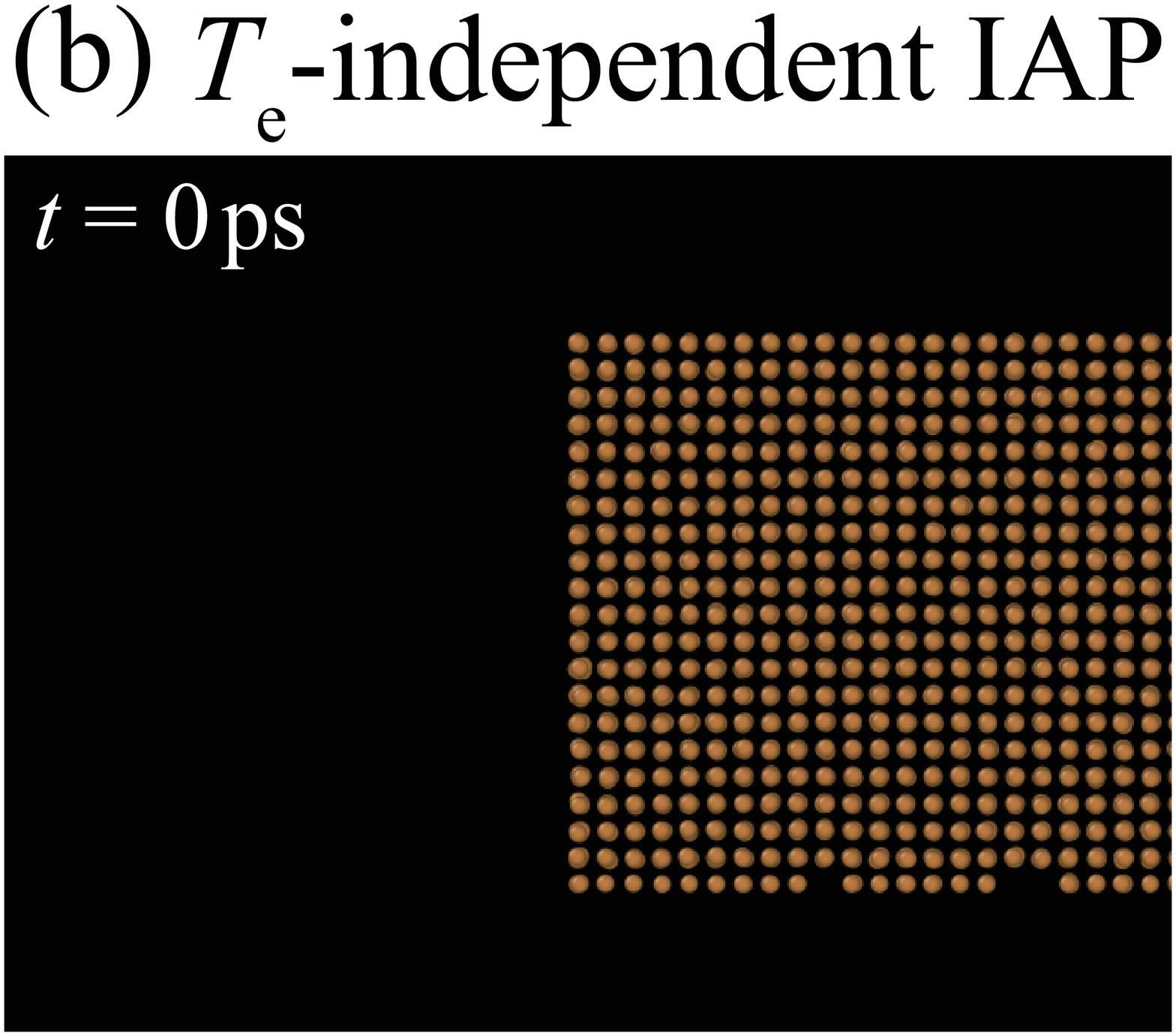}
        \end{center}  
      \end{minipage}
      \\
      \begin{minipage}{0.5\hsize}
        \begin{center}
         \includegraphics[clip, width=4cm]{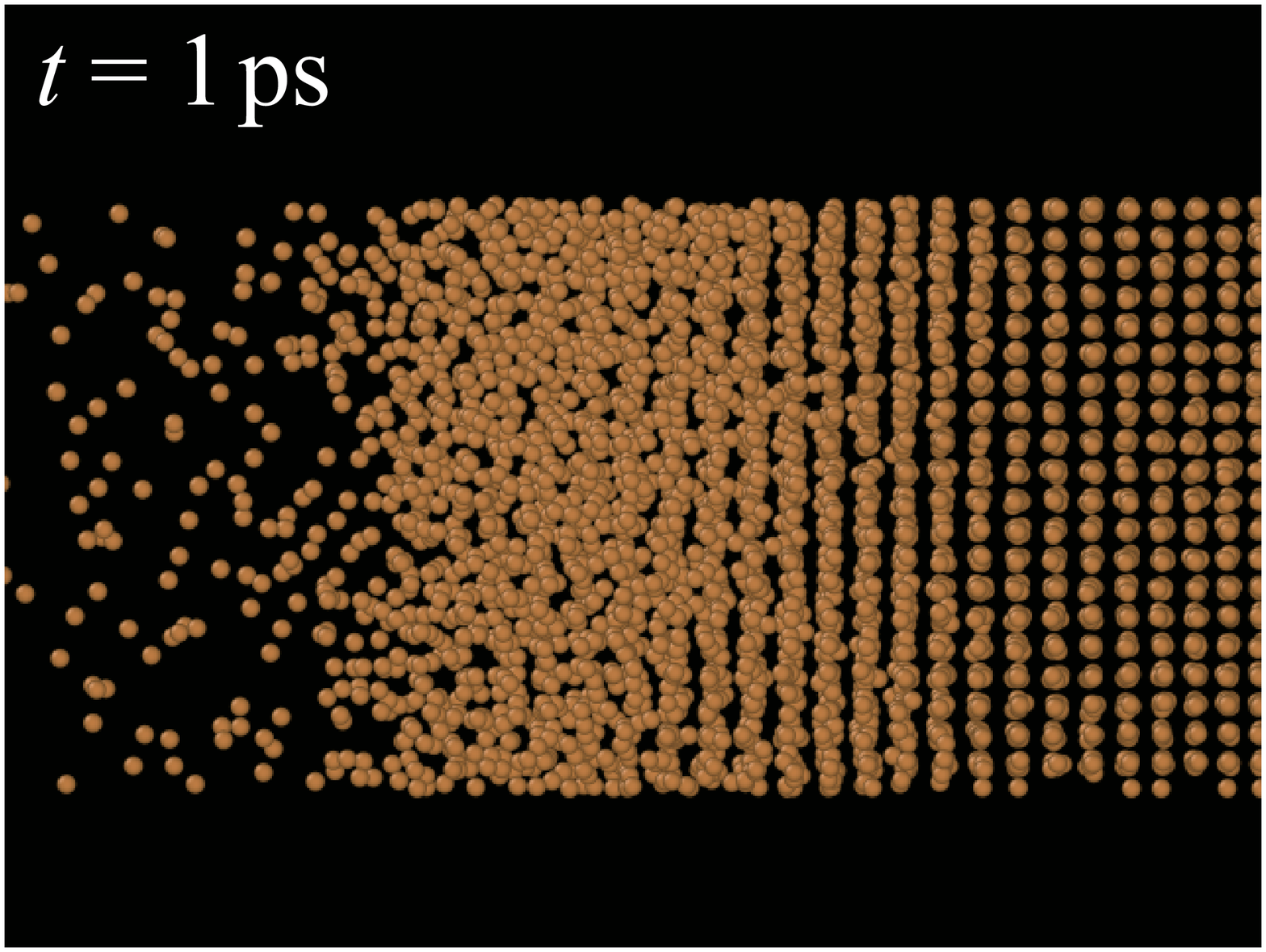}
        \end{center}
      \end{minipage}
      \begin{minipage}{0.5\hsize}
        \begin{center}
          \includegraphics[clip, width=4cm]{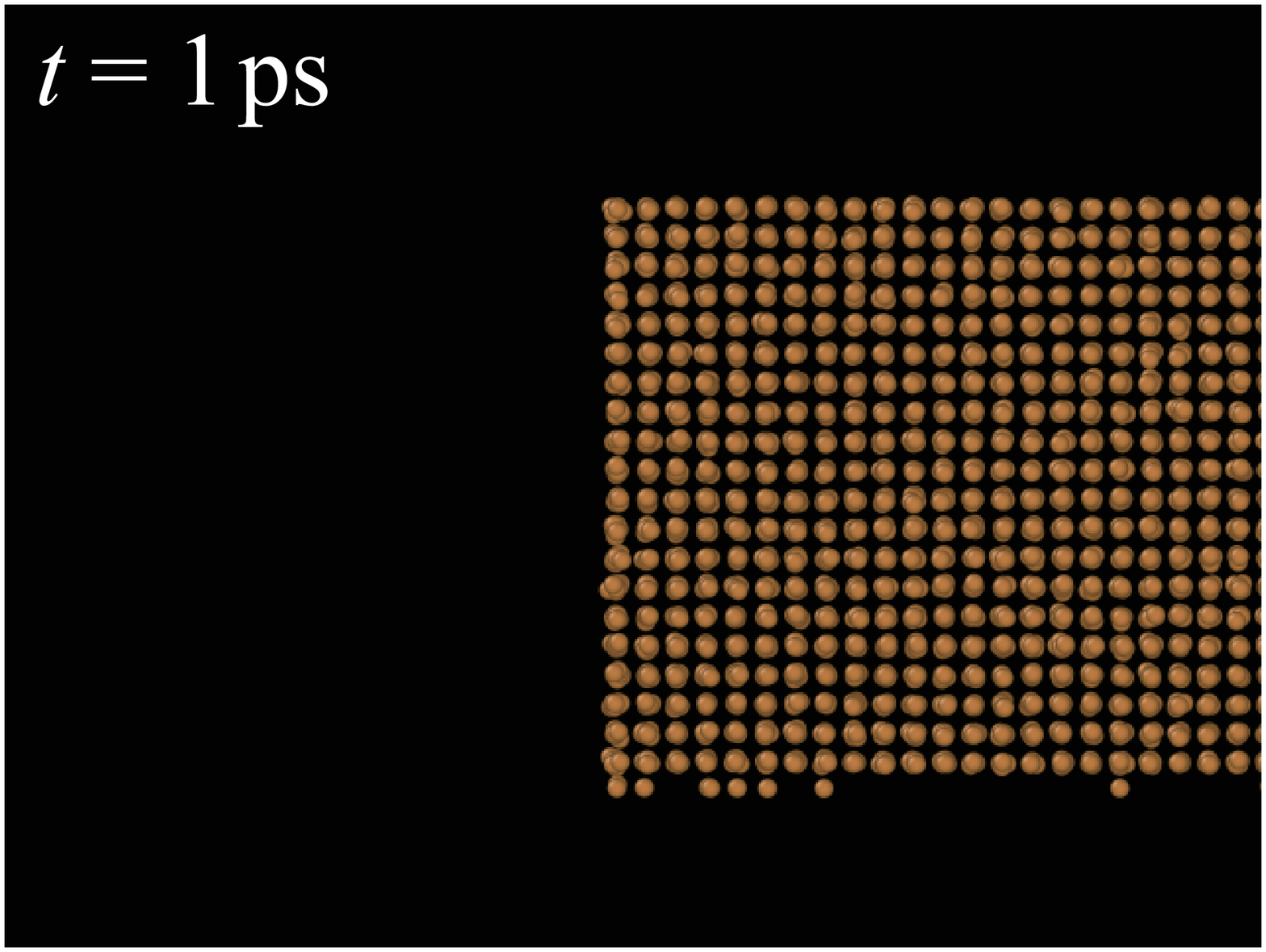}
        \end{center}
      \end{minipage}
     \\
      \begin{minipage}{0.5\hsize}
        \begin{center}
          \includegraphics[clip, width=4cm]{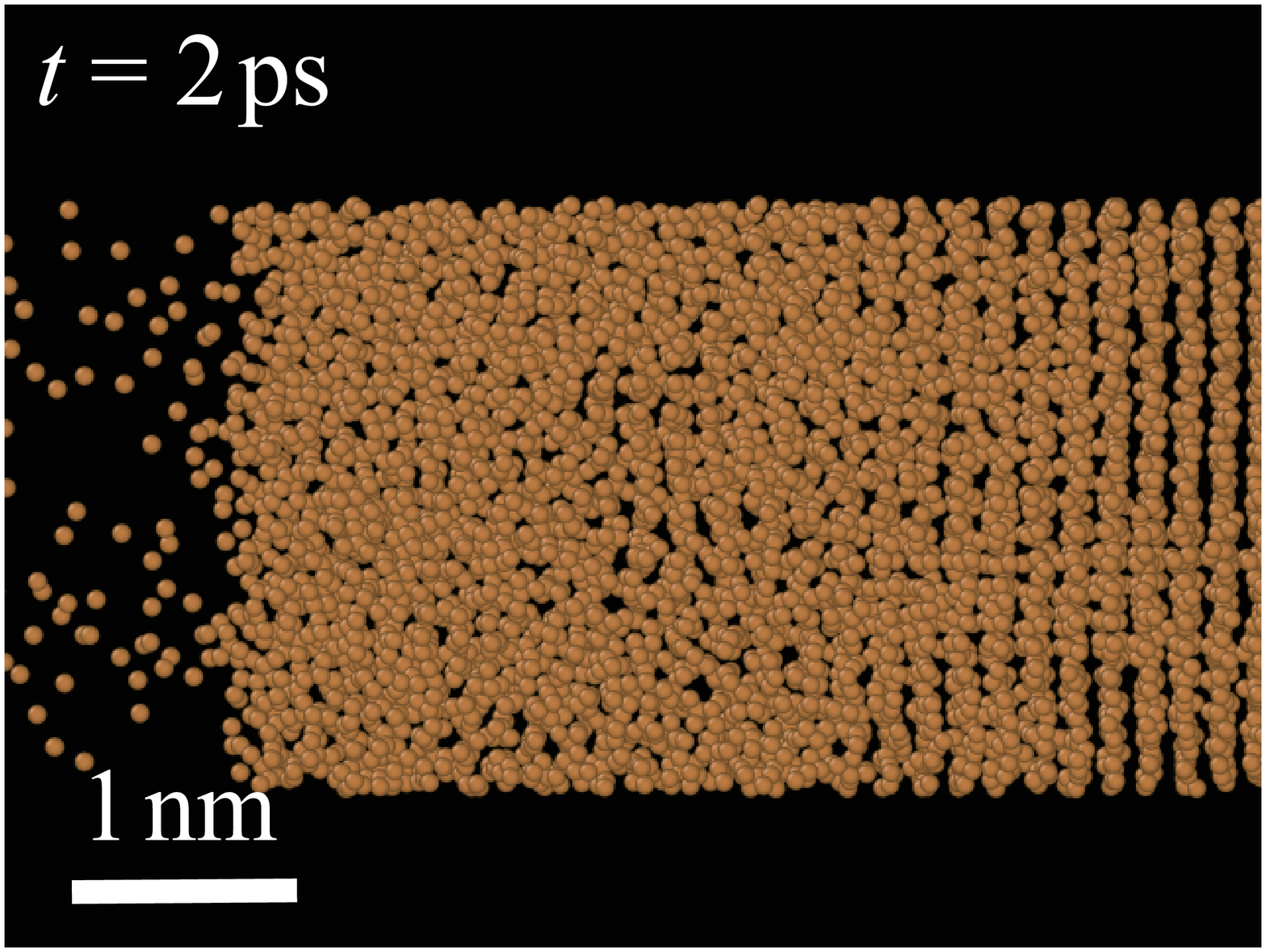}
        \end{center}
      \end{minipage}
      \begin{minipage}{0.5\hsize}
        \begin{center}
          \includegraphics[clip, width=4cm]{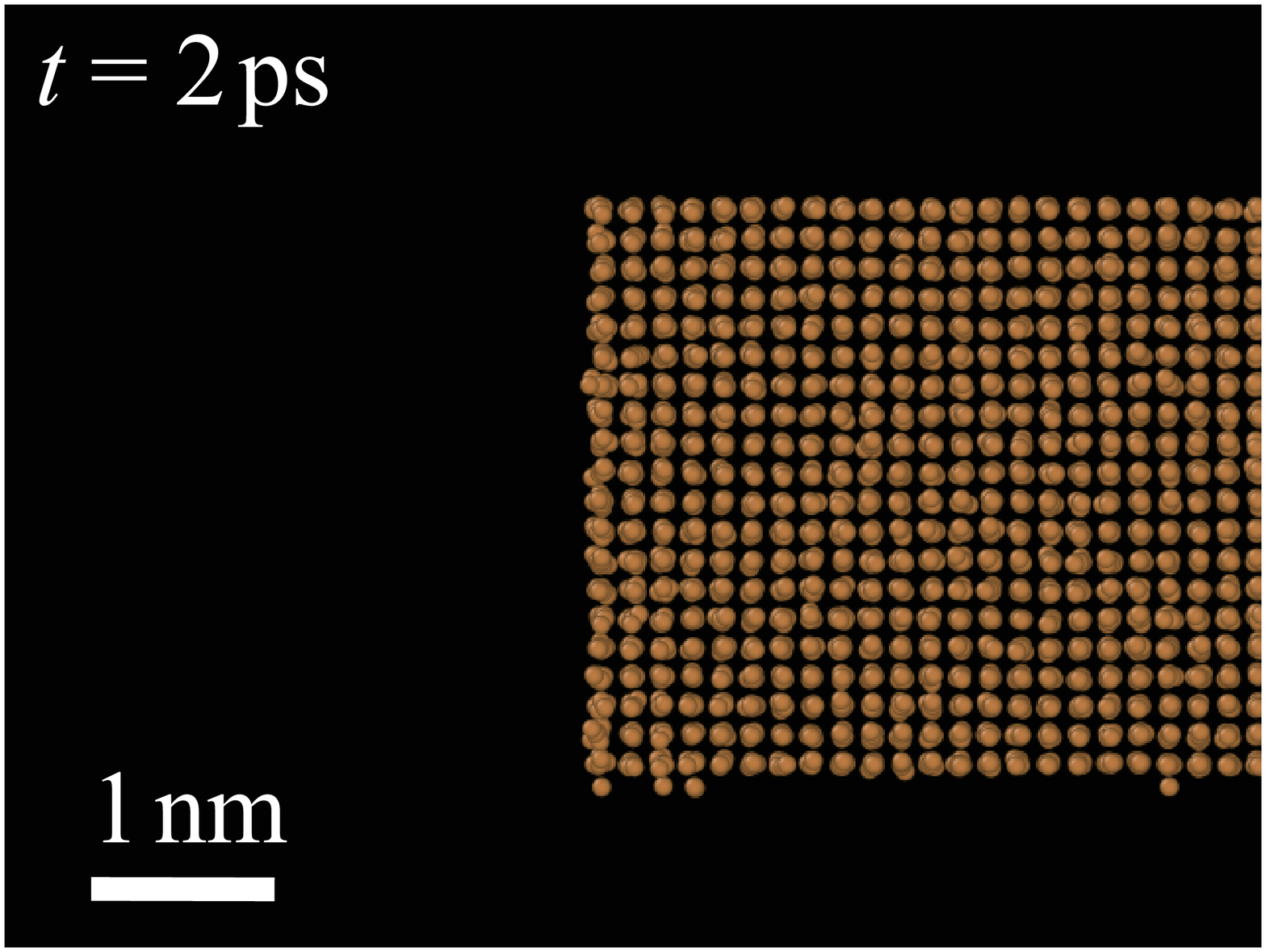}
        \end{center}  
      \end{minipage}

   \end{tabular}
      \caption{Snapshots of atomic configurations near the surface after $0$-$2\,\text{ps}$ irradiation with a  $100\,\text{fs}$ pulse laser of $J_0=0.7\,\text{J}\,\text{cm}^{-2}$.  These simulations are carried out using (a) the $T_e$-dependent IAP and (b) the $T_e$-independent IAP.
             }
    \label{fig:07snaps}
  \end{center}
\end{figure}

\begin{figure}[tbp]
  \begin{center}
    \begin{tabular}{cc}

      \begin{minipage}{1.0\hsize}
        \begin{center}
          \includegraphics[clip, width=5cm]{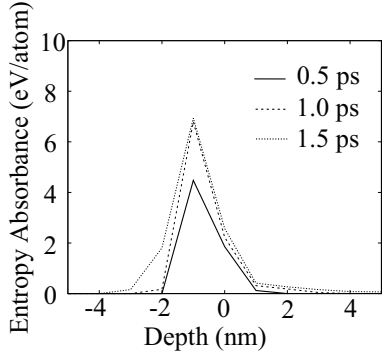}
          \hspace{1.0cm}
       
        \end{center}
      \end{minipage} \\
	
   \end{tabular}
    \caption{Energy absorbed by the electronic entropy effect at each depth.
    The laser fluence is $J_0=0.7\,\text{J}\,\text{cm}^{-2}$.
Solid, dashed, and dotted lines represent the elapsed times $t=0.5$,  $1.0$, and  $1.5\,\text{ps}$, respectively.
The basis of the $x$-axis represents the initial surface position.
              }
    \label{fig:07entropy}
  \end{center}
\end{figure}

 
 \begin{figure}[tbp]
  \begin{center}
    \begin{tabular}{cc}

      \begin{minipage}{0.5\hsize}
        \begin{center}
          \includegraphics[clip, width=2.8cm]{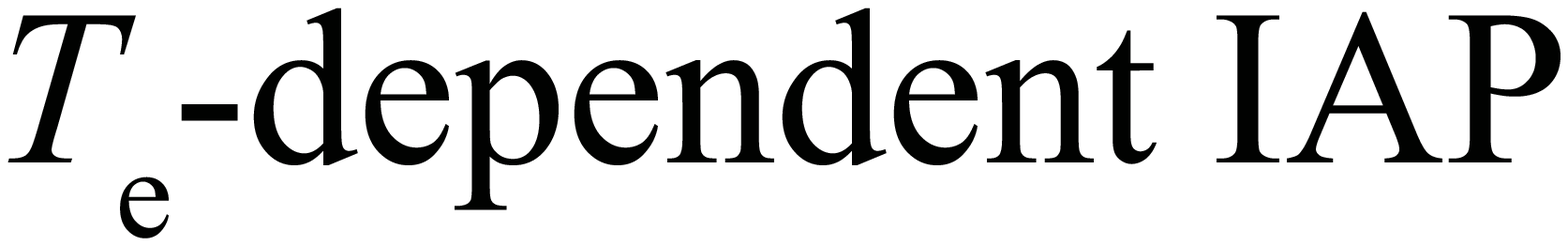}
        \end{center}
      \end{minipage}
  \\

      \begin{minipage}{0.5\hsize}
        \begin{center}
          \includegraphics[clip, width=4cm]{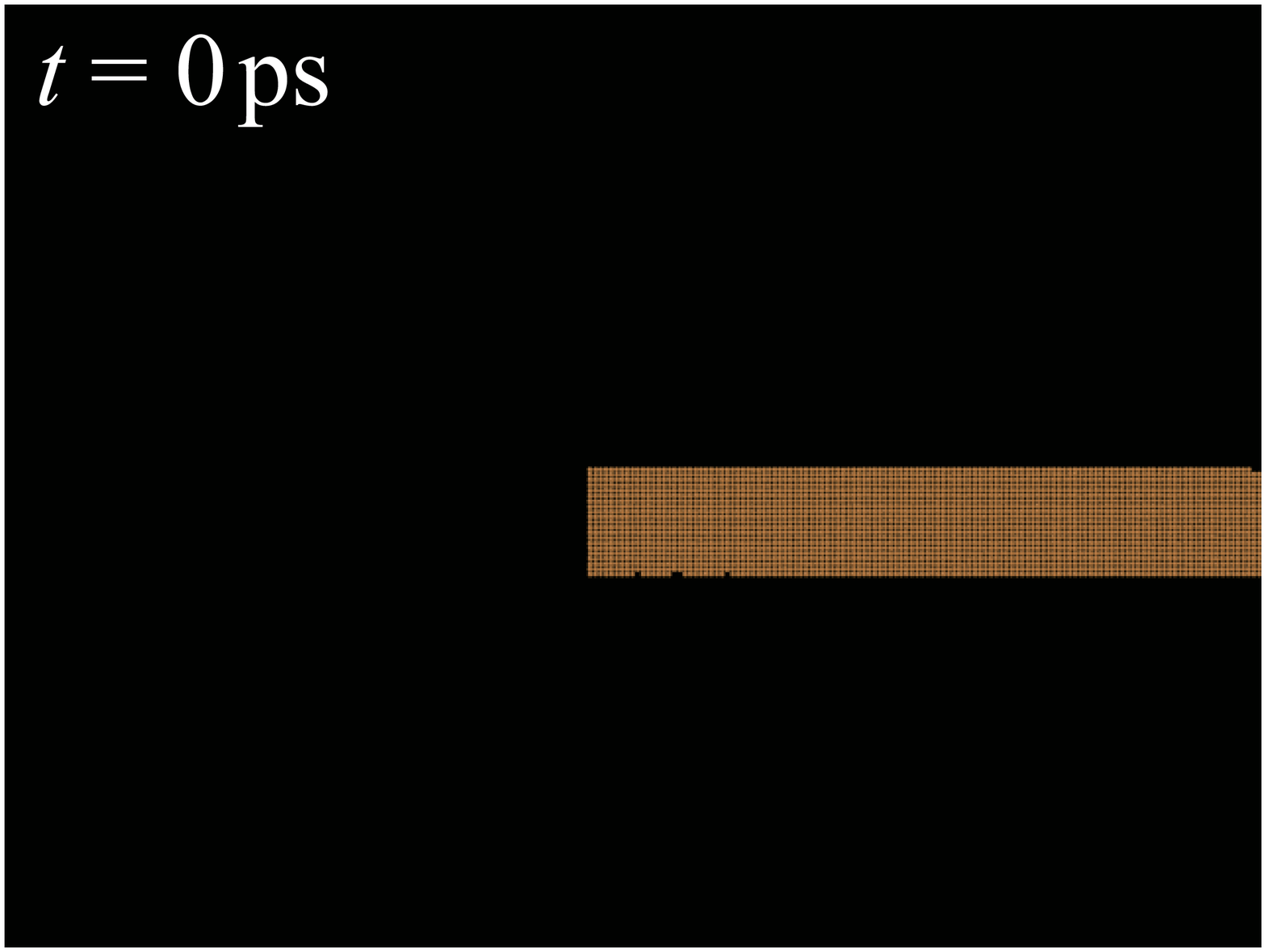}
        \end{center}
      \end{minipage}
      \begin{minipage}{0.5\hsize}
        \begin{center}
          \includegraphics[clip, width=4cm]{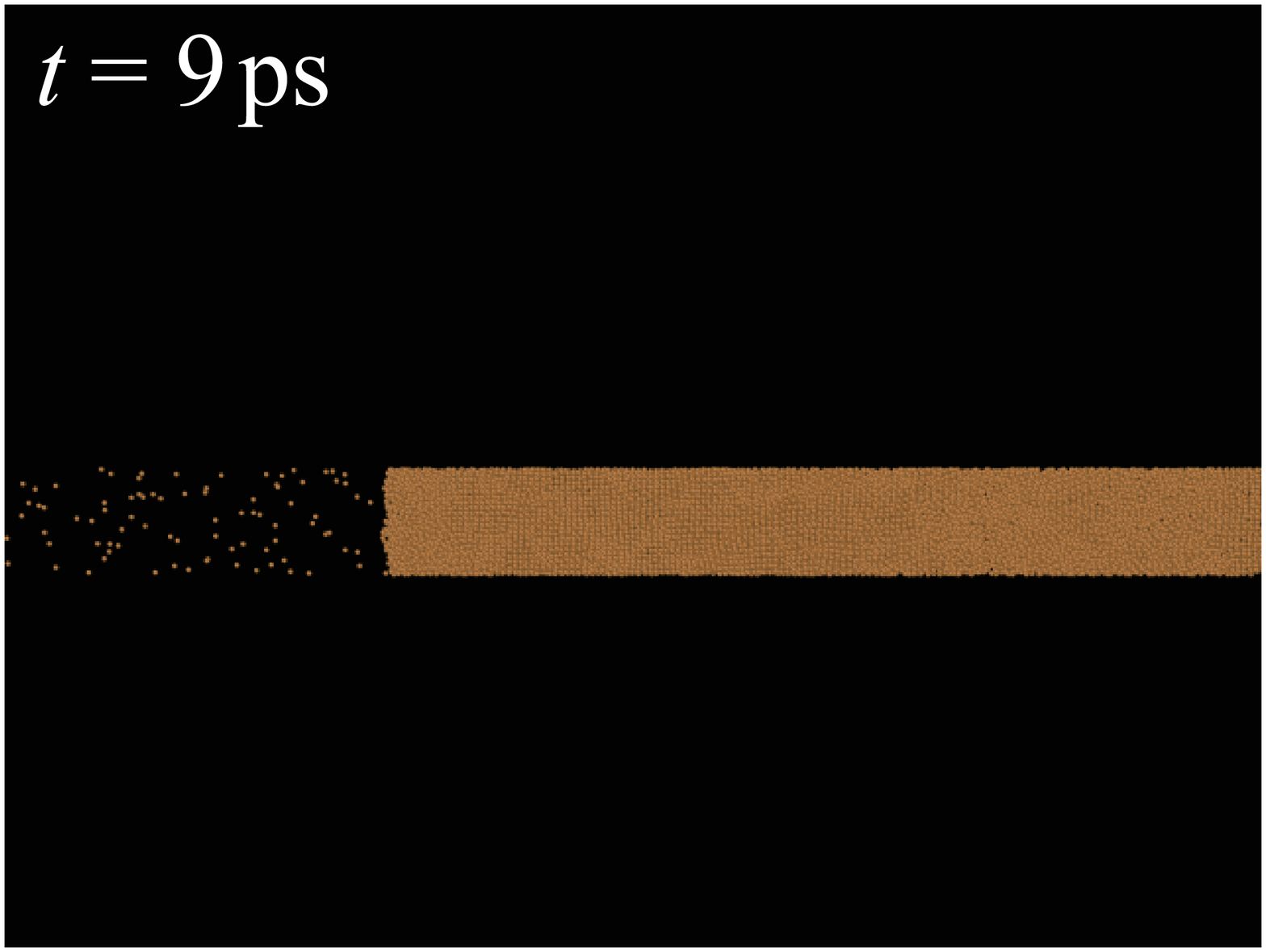}
        \end{center}  
      \end{minipage}
      \\
      \begin{minipage}{0.5\hsize}
        \begin{center}
         \includegraphics[clip, width=4cm]{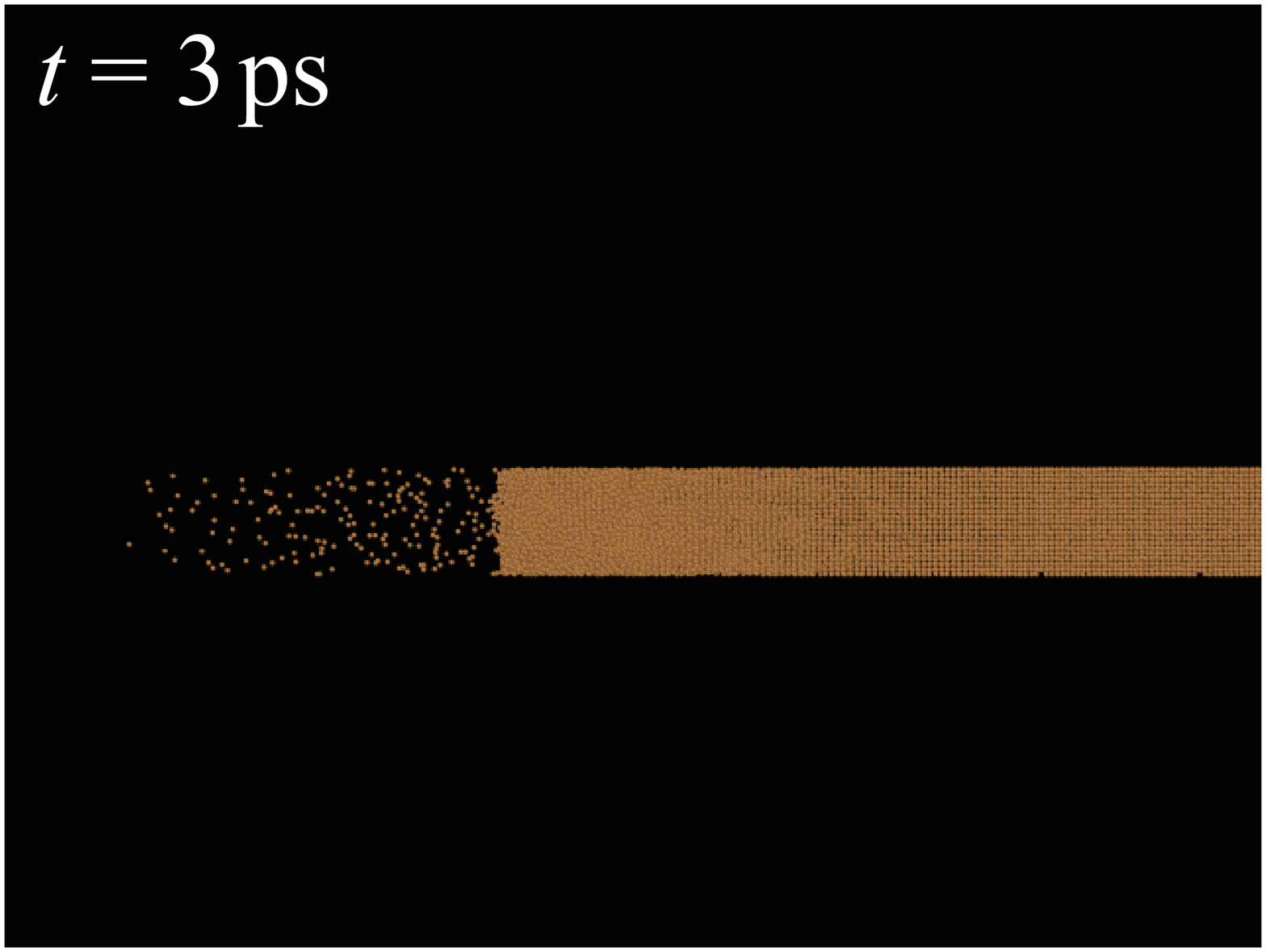}
        \end{center}
      \end{minipage}
      \begin{minipage}{0.5\hsize}
        \begin{center}
          \includegraphics[clip, width=4cm]{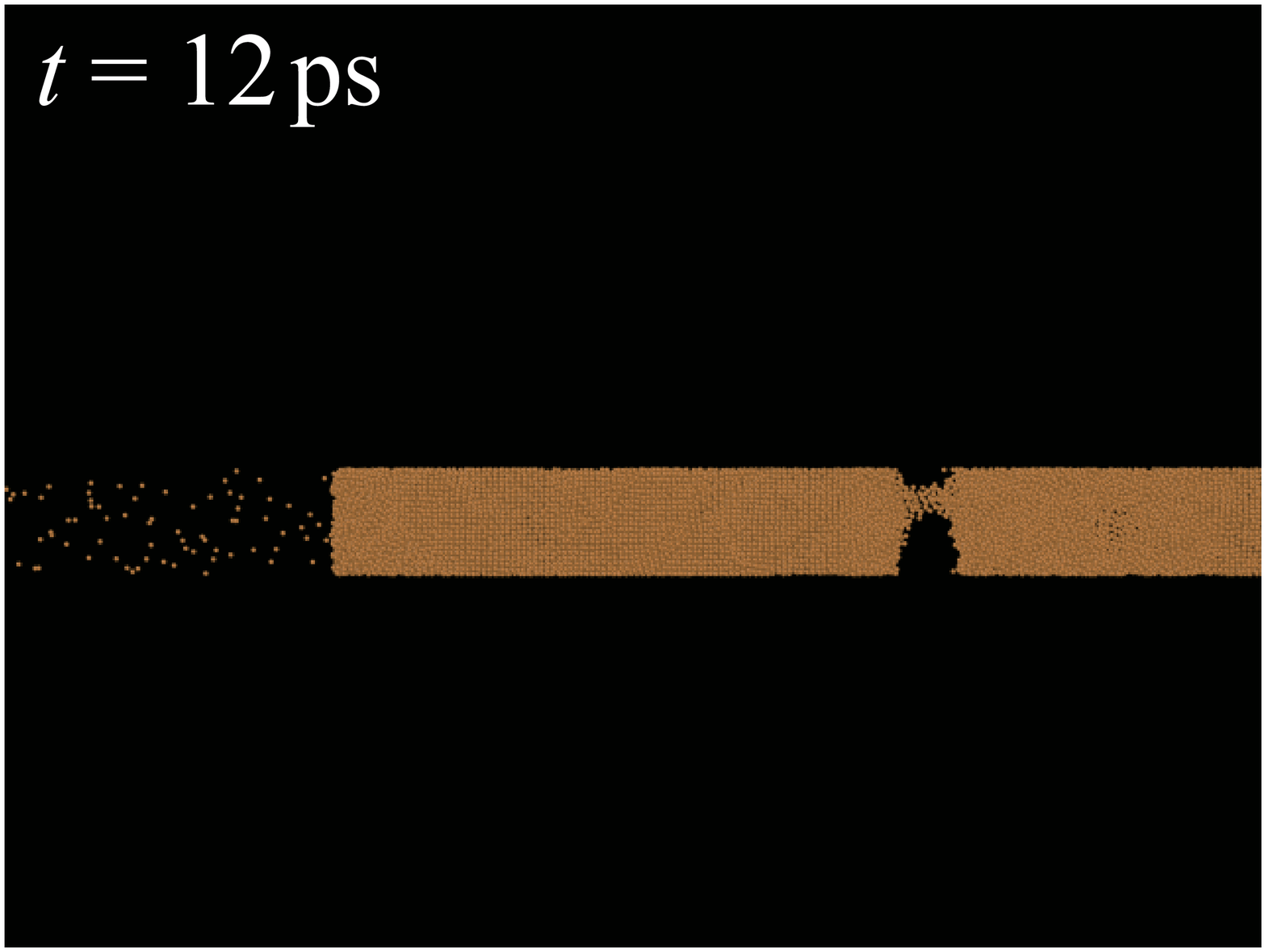}
        \end{center}
      \end{minipage}
     \\
      \begin{minipage}{0.5\hsize}
        \begin{center}
          \includegraphics[clip, width=4cm]{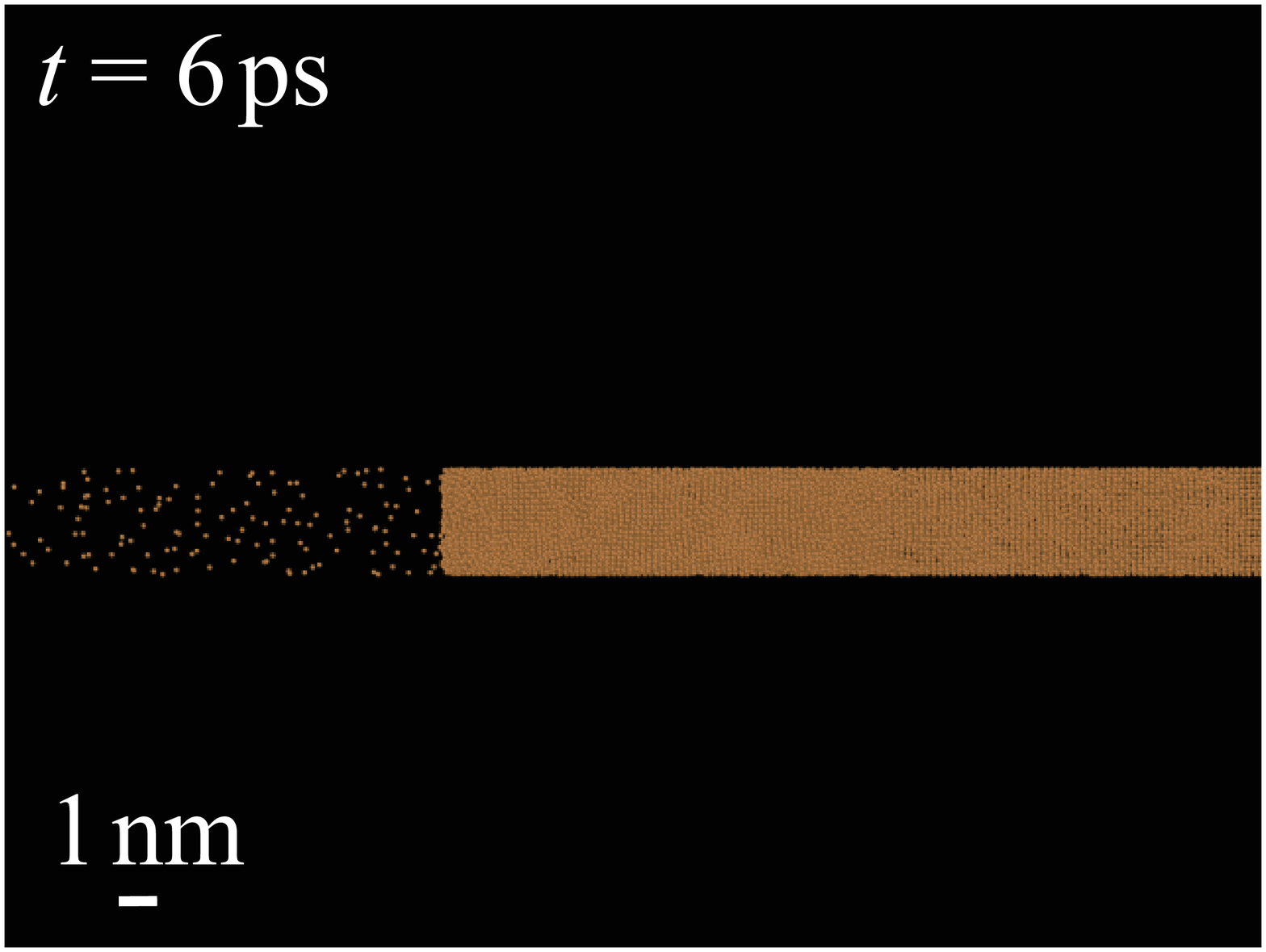}
        \end{center}
      \end{minipage}
      \begin{minipage}{0.5\hsize}
        \begin{center}
          \includegraphics[clip, width=4cm]{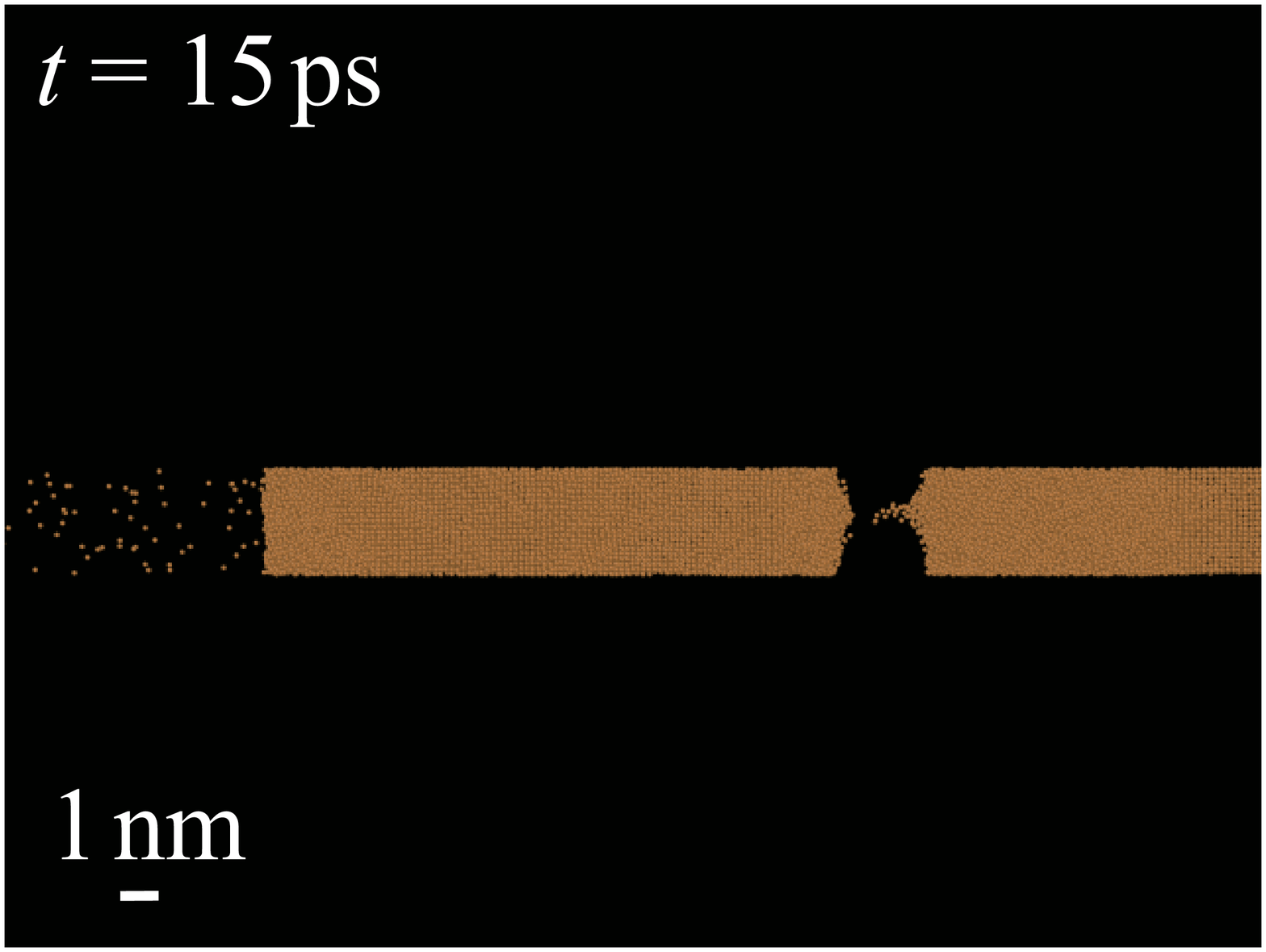}
        \end{center}
      \end{minipage}

   \end{tabular}
      \caption{ Snapshots of atomic configurations near the surface after $0$-$15\,\text{ps}$ irradiation with a  $100\,\text{fs}$-pulse laser of $J_0=0.7\,\text{J}\,\text{cm}^{-2}$.   This simulation is carried out using the $T_e$-dependent IAP.
            }
    \label{fig:07asnaps}
  \end{center}
\end{figure}

 
 \begin{figure}[tb]
  \begin{center}
    \begin{tabular}{cc}

      \begin{minipage}{0.5\hsize}
        \begin{center}
          \includegraphics[clip, width=2.8cm]{Te-dep.eps}
        \end{center}
      \end{minipage}
  \\[0.8ex]

      \begin{minipage}{0.5\hsize}
        \begin{center}
          \includegraphics[clip, width=4cm]{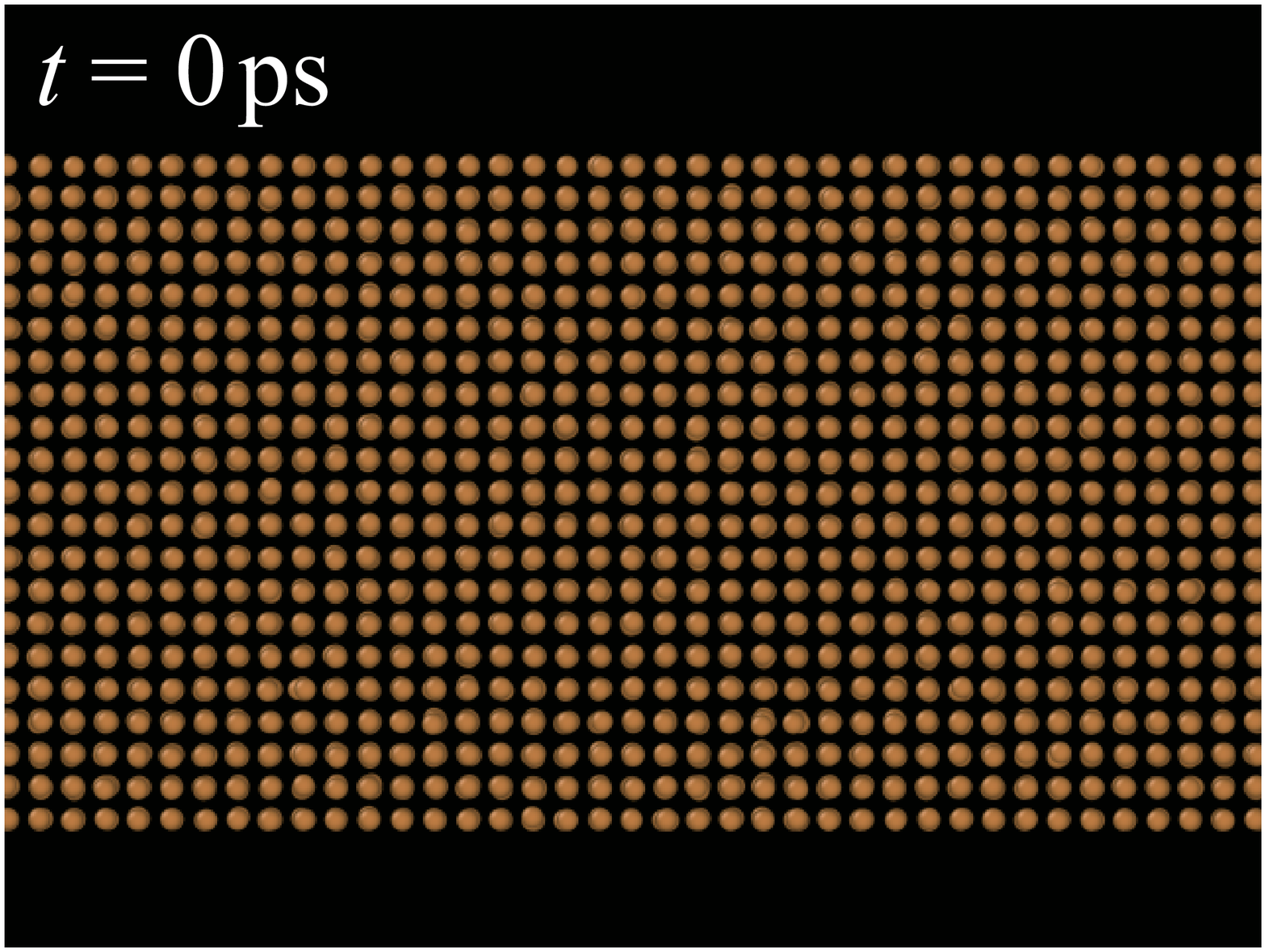}
        \end{center}
      \end{minipage}
      \begin{minipage}{0.5\hsize}
        \begin{center}
          \includegraphics[clip, width=4cm]{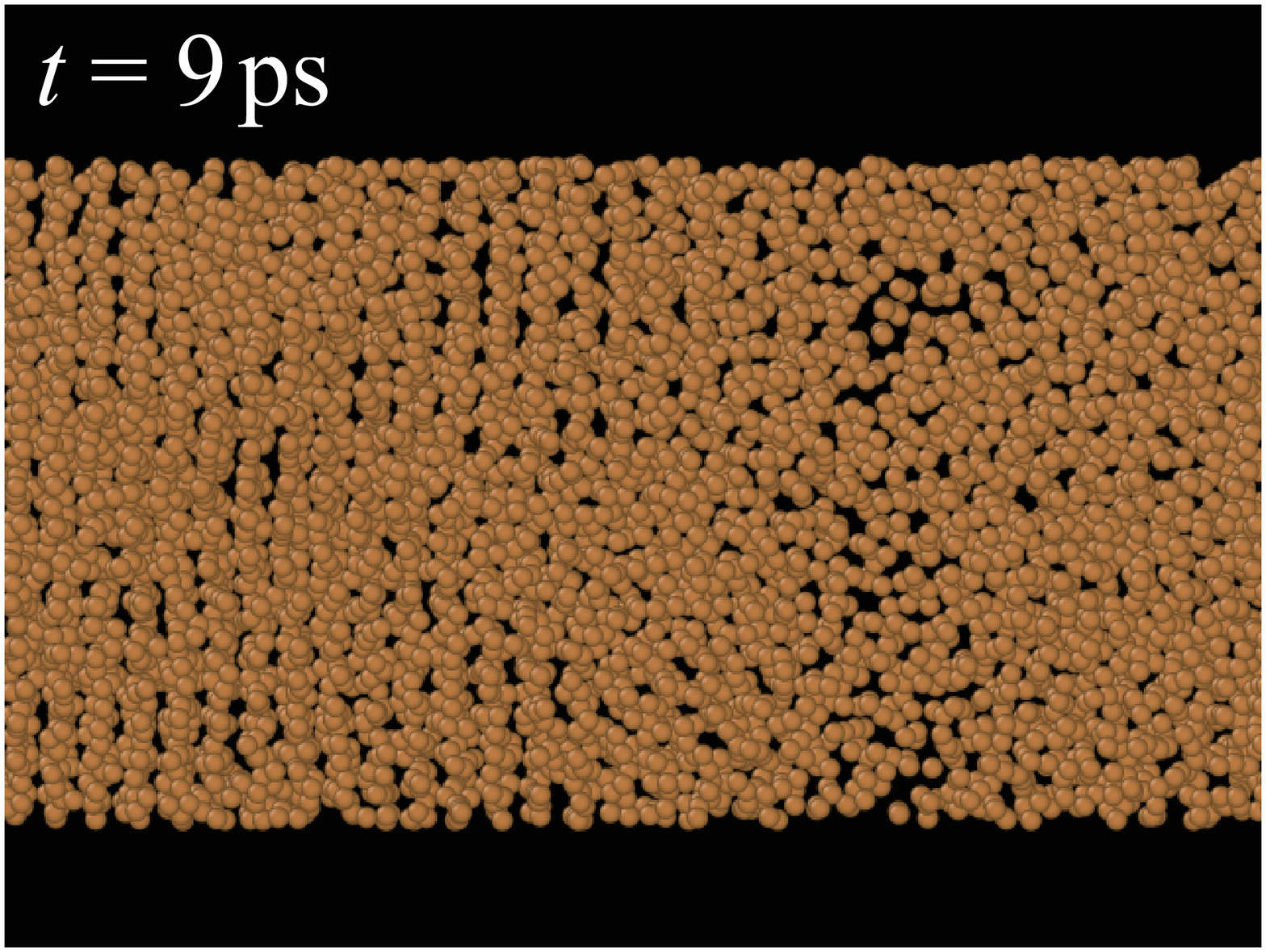}
        \end{center}  
      \end{minipage}
      \\
      \begin{minipage}{0.5\hsize}
        \begin{center}
         \includegraphics[clip, width=4cm]{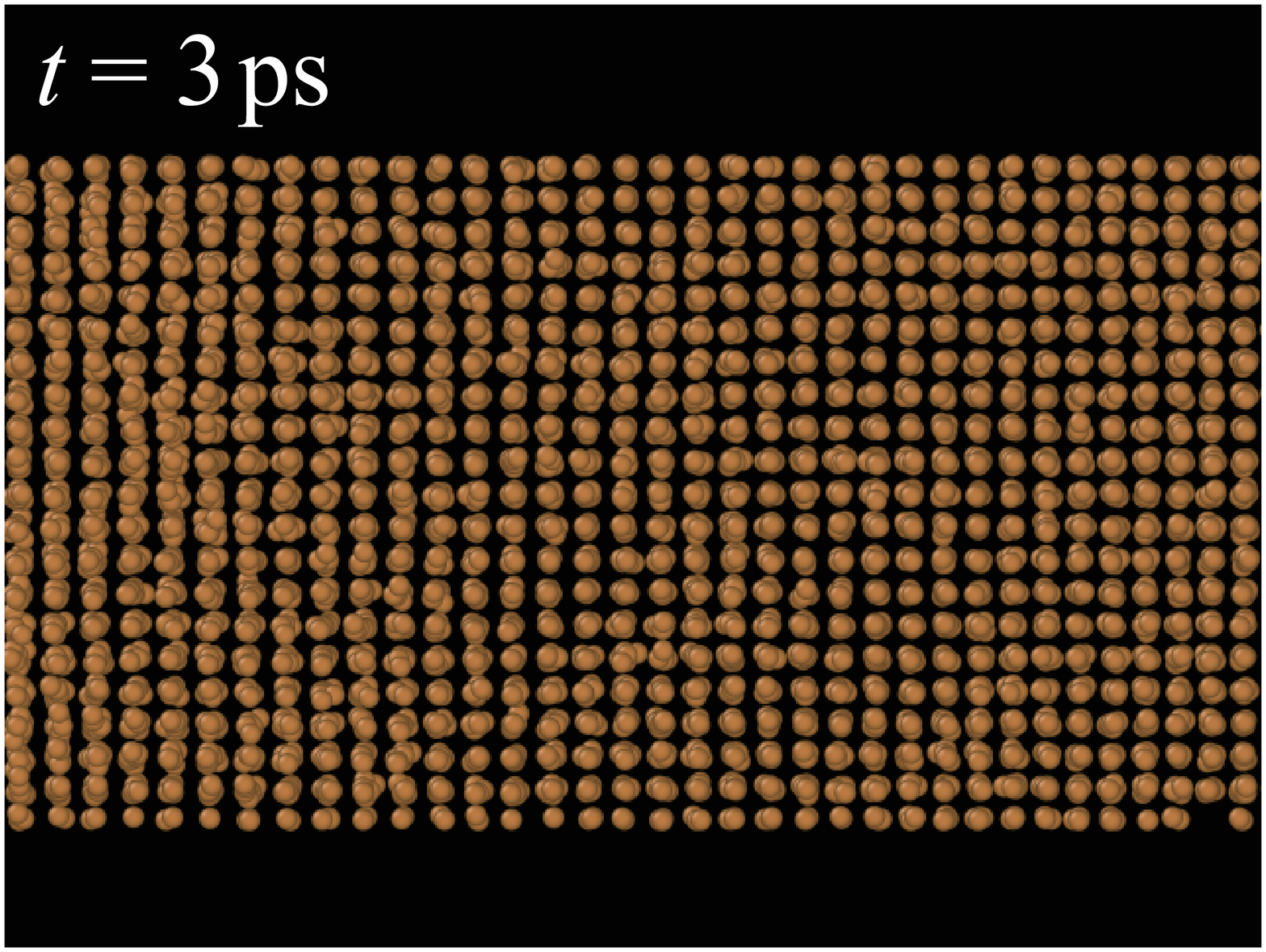}
        \end{center}
      \end{minipage}
      \begin{minipage}{0.5\hsize}
        \begin{center}
          \includegraphics[clip, width=4cm]{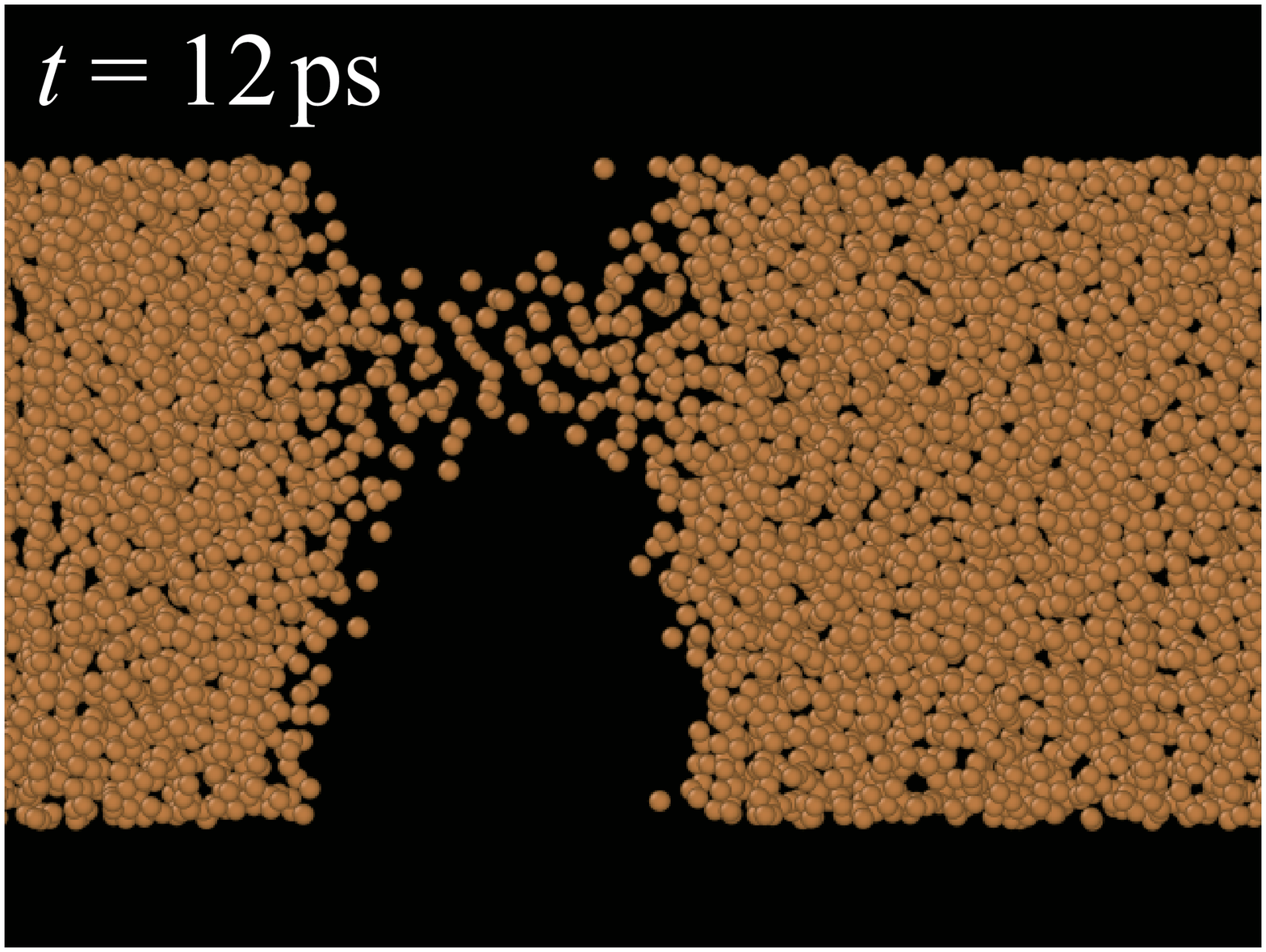}
        \end{center}
      \end{minipage}
     \\
      \begin{minipage}{0.5\hsize}
        \begin{center}
          \includegraphics[clip, width=4cm]{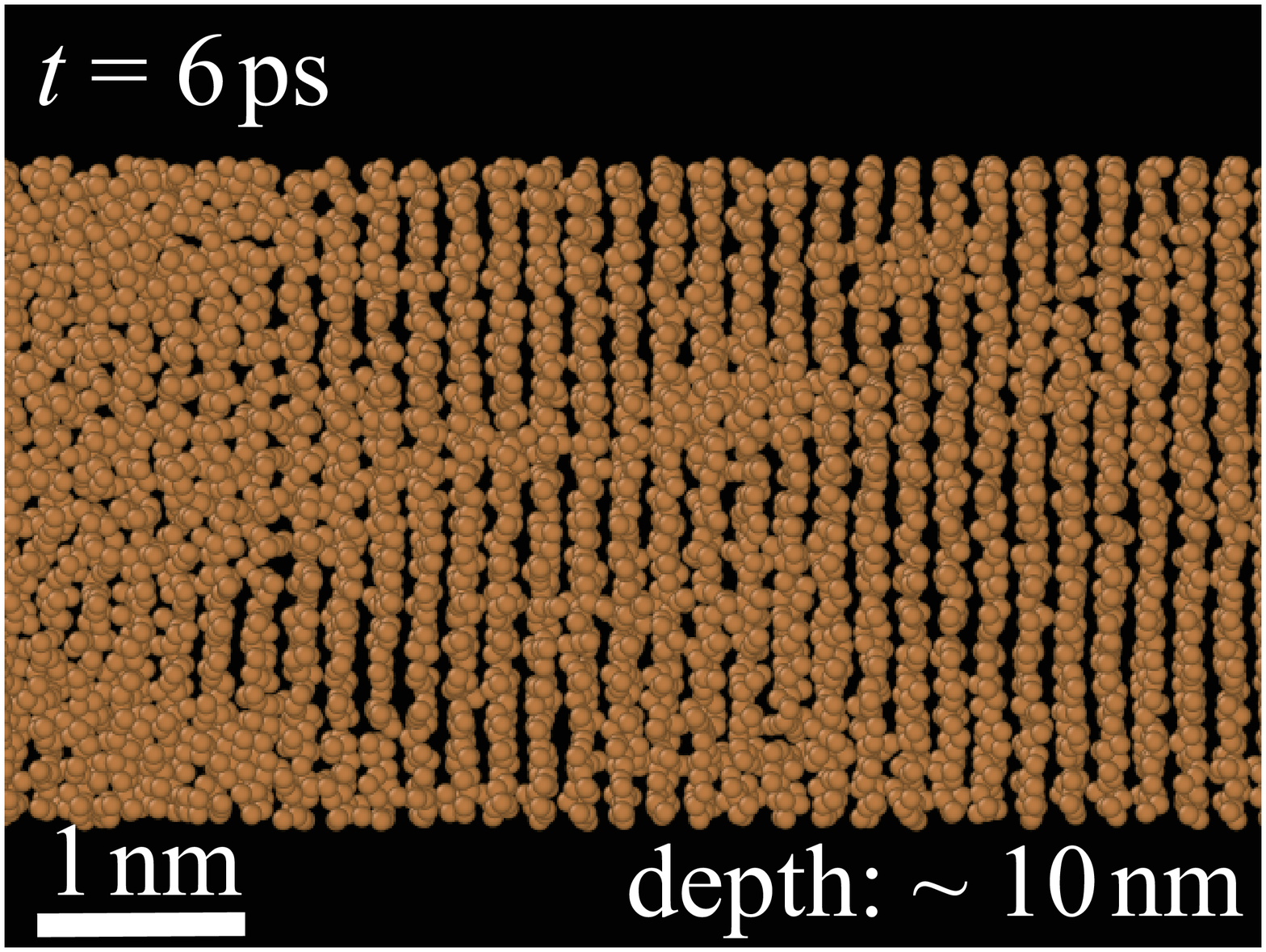}
        \end{center}
      \end{minipage}
      \begin{minipage}{0.5\hsize}
        \begin{center}
          \includegraphics[clip, width=4cm]{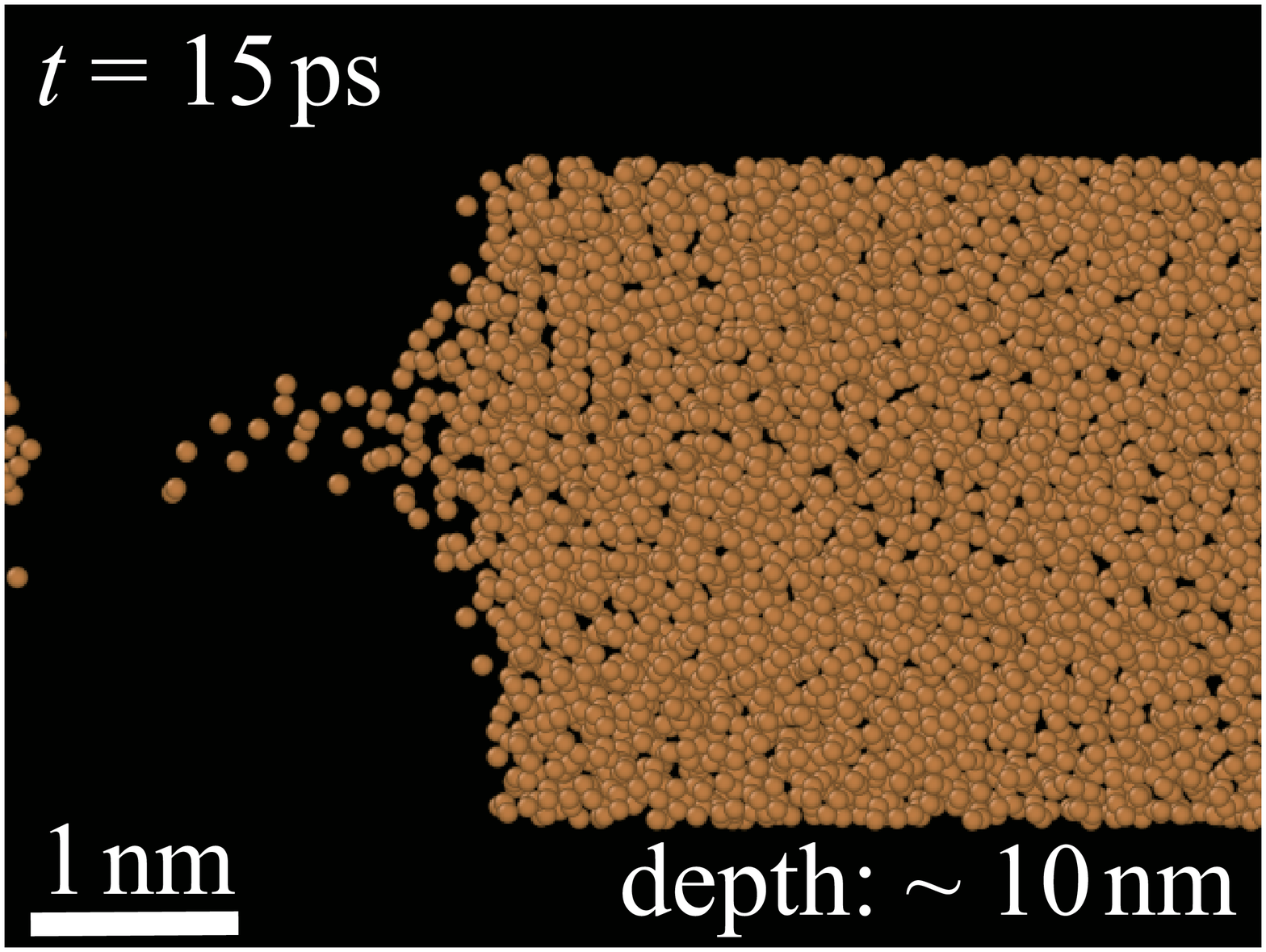}
        \end{center}
      \end{minipage}

   \end{tabular}
      \caption{Snapshots of atomic configurations (depth: $\sim10\,\text{nm}$) of  Fig.~\ref{fig:07asnaps}.
                           }
    \label{fig:07bsnaps}
  \end{center}
\end{figure}

\begin{figure}[h]
  \begin{center}
    \begin{tabular}{cc}

      \begin{minipage}{0.5\hsize}
        \begin{center}
          \includegraphics[clip, width=4.5cm]{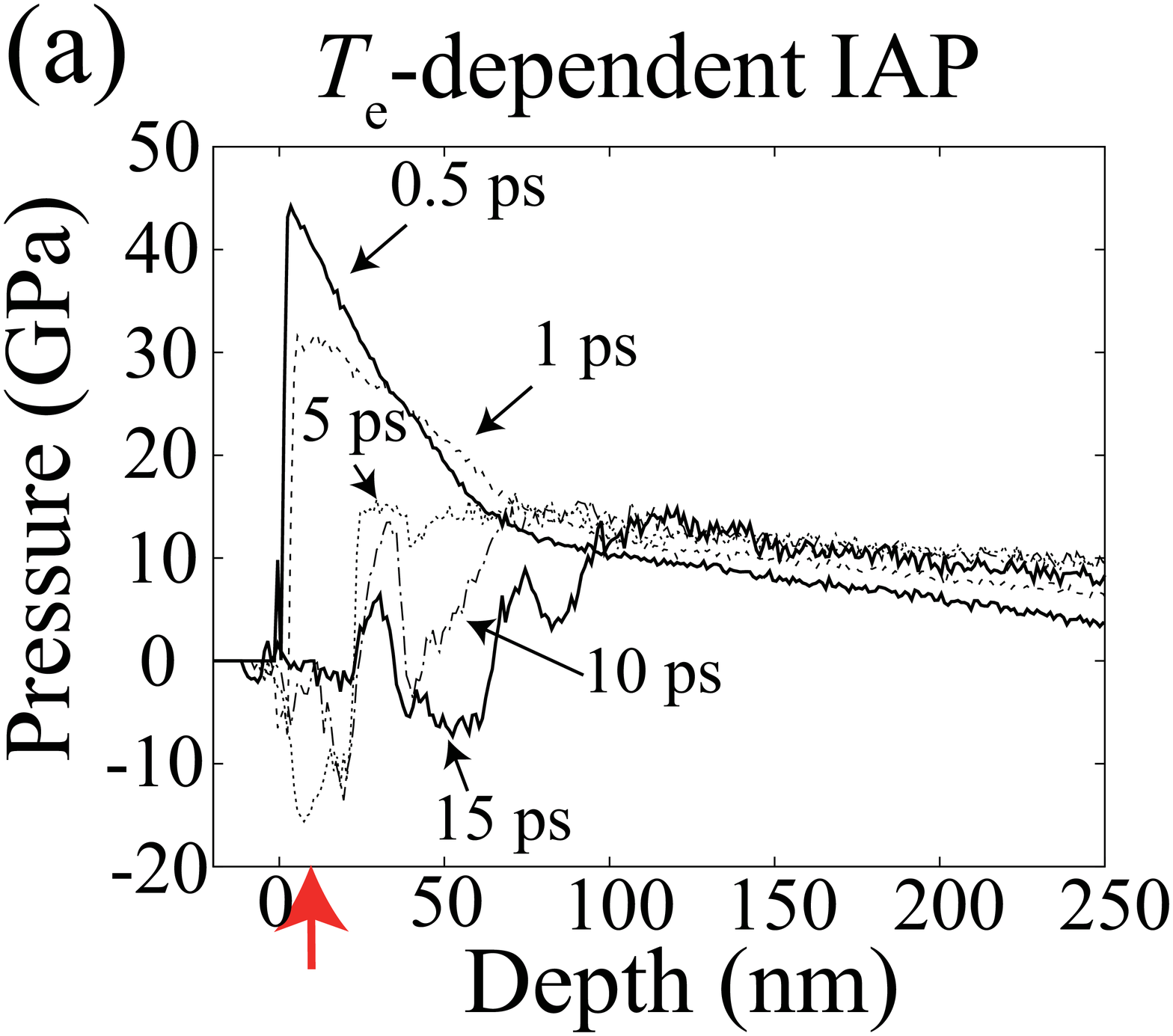}
       
        \end{center}
      \end{minipage}  
      
      \begin{minipage}{0.5\hsize}
        \begin{center}
          \includegraphics[clip, width=4.5cm]{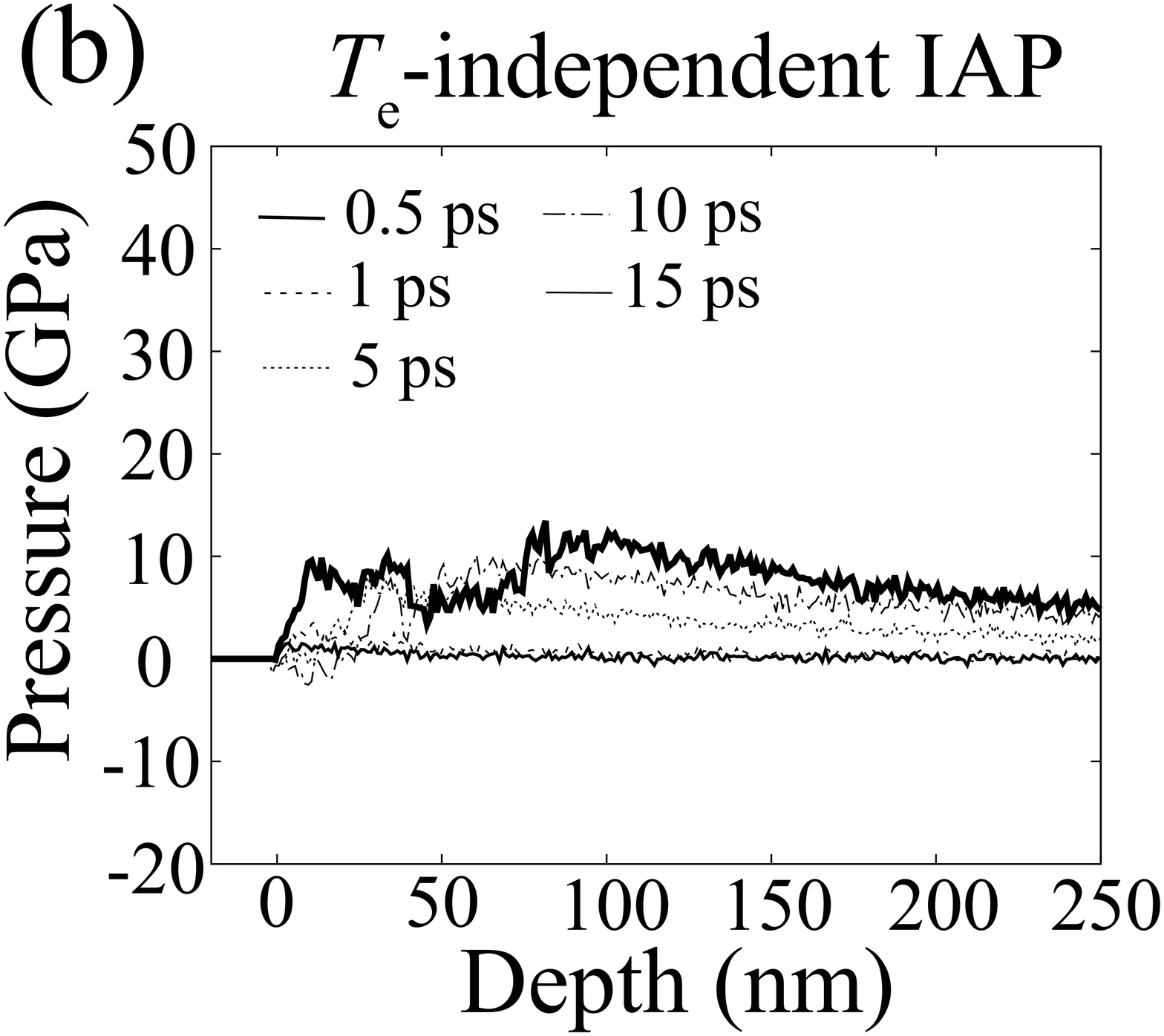}
        \end{center}
      \end{minipage} \\

   \end{tabular}
    \caption{Spatial distribution of the local pressure along the $z$ direction of simulations using (a) the $T_e$-dependent IAP and  (b) the $T_e$-independent IAP.
      Solid, dashed, dotted, chained, and bold lines represent the results at $t=0.5$, $1.0$, $5.0$, $10.0$, and $15.0\,\text{ps}$, respectively. 
      The red arrow in (a) represents the point where spallation occurs.
                    }
    \label{fig:07pres}
  \end{center}
\end{figure}

Here, we exhibit the results of simulations in which the fluence of the applied laser ($J_0=0.7\,\text{J}\,\text{cm}^{-2}$)  is a little higher than the ablation threshold ($J_0=0.55\,\text{J}\,\text{cm}^{-2}$).

Fig.~\ref{fig:07snaps} shows snapshots of the atomic configurations of the Cu film irradiated by the ultrashort-pulse laser. 
As shown in  Fig.~\ref{fig:07snaps}(b), when the $T_e$-independent IAP is used in the simulation, ablation does not occur.
On the other hand, Fig.~\ref{fig:07snaps}(a) shows that ablation occurs when the $T_e$-dependent IAP is used.
In addition,  Fig.~\ref{fig:07entropy} shows that  the absorption due to electronic entropy is more than $4.0\,\text{eV}\,\text{atom}^{-1}$, which is larger than that of simulation at a lower laser irradiation and cohesive energy.
According to the calculation results, the electronic entropy plays an important role in causing atom emission even at this laser fluence.

Fig.~\ref{fig:07asnaps} shows snapshots of the atomic configurations of $0$ to $15\,\text{ps}$ after laser irradiation.
As shown in these figures, spallation is also observed for laser irradiation with $J_0=0.7\,\text{J}\,\text{cm}^{-2}$. 
Part of the atomic configuration (depth: $\sim10\,\text{nm}$) of  Fig.~\ref{fig:07asnaps} is shown in Fig.~\ref{fig:07bsnaps}.
Since it is thought that the trigger for spallation is tensile stress, the time development of local pressure in the TTM-MD simulation is calculated to investigate the electronic entropy contribution. 
The space distribution of the local pressure along the $z$ direction is shown in Fig.~\ref{fig:07pres}(a).
At least within $t=5\,\text{ps}$, the pressure wave passes through a point indicated by the red arrow in Fig.~\ref{fig:07pres}(a), at which the surface layer is spalled, and a large negative pressure is created.
Owing to the negative pressure, a void begins to be formed around $t=9\,\text{ps}$, and as a result, spallation occurs.
In the simulation with the $T_e$-independent IAP, spallation does not occur at least within $100\,\text{ps}$, which is enough time for the recoil pressure created near the surface to reach the MD/CM boundary.
Fig.~\ref{fig:07pres}(b) shows that the negative pressure for the simulation using  the $T_e$-independent IAP is smaller than that using the $T_e$-dependent IAP by one order of magnitude.
Since it has been widely accepted that large negative pressure is the origin of spallation,~\cite{Wu_2013} we consider that one of the reasons that spallation is not caused in the $T_e$-independent IAP simulation is the small negative pressure.
The reason for the small negative pressure is considered to be the lack of atom emission and the small internal pressure due to neglecting the effect of electronic entropy.
From these results,  we conclude that the effect of  electronic entropy enhances not only atom emission but also spallation.

\subsubsection{Ablation a little higher than the ablation threshold: transition to phase explosion}
\label{sec:results_phase}

 
 \begin{figure}[b]
  \begin{center}
    \begin{tabular}{cc}

      \begin{minipage}{0.5\hsize}
        \begin{center}
          \includegraphics[clip, width=2.8cm]{Te-dep.eps}
        \end{center}
      \end{minipage}
  \\[0.8ex]

      \begin{minipage}{0.5\hsize}
        \begin{center}
          \includegraphics[clip, width=4cm]{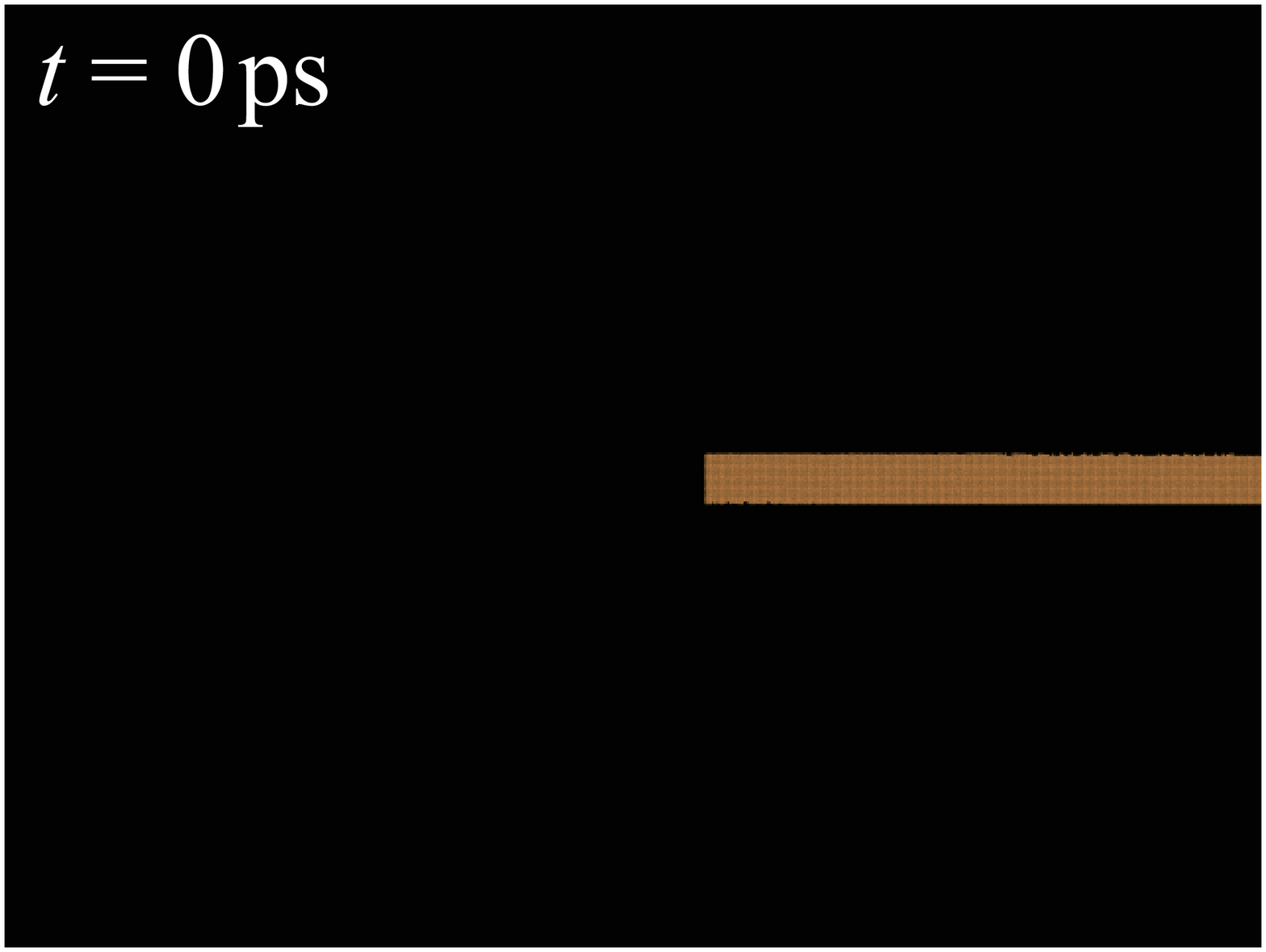}
        \end{center}
      \end{minipage}
      \begin{minipage}{0.5\hsize}
        \begin{center}
          \includegraphics[clip, width=4cm]{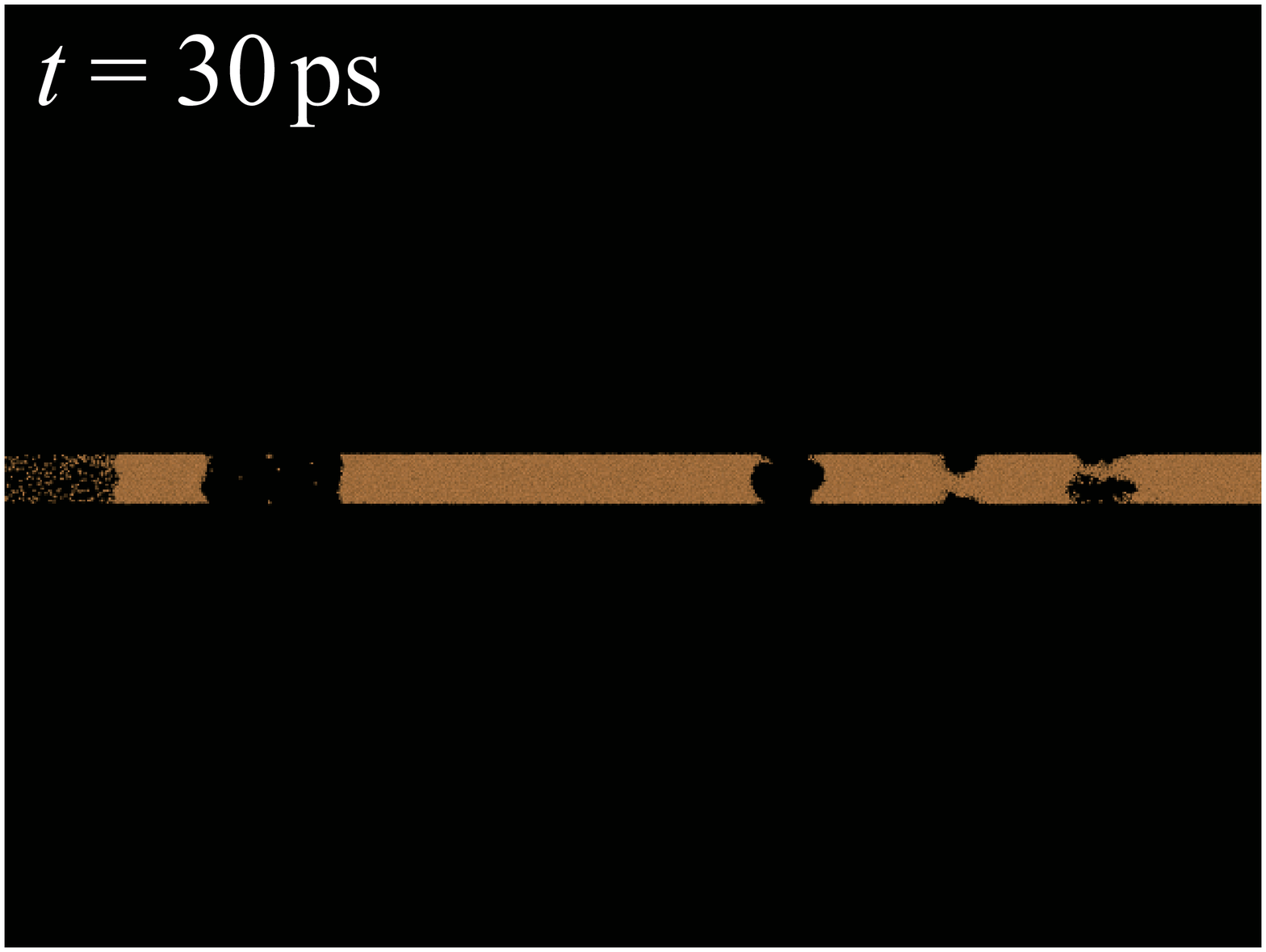}
        \end{center}  
      \end{minipage}
      \\
      \begin{minipage}{0.5\hsize}
        \begin{center}
         \includegraphics[clip, width=4cm]{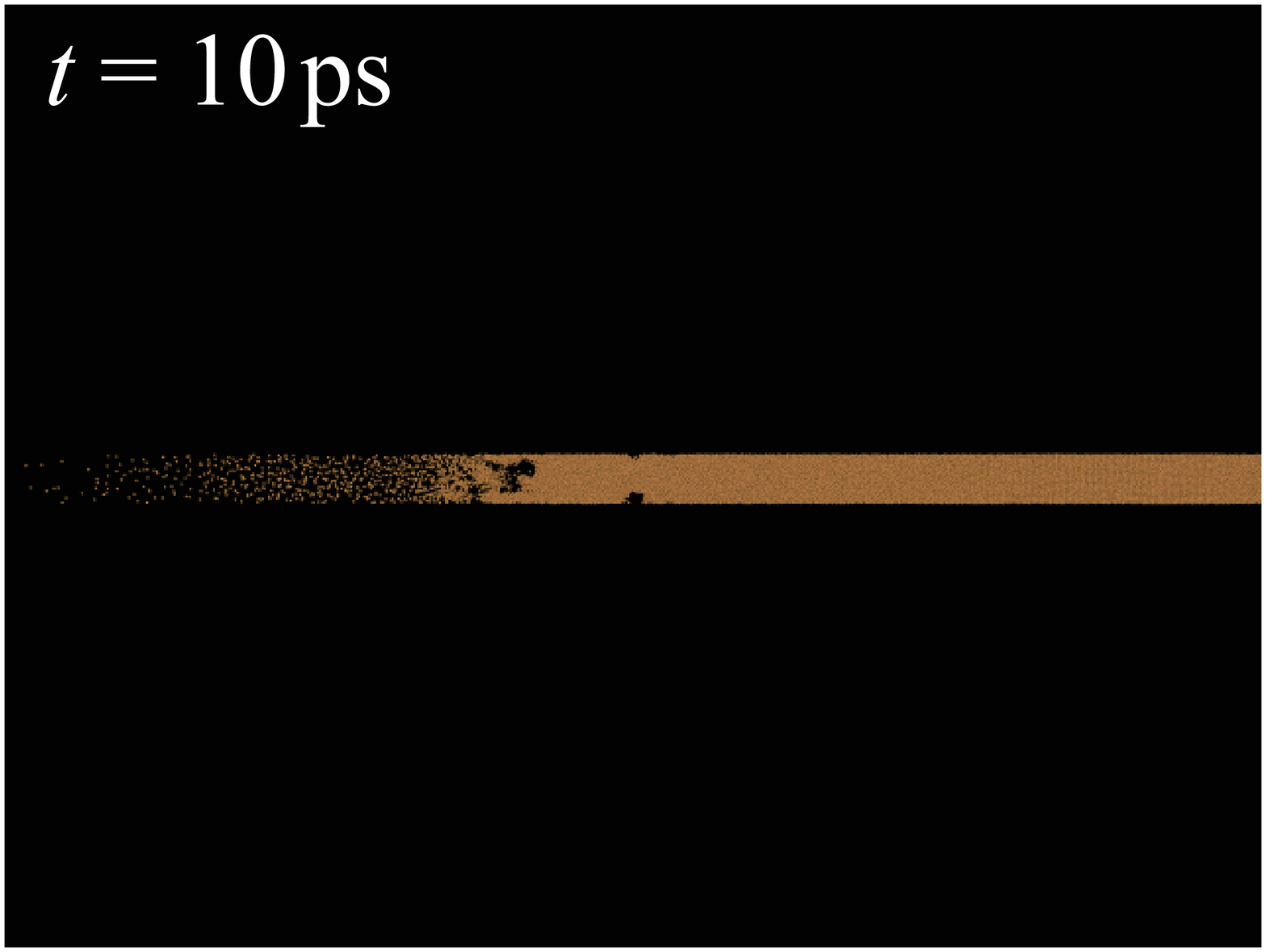}
        \end{center}
      \end{minipage}
      \begin{minipage}{0.5\hsize}
        \begin{center}
          \includegraphics[clip, width=4cm]{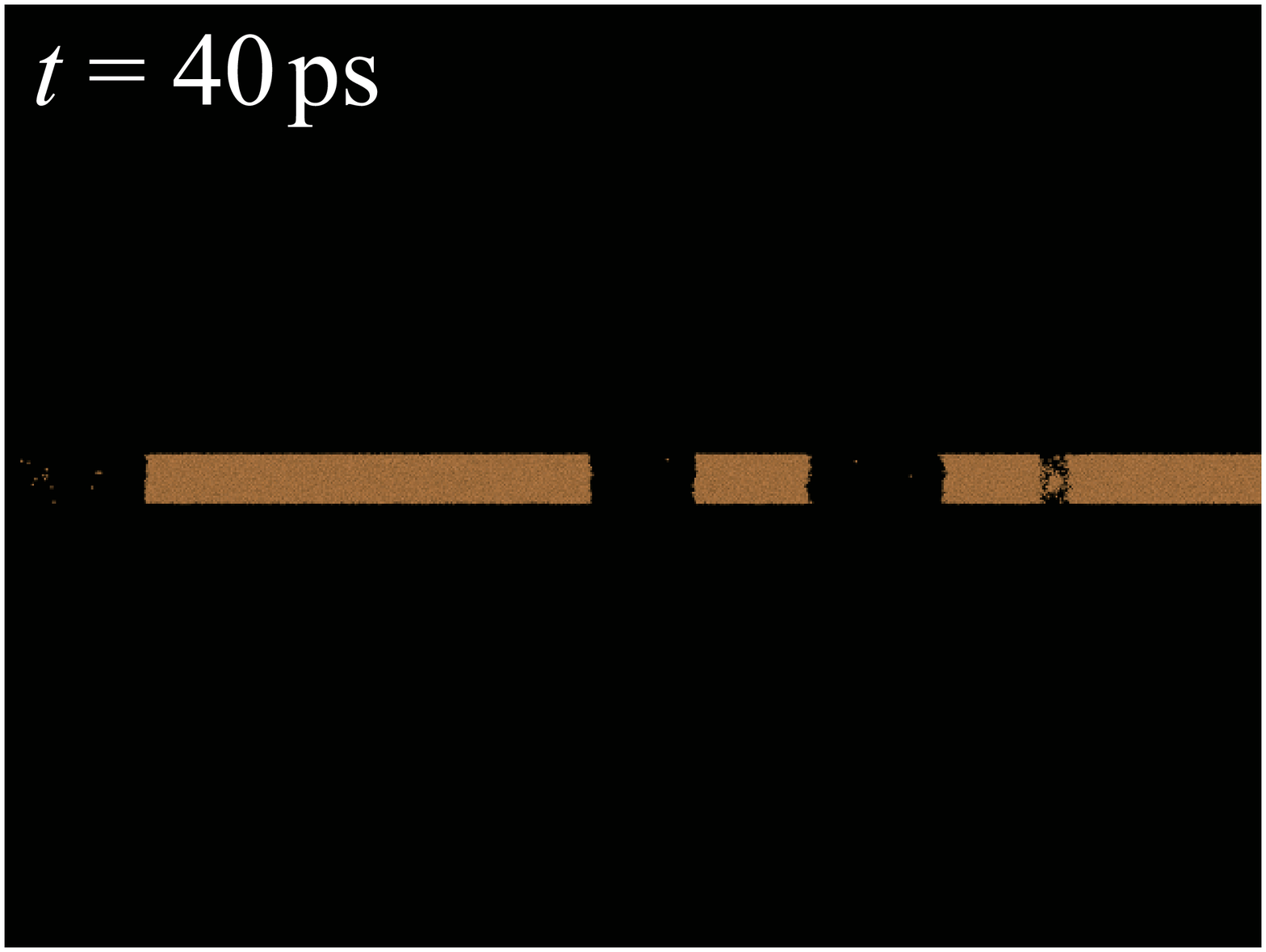}
        \end{center}
      \end{minipage}
     \\
      \begin{minipage}{0.5\hsize}
        \begin{center}
          \includegraphics[clip, width=4cm]{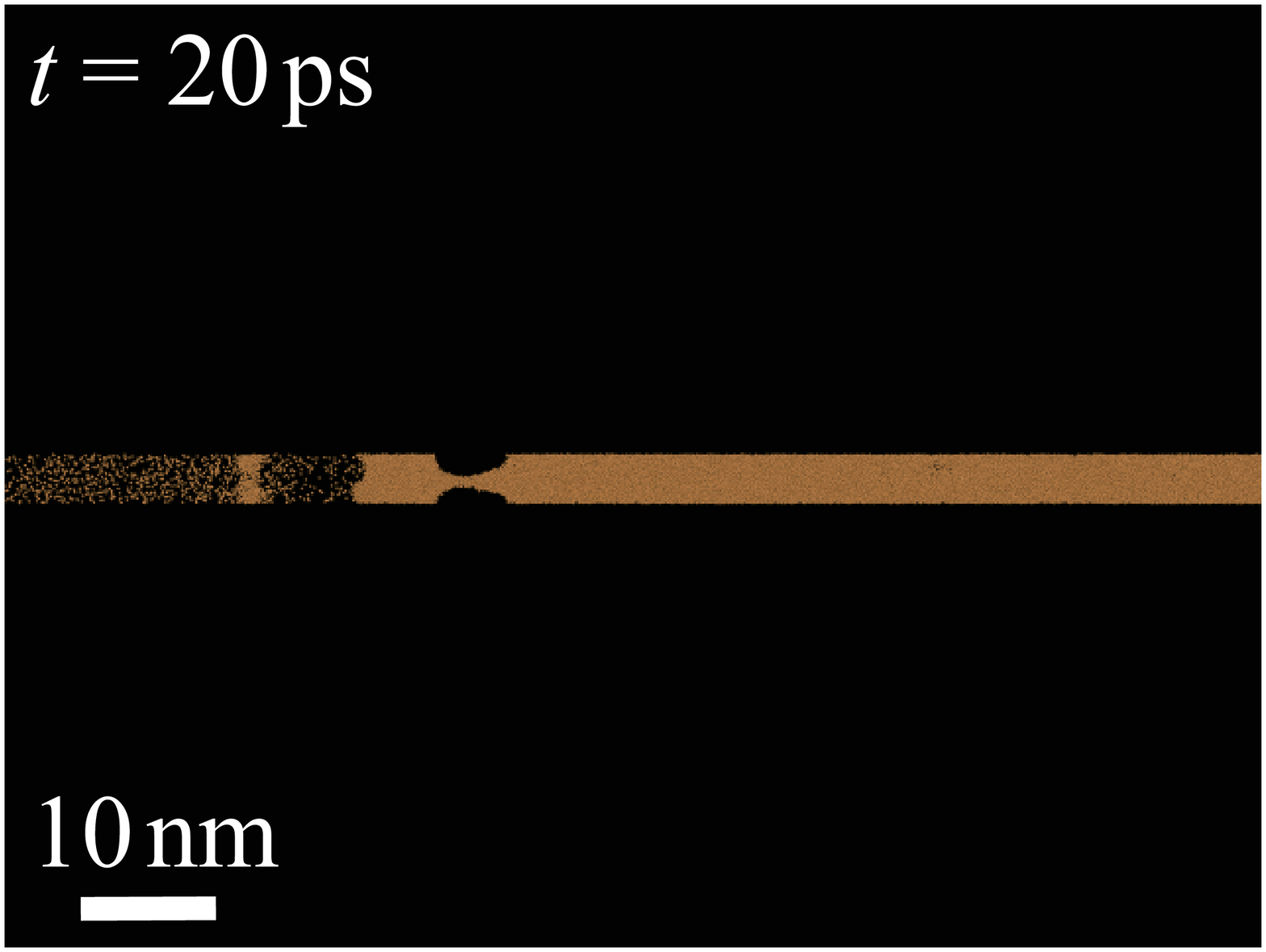}
        \end{center}
      \end{minipage}
      \begin{minipage}{0.5\hsize}
        \begin{center}
          \includegraphics[clip, width=4cm]{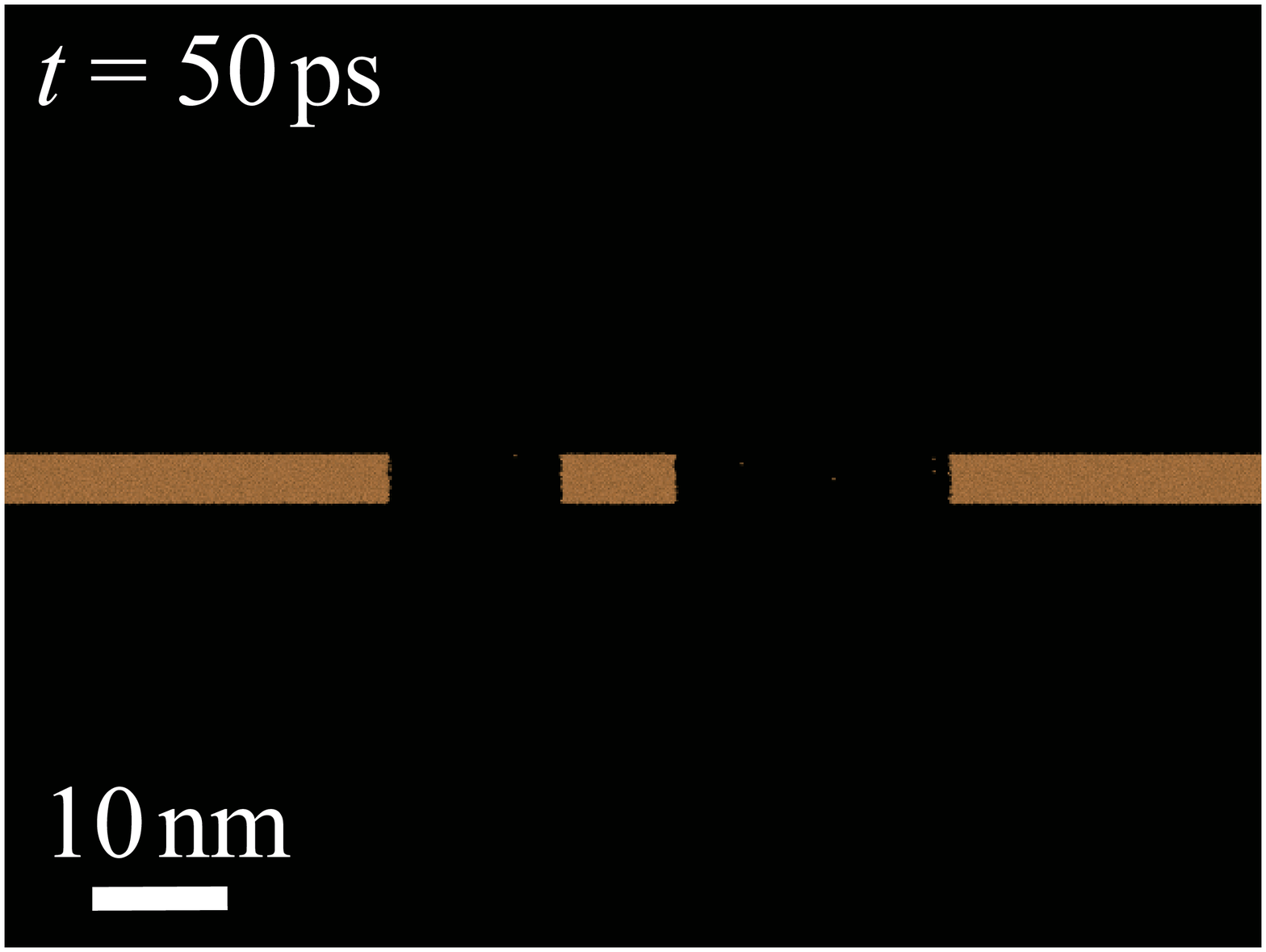}
        \end{center}
      \end{minipage}

   \end{tabular}
      \caption{Snapshots of atomic configurations near the surface after $0$-$50\,\text{ps}$ irradiation with a  $100\,\text{fs}$ pulse laser of $J_0=1.0\,\text{J}\,\text{cm}^{-2}$.   This simulation is carried out using the $T_e$-dependent IAP. }
    \label{fig:10asnaps}
  \end{center}
\end{figure}

Here, we exhibit the results of a simulation in which the fluence of the applied laser is $J_0=1.0\,\text{J}\,\text{cm}^{-2}$.

Fig.~\ref{fig:10asnaps} shows snapshots  of the atomic configurations of the Cu film irradiated by the pulse laser.
In the TTM-MD simulation, the  $T_e$-dependent IAP  is used.
The ablation depth of this simulation is $37.8\,\text{nm}$, which is estimated from the number of atoms emitted before the pressure wave reaches the MD/CM boundary.
Fig.~\ref{fig:10asnaps} shows that homogeneous evaporation (see Fig.~\ref{fig:10asnaps} at $t=10\,\text{ps}$) is observed near the laser-irradiated surface, which is considered to be an indication of phase explosion.
These results indicate that as the laser fluence becomes larger, the ablation process transforms from spallation to phase explosion.
This result is qualitatively consistent with a previous MD simulation~\cite{Wu_2013} in which the $T_e$-independent IAP is used.

\subsection{Ablation depth}
\label{sec:depth}

\begin{figure}[b]
  \begin{center}
    \begin{tabular}{cc}

      \begin{minipage}{1.0\hsize}
        \begin{center}
          \includegraphics[clip, width=8cm]{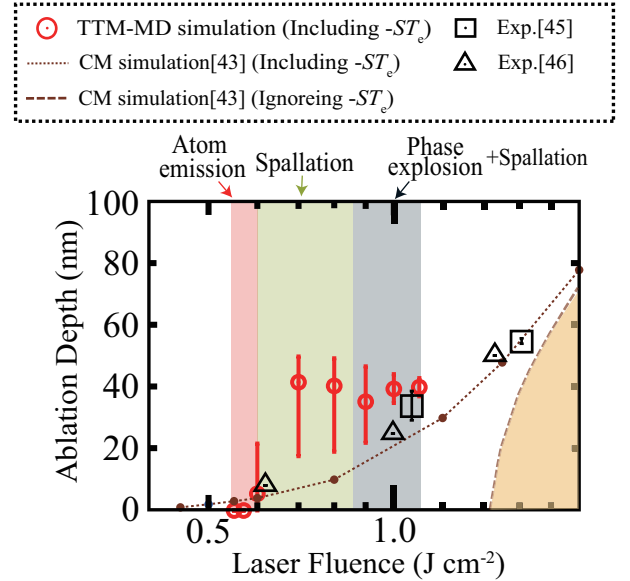}
        \end{center}
      \end{minipage} 
	
   \end{tabular}
    \caption{The developed TTM-MD results for the ablation depth and comparison of the results with  a previous CM simulation~\cite{Tanaka_2018} and experimental results.~\cite{Colombier_2005,Nielsen_2010}
     Circles represent the average ablation depth results from three TTM-MD simulations with different initial thermalization times.
    Dotted and dashed lines represent the results of previous CM calculations including the electronic entropy effect and ignoring the electronic entropy effect, respectively.~\cite{Tanaka_2018} 
Triangles and squares represent experimental results.~\cite{Colombier_2005,Nielsen_2010}
              }
    \label{fig:TTM_depth}
  \end{center}
\end{figure}

Here, we show the simulation results for the ablation depth and comparison of the results with a previous CM simulation~\cite{Tanaka_2018} and experimental results.~\cite{Colombier_2005,Nielsen_2010}
The ablation depth is estimated from the number of atoms emitted before the pressure wave reaches the MD/CM boundary.
Our calculation results for the ablation depth are plotted in Fig.~\ref{fig:TTM_depth}.
The open circles represent the average ablation depth results from three TTM-MD simulations with different initial thermalization times.
The averaged values for the ablation depth at $J_0=0.55$ and $0.57\,\text{J}\,\text{cm}^{-2}$ in the TTM-MD simulation are $0.01$ and $0.30\,\text{nm}$, respectively.
Since spallation occurs sometimes and does not occur at other times at $J_0= 0.60\,\text{J}\,\text{cm}^{-2}$, there is a wide range of ablation depth at this laser fluence,  as shown by the bars accompanying the circles.
Therefore, the ablation depth changes by more than two orders of magnitude around $J_0= 0.60\,\text{J}\text{cm}^{-2}$.
The TTM-MD simulation is qualitatively consistent with experiment,~\cite{Hashida_1999,Hashida_2010}  where similar  large changes in the ablation depth are observed.

The dotted and dashed lines represent the results of the previous CM calculation~\cite{Tanaka_2018}  including the electronic entropy effect and ignoring the electronic entropy effect, respectively.
Triangles and squares represent experimental results.~\cite{Colombier_2005,Nielsen_2010}
As shown in Fig.~\ref{fig:TTM_depth},  our TTM-MD simulations of ablation depth are in qualitative agreement with previous experimental and calculation studies.

\subsection{Pulse-width dependence of ablation threshold}
\label{sec:comp}

 \begin{figure}[b]
  \begin{center}
    \begin{tabular}{cc}

      \begin{minipage}{1.0\hsize}
        \begin{center}
          \includegraphics[clip,width=7cm]{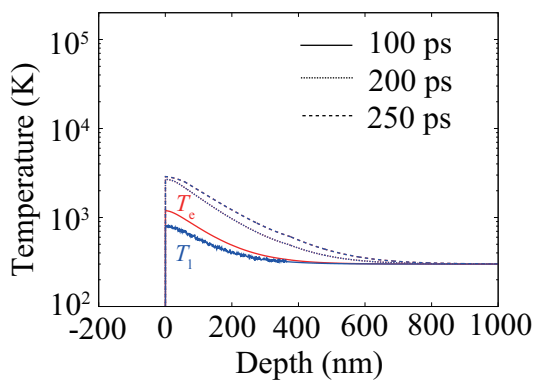}
        \end{center}
      \end{minipage}
     
   \end{tabular}
      \caption{Results of $T_e$(red line) and $T_l$(blue line) space distributions. 
      The duration of the laser pulse is $200\,\text{ps}$ and its fluence is $J_0=0.70\,\text{J}\,\text{cm}^{-2}$.  
      The fluence peak reaches the surface at $t=200\,\text{ps}$. 
       The solid, dashed, and dotted lines represent the results at $t=100$, $200$, and $250\,\text{ps}$, respectively.
       The red and blue dashed lines and red and blue dotted lines overlap with each other.
       }
    \label{fig:07longtemp}
  \end{center}
\end{figure}

Here, the pulse-width dependence of the ablation threshold is investigated.
A previous study~\cite{Hashida_2002} reported that the ablation threshold fluence of an ultrashort-pulse laser is lower  than that of a ps-laser.
We investigate whether our simulation can qualitatively reproduce this experimental result.

Figure~\ref{fig:07longtemp} represents the space distributions of  $T_e$ and $T_l$.
The time duration of the ps-laser pulse is $200\,\text{ps}$ and its fluence is $J_0=0.70\,\text{J}\,\text{cm}^{-2}$.  
The peak  of the laser pulse reaches the surface at $t=200\,\text{ps}$. 
In this simulation,  ablation does not occur, at least within $300\,\text{ps}$.
In the case of ultrashort-pulse laser ($100\,\text{fs}$ laser) irradiation at the same fluence, ablation occurs (see Fig.~\ref{fig:07snaps}).
These calculation results show that the ablation threshold fluence of the ultrashort-pulse laser  is lower  than that of the ps-laser.
This means that the developed TTM-MD simulations can qualitatively reproduce the experimental results of  the pulse-width dependence of the ablation threshold.

The reason for the difference between irradiation with the ultrashort-pulse laser and ps-laser can be explained as follows.
As can be seen from Fig.~\ref{fig:07longtemp}, in the case of ps-laser irradiation, the difference between $T_e$ and $T_l$ is small  compared with the simulation for ultrashort-pulse laser irradiation (see Fig.~\ref{fig:057temp}).
In addition, $T_e$ reaches only about $3,000\,\text{K}$, which is one order of magnitude lower than with the $100\,\text{fs}$ laser irradiation.
At low $T_e$, the electronic entropy effect is small, so atom emission and spallation are suppressed compared with ultrashort-pulse laser irradiation.
Therefore, ablation does not occur for the irradiation with the ps-pulse laser.

\section{Conclusion}
\label{sec:conc}

The microscopic mechanism of metal ablation induced by irradiation with an ultrashort-pulse laser was investigated.

First, a new  TTM-MD scheme was developed considering the electronic entropy effect.
To satisfy  the law of conservation of energy, the correction term [$ - {\sum}_i^{N^n} \bm{v}_i \frac{{\partial}}{{\partial}\bm{r}_i} \left(  S^nT_e^n \right) $] is added to the conventional equation of the TTM-MD scheme.
The energy conservation in the new scheme was verified by simulation of the laser-irradiated Cu film with $T_e$-dependent IAP.

With the TTM-MD simulations, laser ablation of Cu films with an ultrashort laser pulse was investigated.
The TTM-MD simulation predicts  high-energy atom emission and sub-nanometer depth ablation  near the ablation threshold fluence ($J_0=0.55\,\text{J}\,\text{cm}^{-2}$), which were observed in experiments.
This finding  bridges the discrepancy between  experiments and previous theoretical simulations in explaining the physical mechanism of the non-thermal ablation of metals.
Comparing the TTM-MD simulation with the electronic entropy effect and that without this effect, it is found that the electronic entropy plays an important role in atom emission.
In the case of the ultrashort laser pulse with $J_0=0.7\,\text{J}\,\text{cm}^{-2}$, atom emission and spallation were induced only in the case of the $T_e$-dependent IAP, indicating that electronic entropy plays an important role in causing not only atom emission but also spallation.
Moreover,  the TTM-MD results for ablation depth were in harmony with the CM calculation results and the experimental data, qualitatively.
Additionally, the dependence on the pulse width was analyzed.
Ablation does not occur with irradiation by 200 ps laser pulse with $J_0=0.7\,\text{J}\,\text{cm}^{-2}$ since ps-laser irradiation does not realize excessively high $T_e$  in the system.
Hence, the ablation threshold fluence of the ultrashort-pulse laser is found to be lower  than that of the ps-laser, which is consistent with experiment.

In this paper, using the developed TTM-MD scheme, we demonstrated that the electronic entropy effect plays an important role in the ultrashort-pulse laser ablation of metals, and supports the EED mechanism to explain the non-thermal ablation of metals.

\begin{acknowledgments}
  This work was supported in part by the Innovative Center for Coherent Photon Technology (ICCPT) in
Japan and by JST COI Grant Number JPMJCE1313 and also by the Ministry of Education, Culture, Sports, Science and Technology (MEXT) Quantum Leap Flagship Program (MEXT Q-LEAP) Grant No. JPMXS0118067246.
  Y. T. was supported by the Japan Society for the Promotion of Science through the Program for Leading Graduate Schools (MERIT)\@.
 \end{acknowledgments}
 
\bibliography{reference}

\begin{thebibliography}{69}%
\makeatletter
\providecommand \@ifxundefined [1]{%
 \@ifx{#1\undefined}
}%
\providecommand \@ifnum [1]{%
 \ifnum #1\expandafter \@firstoftwo
 \else \expandafter \@secondoftwo
 \fi
}%
\providecommand \@ifx [1]{%
 \ifx #1\expandafter \@firstoftwo
 \else \expandafter \@secondoftwo
 \fi
}%
\providecommand \natexlab [1]{#1}%
\providecommand \enquote  [1]{``#1''}%
\providecommand \bibnamefont  [1]{#1}%
\providecommand \bibfnamefont [1]{#1}%
\providecommand \citenamefont [1]{#1}%
\providecommand \href@noop [0]{\@secondoftwo}%
\providecommand \href [0]{\begingroup \@sanitize@url \@href}%
\providecommand \@href[1]{\@@startlink{#1}\@@href}%
\providecommand \@@href[1]{\endgroup#1\@@endlink}%
\providecommand \@sanitize@url [0]{\catcode `\\12\catcode `\$12\catcode
  `\&12\catcode `\#12\catcode `\^12\catcode `\_12\catcode `\%12\relax}%
\providecommand \@@startlink[1]{}%
\providecommand \@@endlink[0]{}%
\providecommand \url  [0]{\begingroup\@sanitize@url \@url }%
\providecommand \@url [1]{\endgroup\@href {#1}{\urlprefix }}%
\providecommand \urlprefix  [0]{URL }%
\providecommand \Eprint [0]{\href }%
\providecommand \doibase [0]{http://dx.doi.org/}%
\providecommand \selectlanguage [0]{\@gobble}%
\providecommand \bibinfo  [0]{\@secondoftwo}%
\providecommand \bibfield  [0]{\@secondoftwo}%
\providecommand \translation [1]{[#1]}%
\providecommand \BibitemOpen [0]{}%
\providecommand \bibitemStop [0]{}%
\providecommand \bibitemNoStop [0]{.\EOS\space}%
\providecommand \EOS [0]{\spacefactor3000\relax}%
\providecommand \BibitemShut  [1]{\csname bibitem#1\endcsname}%
\let\auto@bib@innerbib\@empty
\bibitem [{\citenamefont {Watanabe}, \citenamefont {Takahashi},\ and\
  \citenamefont {Tsushima}(1998)}]{Watanabe_1998}%
  \BibitemOpen
  \bibfield  {author} {\bibinfo {author} {\bibfnamefont {N.}~\bibnamefont
  {Watanabe}}, \bibinfo {author} {\bibfnamefont {N.}~\bibnamefont {Takahashi}},
  \ and\ \bibinfo {author} {\bibfnamefont {K.}~\bibnamefont {Tsushima}},\
  }\href {\doibase 10.1016/S0254-0584(98)00061-3} {\bibfield  {journal}
  {\bibinfo  {journal} {Mater. Chem. Phys.}\ }\textbf {\bibinfo {volume}
  {54}},\ \bibinfo {pages} {173} (\bibinfo {year} {1998})}\BibitemShut
  {NoStop}%
\bibitem [{\citenamefont {Yoshitake}, \citenamefont {Nagamoto},\ and\
  \citenamefont {Nagayama}(2000)}]{Yoshitake_2000}%
  \BibitemOpen
  \bibfield  {author} {\bibinfo {author} {\bibfnamefont {T.}~\bibnamefont
  {Yoshitake}}, \bibinfo {author} {\bibfnamefont {T.}~\bibnamefont {Nagamoto}},
  \ and\ \bibinfo {author} {\bibfnamefont {K.}~\bibnamefont {Nagayama}},\
  }\href {\doibase 10.1016/S0921-5107(99)00484-5} {\bibfield  {journal}
  {\bibinfo  {journal} {Mater. Sci. Eng. B}\ }\textbf {\bibinfo {volume}
  {72}},\ \bibinfo {pages} {124} (\bibinfo {year} {2000})}\BibitemShut
  {NoStop}%
\bibitem [{\citenamefont {Fojtik}\ and\ \citenamefont
  {Henglein}(1993)}]{Fojtik_1993}%
  \BibitemOpen
  \bibfield  {author} {\bibinfo {author} {\bibfnamefont {A.}~\bibnamefont
  {Fojtik}}\ and\ \bibinfo {author} {\bibfnamefont {A.}~\bibnamefont
  {Henglein}},\ }\href@noop {} {\bibfield  {journal} {\bibinfo  {journal}
  {Phys. Chem.}\ }\textbf {\bibinfo {volume} {97}},\ \bibinfo {pages} {252}
  (\bibinfo {year} {1993})}\BibitemShut {NoStop}%
\bibitem [{\citenamefont {Neddersen}, \citenamefont {Chumanov},\ and\
  \citenamefont {Cotton}(1993)}]{Neddersen_1993}%
  \BibitemOpen
  \bibfield  {author} {\bibinfo {author} {\bibfnamefont {J.}~\bibnamefont
  {Neddersen}}, \bibinfo {author} {\bibfnamefont {G.}~\bibnamefont {Chumanov}},
  \ and\ \bibinfo {author} {\bibfnamefont {T.~M.}\ \bibnamefont {Cotton}},\
  }\href@noop {} {\bibfield  {journal} {\bibinfo  {journal} {Appl. Spectrosc.}\
  }\textbf {\bibinfo {volume} {47}},\ \bibinfo {pages} {1959} (\bibinfo {year}
  {1993})}\BibitemShut {NoStop}%
\bibitem [{\citenamefont {Kobayashi}, \citenamefont {Takahashi},\ and\
  \citenamefont {Tani}(2020)}]{Kobayashi_2020}%
  \BibitemOpen
  \bibfield  {author} {\bibinfo {author} {\bibfnamefont {Y.}~\bibnamefont
  {Kobayashi}}, \bibinfo {author} {\bibfnamefont {T.}~\bibnamefont
  {Takahashi}}, \ and\ \bibinfo {author} {\bibfnamefont {S.}~\bibnamefont
  {Tani}},\ }\href {\doibase 10.11470/oubutsu.89.12_719} {\bibfield  {journal}
  {\bibinfo  {journal} {Oyo Buturi}\ }\textbf {\bibinfo {volume} {89}},\
  \bibinfo {pages} {719} (\bibinfo {year} {2020})}\BibitemShut {NoStop}%
\bibitem [{\citenamefont {Kobayashi}\ \emph {et~al.}(2021)\citenamefont
  {Kobayashi}, \citenamefont {Takahashi}, \citenamefont {Nakazato},
  \citenamefont {Sakurai}, \citenamefont {Tamaru}, \citenamefont {Ishikawa},
  \citenamefont {Sakaue},\ and\ \citenamefont {Tani}}]{Kobayashi_2021}%
  \BibitemOpen
  \bibfield  {author} {\bibinfo {author} {\bibfnamefont {Y.}~\bibnamefont
  {Kobayashi}}, \bibinfo {author} {\bibfnamefont {T.}~\bibnamefont
  {Takahashi}}, \bibinfo {author} {\bibfnamefont {T.}~\bibnamefont {Nakazato}},
  \bibinfo {author} {\bibfnamefont {H.}~\bibnamefont {Sakurai}}, \bibinfo
  {author} {\bibfnamefont {H.}~\bibnamefont {Tamaru}}, \bibinfo {author}
  {\bibfnamefont {K.~L.}\ \bibnamefont {Ishikawa}}, \bibinfo {author}
  {\bibfnamefont {K.}~\bibnamefont {Sakaue}}, \ and\ \bibinfo {author}
  {\bibfnamefont {S.}~\bibnamefont {Tani}},\ }\href {\doibase
  10.1109/JSTQE.2021.3074516} {\bibfield  {journal} {\bibinfo  {journal} {IEEE
  j. sel. top. quantum electron.}\ }\textbf {\bibinfo {volume} {27}},\ \bibinfo
  {pages} {1} (\bibinfo {year} {2021})}\BibitemShut {NoStop}%
\bibitem [{\citenamefont {Chichkov}\ \emph {et~al.}(1996)\citenamefont
  {Chichkov}, \citenamefont {Momma}, \citenamefont {Nolte}, \citenamefont {von
  Alvensleben},\ and\ \citenamefont {T{\"u}nnermann}}]{Chichkov_1996}%
  \BibitemOpen
  \bibfield  {author} {\bibinfo {author} {\bibfnamefont {B.~N.}\ \bibnamefont
  {Chichkov}}, \bibinfo {author} {\bibfnamefont {C.}~\bibnamefont {Momma}},
  \bibinfo {author} {\bibfnamefont {S.}~\bibnamefont {Nolte}}, \bibinfo
  {author} {\bibfnamefont {F.}~\bibnamefont {von Alvensleben}}, \ and\ \bibinfo
  {author} {\bibfnamefont {A.}~\bibnamefont {T{\"u}nnermann}},\ }\href
  {\doibase 10.1007/BF01567637} {\bibfield  {journal} {\bibinfo  {journal}
  {Appl. Phys. A}\ }\textbf {\bibinfo {volume} {63}},\ \bibinfo {pages} {109}
  (\bibinfo {year} {1996})}\BibitemShut {NoStop}%
\bibitem [{\citenamefont {Shaheen}, \citenamefont {Gagnon},\ and\ \citenamefont
  {Fryer}(2013)}]{Shaheen_2013}%
  \BibitemOpen
  \bibfield  {author} {\bibinfo {author} {\bibfnamefont {M.}~\bibnamefont
  {Shaheen}}, \bibinfo {author} {\bibfnamefont {J.}~\bibnamefont {Gagnon}}, \
  and\ \bibinfo {author} {\bibfnamefont {B.}~\bibnamefont {Fryer}},\ }\href
  {\doibase 10.1063/1.4819804} {\bibfield  {journal} {\bibinfo  {journal} {J.
  Appl. Phys.}\ }\textbf {\bibinfo {volume} {114}},\ \bibinfo {pages} {083110}
  (\bibinfo {year} {2013})}\BibitemShut {NoStop}%
\bibitem [{\citenamefont {Hashida}\ \emph {et~al.}(1999)\citenamefont
  {Hashida}, \citenamefont {Semerok}, \citenamefont {Gobert}, \citenamefont
  {Petite},\ and\ \citenamefont {Qgner}}]{Hashida_1999}%
  \BibitemOpen
  \bibfield  {author} {\bibinfo {author} {\bibfnamefont {M.}~\bibnamefont
  {Hashida}}, \bibinfo {author} {\bibfnamefont {A.}~\bibnamefont {Semerok}},
  \bibinfo {author} {\bibfnamefont {O.}~\bibnamefont {Gobert}}, \bibinfo
  {author} {\bibfnamefont {G.}~\bibnamefont {Petite}}, \ and\ \bibinfo {author}
  {\bibfnamefont {J.~F.}\ \bibnamefont {Qgner}},\ }\href@noop {} {\bibfield
  {journal} {\bibinfo  {journal} {Proc. SPIE}\ }\textbf {\bibinfo {volume}
  {4423}},\ \bibinfo {pages} {178} (\bibinfo {year} {1999})}\BibitemShut
  {NoStop}%
\bibitem [{\citenamefont {Hashida}\ \emph {et~al.}(2002)\citenamefont
  {Hashida}, \citenamefont {Semerok}, \citenamefont {Gobert}, \citenamefont
  {Petite}, \citenamefont {Izawa},\ and\ \citenamefont {Qgner}}]{Hashida_2002}%
  \BibitemOpen
  \bibfield  {author} {\bibinfo {author} {\bibfnamefont {M.}~\bibnamefont
  {Hashida}}, \bibinfo {author} {\bibfnamefont {A.}~\bibnamefont {Semerok}},
  \bibinfo {author} {\bibfnamefont {O.}~\bibnamefont {Gobert}}, \bibinfo
  {author} {\bibfnamefont {G.}~\bibnamefont {Petite}}, \bibinfo {author}
  {\bibfnamefont {Y.}~\bibnamefont {Izawa}}, \ and\ \bibinfo {author}
  {\bibfnamefont {J.~F.}\ \bibnamefont {Qgner}},\ }\href@noop {} {\bibfield
  {journal} {\bibinfo  {journal} {Appl. Surf. Sci.}\ }\textbf {\bibinfo
  {volume} {197}},\ \bibinfo {pages} {862} (\bibinfo {year}
  {2002})}\BibitemShut {NoStop}%
\bibitem [{\citenamefont {Miyasaka}\ \emph {et~al.}(2012)\citenamefont
  {Miyasaka}, \citenamefont {Hashida}, \citenamefont {Ikuta}, \citenamefont
  {Otani}, \citenamefont {Tokita},\ and\ \citenamefont
  {Sakabe}}]{Miyasaka_2012}%
  \BibitemOpen
  \bibfield  {author} {\bibinfo {author} {\bibfnamefont {Y.}~\bibnamefont
  {Miyasaka}}, \bibinfo {author} {\bibfnamefont {M.}~\bibnamefont {Hashida}},
  \bibinfo {author} {\bibfnamefont {Y.}~\bibnamefont {Ikuta}}, \bibinfo
  {author} {\bibfnamefont {K.}~\bibnamefont {Otani}}, \bibinfo {author}
  {\bibfnamefont {S.}~\bibnamefont {Tokita}}, \ and\ \bibinfo {author}
  {\bibfnamefont {S.}~\bibnamefont {Sakabe}},\ }\href {\doibase
  10.1103/PhysRevB.86.075431} {\bibfield  {journal} {\bibinfo  {journal} {Phys.
  Rev. B}\ }\textbf {\bibinfo {volume} {86}},\ \bibinfo {pages} {075431}
  (\bibinfo {year} {2012})}\BibitemShut {NoStop}%
\bibitem [{\citenamefont {Hashida}\ \emph {et~al.}(2010)\citenamefont
  {Hashida}, \citenamefont {Namba}, \citenamefont {Okamuro}, \citenamefont
  {Tokita},\ and\ \citenamefont {Sakabe}}]{Hashida_2010}%
  \BibitemOpen
  \bibfield  {author} {\bibinfo {author} {\bibfnamefont {M.}~\bibnamefont
  {Hashida}}, \bibinfo {author} {\bibfnamefont {S.}~\bibnamefont {Namba}},
  \bibinfo {author} {\bibfnamefont {K.}~\bibnamefont {Okamuro}}, \bibinfo
  {author} {\bibfnamefont {S.}~\bibnamefont {Tokita}}, \ and\ \bibinfo {author}
  {\bibfnamefont {S.}~\bibnamefont {Sakabe}},\ }\href@noop {} {\bibfield
  {journal} {\bibinfo  {journal} {Phys. Rev. B}\ }\textbf {\bibinfo {volume}
  {81}},\ \bibinfo {pages} {115442} (\bibinfo {year} {2010})}\BibitemShut
  {NoStop}%
\bibitem [{\citenamefont {Dachraoui}\ and\ \citenamefont
  {Husinsky}(2006)}]{Dachraoui_2006}%
  \BibitemOpen
  \bibfield  {author} {\bibinfo {author} {\bibfnamefont {H.}~\bibnamefont
  {Dachraoui}}\ and\ \bibinfo {author} {\bibfnamefont {W.}~\bibnamefont
  {Husinsky}},\ }\href {\doibase 10.1063/1.2338540} {\bibfield  {journal}
  {\bibinfo  {journal} {Appl. Phys. Lett.}\ }\textbf {\bibinfo {volume} {89}},\
  \bibinfo {pages} {104102} (\bibinfo {year} {2006})}\BibitemShut {NoStop}%
\bibitem [{\citenamefont {Dachraoui}, \citenamefont {Husinsky},\ and\
  \citenamefont {Betz}(2006)}]{Dachraoui_2006_2}%
  \BibitemOpen
  \bibfield  {author} {\bibinfo {author} {\bibfnamefont {H.}~\bibnamefont
  {Dachraoui}}, \bibinfo {author} {\bibfnamefont {W.}~\bibnamefont {Husinsky}},
  \ and\ \bibinfo {author} {\bibfnamefont {G.}~\bibnamefont {Betz}},\ }\href
  {\doibase 10.1007/s00339-006-3499-y} {\bibfield  {journal} {\bibinfo
  {journal} {Appl. Phys. A}\ }\textbf {\bibinfo {volume} {83}},\ \bibinfo
  {pages} {333} (\bibinfo {year} {2006})}\BibitemShut {NoStop}%
\bibitem [{\citenamefont {Momma}\ \emph {et~al.}(1996)\citenamefont {Momma},
  \citenamefont {Chichkov}, \citenamefont {Nolte}, \citenamefont {von
  Alvensleben}, \citenamefont {T{\"u}nnermann}, \citenamefont {Welling},\ and\
  \citenamefont {Wellegehausen}}]{Momma_1996}%
  \BibitemOpen
  \bibfield  {author} {\bibinfo {author} {\bibfnamefont {C.}~\bibnamefont
  {Momma}}, \bibinfo {author} {\bibfnamefont {B.~N.}\ \bibnamefont {Chichkov}},
  \bibinfo {author} {\bibfnamefont {S.}~\bibnamefont {Nolte}}, \bibinfo
  {author} {\bibfnamefont {F.}~\bibnamefont {von Alvensleben}}, \bibinfo
  {author} {\bibfnamefont {A.}~\bibnamefont {T{\"u}nnermann}}, \bibinfo
  {author} {\bibfnamefont {H.}~\bibnamefont {Welling}}, \ and\ \bibinfo
  {author} {\bibfnamefont {B.}~\bibnamefont {Wellegehausen}},\ }\href {\doibase
  10.1016/0030-4018(96)00250-7} {\bibfield  {journal} {\bibinfo  {journal}
  {Opt. Commun.}\ }\textbf {\bibinfo {volume} {129}},\ \bibinfo {pages} {134}
  (\bibinfo {year} {1996})}\BibitemShut {NoStop}%
\bibitem [{\citenamefont {Wu}\ and\ \citenamefont {Zhigilei}(2013)}]{Wu_2013}%
  \BibitemOpen
  \bibfield  {author} {\bibinfo {author} {\bibfnamefont {C.}~\bibnamefont
  {Wu}}\ and\ \bibinfo {author} {\bibfnamefont {L.}~\bibnamefont {Zhigilei}},\
  }\href {\doibase 10.1007/s00339-013-8086-4} {\bibfield  {journal} {\bibinfo
  {journal} {Appl. Phys. A}\ }\textbf {\bibinfo {volume} {114}},\ \bibinfo
  {pages} {11} (\bibinfo {year} {2013})}\BibitemShut {NoStop}%
\bibitem [{\citenamefont {Ji}\ and\ \citenamefont {Zhang}(2017)}]{Ji_2017}%
  \BibitemOpen
  \bibfield  {author} {\bibinfo {author} {\bibfnamefont {P.}~\bibnamefont
  {Ji}}\ and\ \bibinfo {author} {\bibfnamefont {Y.}~\bibnamefont {Zhang}},\
  }\href {\doibase 10.1007/s00339-017-1269-7} {\bibfield  {journal} {\bibinfo
  {journal} {Appl. Phys. A}\ }\textbf {\bibinfo {volume} {123}},\ \bibinfo
  {pages} {671} (\bibinfo {year} {2017})}\BibitemShut {NoStop}%
\bibitem [{\citenamefont {Foumani}\ and\ \citenamefont
  {Niknam}(2018)}]{Foumani_2018}%
  \BibitemOpen
  \bibfield  {author} {\bibinfo {author} {\bibfnamefont {A.~A.}\ \bibnamefont
  {Foumani}}\ and\ \bibinfo {author} {\bibfnamefont {A.~R.}\ \bibnamefont
  {Niknam}},\ }\href@noop {} {\bibfield  {journal} {\bibinfo  {journal} {J.
  Appl. Phys.}\ }\textbf {\bibinfo {volume} {123}},\ \bibinfo {pages} {043106}
  (\bibinfo {year} {2018})}\BibitemShut {NoStop}%
\bibitem [{\citenamefont {Sch\"afer}, \citenamefont {Urbassek},\ and\
  \citenamefont {Zhigilei}(2002)}]{Schafer_2002}%
  \BibitemOpen
  \bibfield  {author} {\bibinfo {author} {\bibfnamefont {C.}~\bibnamefont
  {Sch\"afer}}, \bibinfo {author} {\bibfnamefont {H.~M.}\ \bibnamefont
  {Urbassek}}, \ and\ \bibinfo {author} {\bibfnamefont {L.~V.}\ \bibnamefont
  {Zhigilei}},\ }\href@noop {} {\bibfield  {journal} {\bibinfo  {journal}
  {Phys. Rev. B}\ }\textbf {\bibinfo {volume} {66}},\ \bibinfo {pages} {115404}
  (\bibinfo {year} {2002})}\BibitemShut {NoStop}%
\bibitem [{\citenamefont {Ivanov}\ and\ \citenamefont
  {Zhigilei}(2003)}]{Zhiglei_2003}%
  \BibitemOpen
  \bibfield  {author} {\bibinfo {author} {\bibfnamefont {D.~S.}\ \bibnamefont
  {Ivanov}}\ and\ \bibinfo {author} {\bibfnamefont {L.~V.}\ \bibnamefont
  {Zhigilei}},\ }\href@noop {} {\bibfield  {journal} {\bibinfo  {journal}
  {Phys. Rev. B}\ }\textbf {\bibinfo {volume} {68}},\ \bibinfo {pages} {064114}
  (\bibinfo {year} {2003})}\BibitemShut {NoStop}%
\bibitem [{\citenamefont {Rouleau}\ \emph {et~al.}(2014)\citenamefont
  {Rouleau}, \citenamefont {Shih}, \citenamefont {Wu}, \citenamefont
  {Zhigilei}, \citenamefont {Puretzky},\ and\ \citenamefont
  {Geohegan}}]{Rouleau_2014}%
  \BibitemOpen
  \bibfield  {author} {\bibinfo {author} {\bibfnamefont {C.~M.}\ \bibnamefont
  {Rouleau}}, \bibinfo {author} {\bibfnamefont {C.-Y.}\ \bibnamefont {Shih}},
  \bibinfo {author} {\bibfnamefont {C.}~\bibnamefont {Wu}}, \bibinfo {author}
  {\bibfnamefont {L.~V.}\ \bibnamefont {Zhigilei}}, \bibinfo {author}
  {\bibfnamefont {A.~A.}\ \bibnamefont {Puretzky}}, \ and\ \bibinfo {author}
  {\bibfnamefont {D.~B.}\ \bibnamefont {Geohegan}},\ }\href@noop {} {\bibfield
  {journal} {\bibinfo  {journal} {Appl. Phys. Lett.}\ }\textbf {\bibinfo
  {volume} {104}},\ \bibinfo {pages} {193106} (\bibinfo {year}
  {2014})}\BibitemShut {NoStop}%
\bibitem [{\citenamefont {Zhigilei}, \citenamefont {Lin},\ and\ \citenamefont
  {Ivanov}(2009)}]{Zhigilei_2009}%
  \BibitemOpen
  \bibfield  {author} {\bibinfo {author} {\bibfnamefont {L.}~\bibnamefont
  {Zhigilei}}, \bibinfo {author} {\bibfnamefont {Z.}~\bibnamefont {Lin}}, \
  and\ \bibinfo {author} {\bibfnamefont {D.}~\bibnamefont {Ivanov}},\ }\href
  {\doibase 10.1021/jp902294m} {\bibfield  {journal} {\bibinfo  {journal} {J
  Phys Chem C}\ }\textbf {\bibinfo {volume} {113}},\ \bibinfo {pages} {11892}
  (\bibinfo {year} {2009})}\BibitemShut {NoStop}%
\bibitem [{\citenamefont {Gan}\ and\ \citenamefont {Chen}(2009)}]{Gan_2009}%
  \BibitemOpen
  \bibfield  {author} {\bibinfo {author} {\bibfnamefont {Y.}~\bibnamefont
  {Gan}}\ and\ \bibinfo {author} {\bibfnamefont {J.~K.}\ \bibnamefont {Chen}},\
  }\href@noop {} {\bibfield  {journal} {\bibinfo  {journal} {Appl. Phys.
  Lett.}\ }\textbf {\bibinfo {volume} {94}},\ \bibinfo {pages} {201116}
  (\bibinfo {year} {2009})}\BibitemShut {NoStop}%
\bibitem [{\citenamefont {von~der Linde}, \citenamefont {Sokolowski-Tinten},\
  and\ \citenamefont {Bialkowski}(1997)}]{Linde_1997}%
  \BibitemOpen
  \bibfield  {author} {\bibinfo {author} {\bibfnamefont {D.}~\bibnamefont
  {von~der Linde}}, \bibinfo {author} {\bibfnamefont {K.}~\bibnamefont
  {Sokolowski-Tinten}}, \ and\ \bibinfo {author} {\bibfnamefont
  {J.}~\bibnamefont {Bialkowski}},\ }\href {\doibase
  https://doi.org/10.1016/S0169-4332(96)00611-3} {\bibfield  {journal}
  {\bibinfo  {journal} {Appl. Surf. Sci.}\ }\textbf {\bibinfo {volume}
  {109-110}},\ \bibinfo {pages} {1} (\bibinfo {year} {1997})}\BibitemShut
  {NoStop}%
\bibitem [{\citenamefont {Sokolowski-Tinten}\ \emph
  {et~al.}(1998{\natexlab{a}})\citenamefont {Sokolowski-Tinten}, \citenamefont
  {Bialkowski}, \citenamefont {Cavalleri}, \citenamefont {Boring},
  \citenamefont {Sch{\"u}ler},\ and\ \citenamefont {von~der
  Linde}}]{Sokolowski_1998}%
  \BibitemOpen
  \bibfield  {author} {\bibinfo {author} {\bibfnamefont {K.}~\bibnamefont
  {Sokolowski-Tinten}}, \bibinfo {author} {\bibfnamefont {J.}~\bibnamefont
  {Bialkowski}}, \bibinfo {author} {\bibfnamefont {A.}~\bibnamefont
  {Cavalleri}}, \bibinfo {author} {\bibfnamefont {M.}~\bibnamefont {Boring}},
  \bibinfo {author} {\bibfnamefont {H.}~\bibnamefont {Sch{\"u}ler}}, \ and\
  \bibinfo {author} {\bibfnamefont {D.}~\bibnamefont {von~der Linde}},\
  }\href@noop {} {\bibfield  {journal} {\bibinfo  {journal} {Proc. SPIE}\
  }\textbf {\bibinfo {volume} {3343}},\ \bibinfo {pages} {46} (\bibinfo {year}
  {1998}{\natexlab{a}})}\BibitemShut {NoStop}%
\bibitem [{\citenamefont {Sokolowski-Tinten}\ \emph
  {et~al.}(1998{\natexlab{b}})\citenamefont {Sokolowski-Tinten}, \citenamefont
  {Bialkowski}, \citenamefont {Cavalleri}, \citenamefont {von~der Linde},
  \citenamefont {Oparin}, \citenamefont {{Meyer-ter-Vehn}},\ and\ \citenamefont
  {Anisimov}}]{Sokolowski2_1998}%
  \BibitemOpen
  \bibfield  {author} {\bibinfo {author} {\bibfnamefont {K.}~\bibnamefont
  {Sokolowski-Tinten}}, \bibinfo {author} {\bibfnamefont {J.}~\bibnamefont
  {Bialkowski}}, \bibinfo {author} {\bibfnamefont {A.}~\bibnamefont
  {Cavalleri}}, \bibinfo {author} {\bibfnamefont {D.}~\bibnamefont {von~der
  Linde}}, \bibinfo {author} {\bibfnamefont {A.}~\bibnamefont {Oparin}},
  \bibinfo {author} {\bibfnamefont {J.}~\bibnamefont {{Meyer-ter-Vehn}}}, \
  and\ \bibinfo {author} {\bibfnamefont {S.~I.}\ \bibnamefont {Anisimov}},\
  }\href@noop {} {\bibfield  {journal} {\bibinfo  {journal} {Phys. Rev. Lett.}\
  }\textbf {\bibinfo {volume} {81}},\ \bibinfo {pages} {224} (\bibinfo {year}
  {1998}{\natexlab{b}})}\BibitemShut {NoStop}%
\bibitem [{\citenamefont {Sokolowski-Tinten}\ and\ \citenamefont {von~der
  Linde}(2000)}]{Libde_2000}%
  \BibitemOpen
  \bibfield  {author} {\bibinfo {author} {\bibfnamefont {K.}~\bibnamefont
  {Sokolowski-Tinten}}\ and\ \bibinfo {author} {\bibfnamefont {D.}~\bibnamefont
  {von~der Linde}},\ }\href@noop {} {\bibfield  {journal} {\bibinfo  {journal}
  {Appl. Surf. Sci.}\ }\textbf {\bibinfo {volume} {154-155}},\ \bibinfo {pages}
  {1} (\bibinfo {year} {2000})}\BibitemShut {NoStop}%
\bibitem [{\citenamefont {Miotello}\ and\ \citenamefont
  {Kelly}(1999)}]{Miotello_1999}%
  \BibitemOpen
  \bibfield  {author} {\bibinfo {author} {\bibfnamefont {A.}~\bibnamefont
  {Miotello}}\ and\ \bibinfo {author} {\bibfnamefont {R.}~\bibnamefont
  {Kelly}},\ }\href {\doibase 10.1007/s003399900296} {\bibfield  {journal}
  {\bibinfo  {journal} {Appl. Phys. A}\ }\textbf {\bibinfo {volume} {69}},\
  \bibinfo {pages} {S67} (\bibinfo {year} {1999})}\BibitemShut {NoStop}%
\bibitem [{\citenamefont {Bulgakova}\ and\ \citenamefont
  {Bulgakov}(2001)}]{Miotello_2001}%
  \BibitemOpen
  \bibfield  {author} {\bibinfo {author} {\bibfnamefont {N.}~\bibnamefont
  {Bulgakova}}\ and\ \bibinfo {author} {\bibfnamefont {A.}~\bibnamefont
  {Bulgakov}},\ }\href {\doibase https://doi.org/10.1007/s003390000686}
  {\bibfield  {journal} {\bibinfo  {journal} {Appl. Phys. A}\ }\textbf
  {\bibinfo {volume} {73}},\ \bibinfo {pages} {199} (\bibinfo {year}
  {2001})}\BibitemShut {NoStop}%
\bibitem [{\citenamefont {Tao}\ and\ \citenamefont {Wu}(2014)}]{Tao_2014}%
  \BibitemOpen
  \bibfield  {author} {\bibinfo {author} {\bibfnamefont {S.}~\bibnamefont
  {Tao}}\ and\ \bibinfo {author} {\bibfnamefont {B.}~\bibnamefont {Wu}},\
  }\href@noop {} {\bibfield  {journal} {\bibinfo  {journal} {Appl. Surf. Sci.}\
  }\textbf {\bibinfo {volume} {298}},\ \bibinfo {pages} {90} (\bibinfo {year}
  {2014})}\BibitemShut {NoStop}%
\bibitem [{\citenamefont {Li}\ \emph {et~al.}(2015)\citenamefont {Li},
  \citenamefont {Li}, \citenamefont {Zhang}, \citenamefont {Tian},
  \citenamefont {Li}, \citenamefont {Liu}, \citenamefont {Jiang}, \citenamefont
  {Chen},\ and\ \citenamefont {Jin}}]{Li_2015}%
  \BibitemOpen
  \bibfield  {author} {\bibinfo {author} {\bibfnamefont {S.}~\bibnamefont
  {Li}}, \bibinfo {author} {\bibfnamefont {S.}~\bibnamefont {Li}}, \bibinfo
  {author} {\bibfnamefont {F.}~\bibnamefont {Zhang}}, \bibinfo {author}
  {\bibfnamefont {D.}~\bibnamefont {Tian}}, \bibinfo {author} {\bibfnamefont
  {H.}~\bibnamefont {Li}}, \bibinfo {author} {\bibfnamefont {D.}~\bibnamefont
  {Liu}}, \bibinfo {author} {\bibfnamefont {Y.}~\bibnamefont {Jiang}}, \bibinfo
  {author} {\bibfnamefont {A.}~\bibnamefont {Chen}}, \ and\ \bibinfo {author}
  {\bibfnamefont {M.}~\bibnamefont {Jin}},\ }\href@noop {} {\bibfield
  {journal} {\bibinfo  {journal} {Appl. Surf. Sci.}\ }\textbf {\bibinfo
  {volume} {355}},\ \bibinfo {pages} {681} (\bibinfo {year}
  {2015})}\BibitemShut {NoStop}%
\bibitem [{\citenamefont {Norman}, \citenamefont {Starikov},\ and\
  \citenamefont {Stegailov}(2012)}]{Norman_2012}%
  \BibitemOpen
  \bibfield  {author} {\bibinfo {author} {\bibfnamefont {G.~E.}\ \bibnamefont
  {Norman}}, \bibinfo {author} {\bibfnamefont {S.~V.}\ \bibnamefont
  {Starikov}}, \ and\ \bibinfo {author} {\bibfnamefont {V.~V.}\ \bibnamefont
  {Stegailov}},\ }\href {\doibase 10.1134/S1063776112040115} {\bibfield
  {journal} {\bibinfo  {journal} {J. Exp. Theor. Phys.}\ }\textbf {\bibinfo
  {volume} {114}},\ \bibinfo {pages} {792} (\bibinfo {year}
  {2012})}\BibitemShut {NoStop}%
\bibitem [{\citenamefont {Norman}\ \emph {et~al.}(2013)\citenamefont {Norman},
  \citenamefont {Starikov}, \citenamefont {Stegailov}, \citenamefont {Saitov},\
  and\ \citenamefont {Zhilyaev}}]{Norman_2013}%
  \BibitemOpen
  \bibfield  {author} {\bibinfo {author} {\bibfnamefont {G.~E.}\ \bibnamefont
  {Norman}}, \bibinfo {author} {\bibfnamefont {S.~V.}\ \bibnamefont
  {Starikov}}, \bibinfo {author} {\bibfnamefont {V.~V.}\ \bibnamefont
  {Stegailov}}, \bibinfo {author} {\bibfnamefont {I.}~\bibnamefont {Saitov}}, \
  and\ \bibinfo {author} {\bibfnamefont {P.~A.}\ \bibnamefont {Zhilyaev}},\
  }\href {\doibase 10.1002/ctpp.201310025} {\bibfield  {journal} {\bibinfo
  {journal} {Contrib. Plasma Phys.}\ }\textbf {\bibinfo {volume} {53}},\
  \bibinfo {pages} {129} (\bibinfo {year} {2013})}\BibitemShut {NoStop}%
\bibitem [{\citenamefont {Stegailov}\ and\ \citenamefont
  {Zhilyaev}(2015)}]{Stegailov_2015}%
  \BibitemOpen
  \bibfield  {author} {\bibinfo {author} {\bibfnamefont {V.}~\bibnamefont
  {Stegailov}}\ and\ \bibinfo {author} {\bibfnamefont {P.}~\bibnamefont
  {Zhilyaev}},\ }\href {\doibase 10.1002/ctpp.201400103} {\bibfield  {journal}
  {\bibinfo  {journal} {Contrib. Plasma Phys.}\ }\textbf {\bibinfo {volume}
  {55}},\ \bibinfo {pages} {164} (\bibinfo {year} {2015})}\BibitemShut
  {NoStop}%
\bibitem [{\citenamefont {Stegailov}\ and\ \citenamefont
  {Zhilyaev}(2016)}]{Stegailov_2016}%
  \BibitemOpen
  \bibfield  {author} {\bibinfo {author} {\bibfnamefont {V.}~\bibnamefont
  {Stegailov}}\ and\ \bibinfo {author} {\bibfnamefont {P.}~\bibnamefont
  {Zhilyaev}},\ }\href {\doibase 10.1080/00268976.2015.1105390} {\bibfield
  {journal} {\bibinfo  {journal} {Mol. Phys.}\ }\textbf {\bibinfo {volume}
  {114}},\ \bibinfo {pages} {509} (\bibinfo {year} {2016})}\BibitemShut
  {NoStop}%
\bibitem [{\citenamefont {Ilnitsky}\ \emph {et~al.}(2016)\citenamefont
  {Ilnitsky}, \citenamefont {Khokhlov}, \citenamefont {Zhakhovsky},
  \citenamefont {Petrov}, \citenamefont {Migdal},\ and\ \citenamefont
  {Inogamov}}]{Ilnitsky_2016}%
  \BibitemOpen
  \bibfield  {author} {\bibinfo {author} {\bibfnamefont {D.~K.}\ \bibnamefont
  {Ilnitsky}}, \bibinfo {author} {\bibfnamefont {V.~A.}\ \bibnamefont
  {Khokhlov}}, \bibinfo {author} {\bibfnamefont {V.~V.}\ \bibnamefont
  {Zhakhovsky}}, \bibinfo {author} {\bibfnamefont {Y.~V.}\ \bibnamefont
  {Petrov}}, \bibinfo {author} {\bibfnamefont {K.~P.}\ \bibnamefont {Migdal}},
  \ and\ \bibinfo {author} {\bibfnamefont {N.~A.}\ \bibnamefont {Inogamov}},\
  }\href {\doibase 10.1088/1742-6596/774/1/012101} {\bibfield  {journal}
  {\bibinfo  {journal} {Journal of Physics: Conference Series}\ }\textbf
  {\bibinfo {volume} {774}},\ \bibinfo {pages} {012101} (\bibinfo {year}
  {2016})}\BibitemShut {NoStop}%
\bibitem [{\citenamefont {Zhao}\ and\ \citenamefont {Shin}(2013)}]{Zhao_2013}%
  \BibitemOpen
  \bibfield  {author} {\bibinfo {author} {\bibfnamefont {X.}~\bibnamefont
  {Zhao}}\ and\ \bibinfo {author} {\bibfnamefont {Y.~C.}\ \bibnamefont
  {Shin}},\ }\href {\doibase 10.1088/0022-3727/46/33/335501} {\bibfield
  {journal} {\bibinfo  {journal} {J. Phys. D: Appl. Phys.}\ }\textbf {\bibinfo
  {volume} {46}},\ \bibinfo {pages} {335501} (\bibinfo {year}
  {2013})}\BibitemShut {NoStop}%
\bibitem [{\citenamefont {Stoian}\ \emph
  {et~al.}(2000{\natexlab{a}})\citenamefont {Stoian}, \citenamefont
  {Ashkenasi}, \citenamefont {Rosenfeld}, \citenamefont {Wittmann},
  \citenamefont {Kelly},\ and\ \citenamefont {Campbell}}]{Stoian_2000}%
  \BibitemOpen
  \bibfield  {author} {\bibinfo {author} {\bibfnamefont {R.}~\bibnamefont
  {Stoian}}, \bibinfo {author} {\bibfnamefont {D.}~\bibnamefont {Ashkenasi}},
  \bibinfo {author} {\bibfnamefont {A.}~\bibnamefont {Rosenfeld}}, \bibinfo
  {author} {\bibfnamefont {M.}~\bibnamefont {Wittmann}}, \bibinfo {author}
  {\bibfnamefont {R.}~\bibnamefont {Kelly}}, \ and\ \bibinfo {author}
  {\bibfnamefont {E.~E.~B.}\ \bibnamefont {Campbell}},\ }\href {\doibase
  10.1016/S0168-583X(99)00725-9} {\bibfield  {journal} {\bibinfo  {journal}
  {Nucl. Instrum. Methods Phys. Res. B}\ }\textbf {\bibinfo {volume} {166}},\
  \bibinfo {pages} {682} (\bibinfo {year} {2000}{\natexlab{a}})}\BibitemShut
  {NoStop}%
\bibitem [{\citenamefont {Stoian}\ \emph
  {et~al.}(2000{\natexlab{b}})\citenamefont {Stoian}, \citenamefont
  {Ashkenasi}, \citenamefont {Rosenfeld},\ and\ \citenamefont
  {Campbell}}]{Stoian_2000_2}%
  \BibitemOpen
  \bibfield  {author} {\bibinfo {author} {\bibfnamefont {R.}~\bibnamefont
  {Stoian}}, \bibinfo {author} {\bibfnamefont {D.}~\bibnamefont {Ashkenasi}},
  \bibinfo {author} {\bibfnamefont {A.}~\bibnamefont {Rosenfeld}}, \ and\
  \bibinfo {author} {\bibfnamefont {E.~E.~B.}\ \bibnamefont {Campbell}},\
  }\href {\doibase 10.1103/PhysRevB.62.13167} {\bibfield  {journal} {\bibinfo
  {journal} {Phys. Rev. B}\ }\textbf {\bibinfo {volume} {62}},\ \bibinfo
  {pages} {13167} (\bibinfo {year} {2000}{\natexlab{b}})}\BibitemShut {NoStop}%
\bibitem [{\citenamefont {Sato}\ \emph {et~al.}(2008)\citenamefont {Sato},
  \citenamefont {Okino}, \citenamefont {Yamanouchi}, \citenamefont {Yagishita},
  \citenamefont {Kannari}, \citenamefont {Yamakawa}, \citenamefont
  {Midorikawa}, \citenamefont {Nakano}, \citenamefont {Yabashi}, \citenamefont
  {Nagasono},\ and\ \citenamefont {Ishikawa}}]{Sato_2008}%
  \BibitemOpen
  \bibfield  {author} {\bibinfo {author} {\bibfnamefont {T.}~\bibnamefont
  {Sato}}, \bibinfo {author} {\bibfnamefont {T.}~\bibnamefont {Okino}},
  \bibinfo {author} {\bibfnamefont {K.}~\bibnamefont {Yamanouchi}}, \bibinfo
  {author} {\bibfnamefont {A.}~\bibnamefont {Yagishita}}, \bibinfo {author}
  {\bibfnamefont {F.}~\bibnamefont {Kannari}}, \bibinfo {author} {\bibfnamefont
  {K.}~\bibnamefont {Yamakawa}}, \bibinfo {author} {\bibfnamefont
  {K.}~\bibnamefont {Midorikawa}}, \bibinfo {author} {\bibfnamefont
  {H.}~\bibnamefont {Nakano}}, \bibinfo {author} {\bibfnamefont
  {M.}~\bibnamefont {Yabashi}}, \bibinfo {author} {\bibfnamefont
  {M.}~\bibnamefont {Nagasono}}, \ and\ \bibinfo {author} {\bibfnamefont
  {T.}~\bibnamefont {Ishikawa}},\ }\href {\doibase 10.1063/1.2911742}
  {\bibfield  {journal} {\bibinfo  {journal} {Appl. Phys. Lett.}\ }\textbf
  {\bibinfo {volume} {92}},\ \bibinfo {pages} {154103} (\bibinfo {year}
  {2008})}\BibitemShut {NoStop}%
\bibitem [{\citenamefont {Li}\ \emph {et~al.}(2011)\citenamefont {Li},
  \citenamefont {Wang}, \citenamefont {Chen}, \citenamefont {Zhou},
  \citenamefont {Mao},\ and\ \citenamefont {Cao}}]{Li_2011}%
  \BibitemOpen
  \bibfield  {author} {\bibinfo {author} {\bibfnamefont {J.}~\bibnamefont
  {Li}}, \bibinfo {author} {\bibfnamefont {X.}~\bibnamefont {Wang}}, \bibinfo
  {author} {\bibfnamefont {Z.}~\bibnamefont {Chen}}, \bibinfo {author}
  {\bibfnamefont {J.}~\bibnamefont {Zhou}}, \bibinfo {author} {\bibfnamefont
  {S.~S.}\ \bibnamefont {Mao}}, \ and\ \bibinfo {author} {\bibfnamefont
  {J.}~\bibnamefont {Cao}},\ }\href {\doibase 10.1063/1.3533811} {\bibfield
  {journal} {\bibinfo  {journal} {Appl. Phys. Lett.}\ }\textbf {\bibinfo
  {volume} {98}},\ \bibinfo {pages} {011501} (\bibinfo {year}
  {2011})}\BibitemShut {NoStop}%
\bibitem [{\citenamefont {Lin}\ \emph {et~al.}(2012)\citenamefont {Lin},
  \citenamefont {Chen}, \citenamefont {Jiang},\ and\ \citenamefont
  {Zhang}}]{Lin_2012}%
  \BibitemOpen
  \bibfield  {author} {\bibinfo {author} {\bibfnamefont {X.}~\bibnamefont
  {Lin}}, \bibinfo {author} {\bibfnamefont {H.}~\bibnamefont {Chen}}, \bibinfo
  {author} {\bibfnamefont {S.}~\bibnamefont {Jiang}}, \ and\ \bibinfo {author}
  {\bibfnamefont {C.}~\bibnamefont {Zhang}},\ }\href {\doibase
  0.1007/s11431-011-4702-8} {\bibfield  {journal} {\bibinfo  {journal} {Sci.
  China Technol. Sci.}\ }\textbf {\bibinfo {volume} {55}},\ \bibinfo {pages}
  {694} (\bibinfo {year} {2012})}\BibitemShut {NoStop}%
\bibitem [{\citenamefont {Tanaka}\ and\ \citenamefont
  {Tsuneyuki}(2018)}]{Tanaka_2018}%
  \BibitemOpen
  \bibfield  {author} {\bibinfo {author} {\bibfnamefont {Y.}~\bibnamefont
  {Tanaka}}\ and\ \bibinfo {author} {\bibfnamefont {S.}~\bibnamefont
  {Tsuneyuki}},\ }\href {\doibase https://doi.org/10.7567/APEX.11.046701}
  {\bibfield  {journal} {\bibinfo  {journal} {Appl. Phys. Exp.}\ }\textbf
  {\bibinfo {volume} {11}},\ \bibinfo {pages} {046701} (\bibinfo {year}
  {2018})}\BibitemShut {NoStop}%
\bibitem [{\citenamefont {Anisimov}, \citenamefont {Kapeliovich},\ and\
  \citenamefont {Perel'man}(1974)}]{Anisimov_1973}%
  \BibitemOpen
  \bibfield  {author} {\bibinfo {author} {\bibfnamefont {S.~I.}\ \bibnamefont
  {Anisimov}}, \bibinfo {author} {\bibfnamefont {B.~L.}\ \bibnamefont
  {Kapeliovich}}, \ and\ \bibinfo {author} {\bibfnamefont {T.~L.}\ \bibnamefont
  {Perel'man}},\ }\href@noop {} {\bibfield  {journal} {\bibinfo  {journal}
  {Sov. Phys. -JETP}\ }\textbf {\bibinfo {volume} {39}},\ \bibinfo {pages}
  {375} (\bibinfo {year} {1974})}\BibitemShut {NoStop}%
\bibitem [{\citenamefont {Colombier}\ \emph {et~al.}(2005)\citenamefont
  {Colombier}, \citenamefont {Combis}, \citenamefont {Bonneau}, \citenamefont
  {Le~Harzic},\ and\ \citenamefont {Audouard}}]{Colombier_2005}%
  \BibitemOpen
  \bibfield  {author} {\bibinfo {author} {\bibfnamefont {J.~P.}\ \bibnamefont
  {Colombier}}, \bibinfo {author} {\bibfnamefont {P.}~\bibnamefont {Combis}},
  \bibinfo {author} {\bibfnamefont {F.}~\bibnamefont {Bonneau}}, \bibinfo
  {author} {\bibfnamefont {R.}~\bibnamefont {Le~Harzic}}, \ and\ \bibinfo
  {author} {\bibfnamefont {E.}~\bibnamefont {Audouard}},\ }\href {\doibase
  10.1103/PhysRevB.71.165406} {\bibfield  {journal} {\bibinfo  {journal} {Phys.
  Rev. B}\ }\textbf {\bibinfo {volume} {71}},\ \bibinfo {pages} {165406}
  (\bibinfo {year} {2005})}\BibitemShut {NoStop}%
\bibitem [{\citenamefont {Byskov-Nielsen}\ \emph {et~al.}(2010)\citenamefont
  {Byskov-Nielsen}, \citenamefont {Savolainen}, \citenamefont {Christensen},\
  and\ \citenamefont {Balling}}]{Nielsen_2010}%
  \BibitemOpen
  \bibfield  {author} {\bibinfo {author} {\bibfnamefont {J.}~\bibnamefont
  {Byskov-Nielsen}}, \bibinfo {author} {\bibfnamefont {J.~M.}\ \bibnamefont
  {Savolainen}}, \bibinfo {author} {\bibfnamefont {M.~S.}\ \bibnamefont
  {Christensen}}, \ and\ \bibinfo {author} {\bibfnamefont {P.}~\bibnamefont
  {Balling}},\ }\href {\doibase 10.1007/s00339-010-5766-1} {\bibfield
  {journal} {\bibinfo  {journal} {Appl. Phys. A}\ }\textbf {\bibinfo {volume}
  {101}},\ \bibinfo {pages} {97} (\bibinfo {year} {2010})}\BibitemShut
  {NoStop}%
\bibitem [{\citenamefont {Murphy}\ \emph {et~al.}(2015)\citenamefont {Murphy},
  \citenamefont {Daraszewicz}, \citenamefont {Giret}, \citenamefont {Watkins},
  \citenamefont {Shluger}, \citenamefont {Tanimura},\ and\ \citenamefont
  {Duffy}}]{Murphy_2015}%
  \BibitemOpen
  \bibfield  {author} {\bibinfo {author} {\bibfnamefont {S.~T.}\ \bibnamefont
  {Murphy}}, \bibinfo {author} {\bibfnamefont {S.~L.}\ \bibnamefont
  {Daraszewicz}}, \bibinfo {author} {\bibfnamefont {Y.}~\bibnamefont {Giret}},
  \bibinfo {author} {\bibfnamefont {M.}~\bibnamefont {Watkins}}, \bibinfo
  {author} {\bibfnamefont {A.~L.}\ \bibnamefont {Shluger}}, \bibinfo {author}
  {\bibfnamefont {K.}~\bibnamefont {Tanimura}}, \ and\ \bibinfo {author}
  {\bibfnamefont {D.~M.}\ \bibnamefont {Duffy}},\ }\href {\doibase
  10.1103/PhysRevB.92.134110} {\bibfield  {journal} {\bibinfo  {journal} {Phys.
  Rev. B}\ }\textbf {\bibinfo {volume} {92}},\ \bibinfo {pages} {134110}
  (\bibinfo {year} {2015})}\BibitemShut {NoStop}%
\bibitem [{\citenamefont {Murphy}\ \emph {et~al.}(2016)\citenamefont {Murphy},
  \citenamefont {Giret}, \citenamefont {Daraszewicz}, \citenamefont {Lim},
  \citenamefont {Shluger}, \citenamefont {Tanimura},\ and\ \citenamefont
  {Duffy}}]{Murphy_2016}%
  \BibitemOpen
  \bibfield  {author} {\bibinfo {author} {\bibfnamefont {S.~T.}\ \bibnamefont
  {Murphy}}, \bibinfo {author} {\bibfnamefont {Y.}~\bibnamefont {Giret}},
  \bibinfo {author} {\bibfnamefont {S.~L.}\ \bibnamefont {Daraszewicz}},
  \bibinfo {author} {\bibfnamefont {A.~C.}\ \bibnamefont {Lim}}, \bibinfo
  {author} {\bibfnamefont {A.~L.}\ \bibnamefont {Shluger}}, \bibinfo {author}
  {\bibfnamefont {K.}~\bibnamefont {Tanimura}}, \ and\ \bibinfo {author}
  {\bibfnamefont {D.~M.}\ \bibnamefont {Duffy}},\ }\href {\doibase
  10.1103/PhysRevB.93.104105} {\bibfield  {journal} {\bibinfo  {journal} {Phys.
  Rev. B}\ }\textbf {\bibinfo {volume} {93}},\ \bibinfo {pages} {104105}
  (\bibinfo {year} {2016})}\BibitemShut {NoStop}%
\bibitem [{\citenamefont {Daraszewicz}\ \emph {et~al.}(2013)\citenamefont
  {Daraszewicz}, \citenamefont {Giret}, \citenamefont {Naruse}, \citenamefont
  {Murooka}, \citenamefont {Yang}, \citenamefont {Duffy}, \citenamefont
  {Shluger},\ and\ \citenamefont {Tanimura}}]{Daraszewicz_2013}%
  \BibitemOpen
  \bibfield  {author} {\bibinfo {author} {\bibfnamefont {S.~L.}\ \bibnamefont
  {Daraszewicz}}, \bibinfo {author} {\bibfnamefont {Y.}~\bibnamefont {Giret}},
  \bibinfo {author} {\bibfnamefont {N.}~\bibnamefont {Naruse}}, \bibinfo
  {author} {\bibfnamefont {Y.}~\bibnamefont {Murooka}}, \bibinfo {author}
  {\bibfnamefont {J.}~\bibnamefont {Yang}}, \bibinfo {author} {\bibfnamefont
  {D.~M.}\ \bibnamefont {Duffy}}, \bibinfo {author} {\bibfnamefont {A.~L.}\
  \bibnamefont {Shluger}}, \ and\ \bibinfo {author} {\bibfnamefont
  {K.}~\bibnamefont {Tanimura}},\ }\href {\doibase 10.1103/PhysRevB.88.184101}
  {\bibfield  {journal} {\bibinfo  {journal} {Phys. Rev. B}\ }\textbf {\bibinfo
  {volume} {88}},\ \bibinfo {pages} {184101} (\bibinfo {year}
  {2013})}\BibitemShut {NoStop}%
\bibitem [{\citenamefont {Ernstorfer}\ \emph {et~al.}(2009)\citenamefont
  {Ernstorfer}, \citenamefont {Harb}, \citenamefont {Hebeisen}, \citenamefont
  {Sciaini}, \citenamefont {Dartigalongue},\ and\ \citenamefont
  {Dwayne~Miller}}]{Ernstorfer_2009}%
  \BibitemOpen
  \bibfield  {author} {\bibinfo {author} {\bibfnamefont {R.}~\bibnamefont
  {Ernstorfer}}, \bibinfo {author} {\bibfnamefont {M.}~\bibnamefont {Harb}},
  \bibinfo {author} {\bibfnamefont {C.~T.}\ \bibnamefont {Hebeisen}}, \bibinfo
  {author} {\bibfnamefont {G.}~\bibnamefont {Sciaini}}, \bibinfo {author}
  {\bibfnamefont {T.}~\bibnamefont {Dartigalongue}}, \ and\ \bibinfo {author}
  {\bibfnamefont {R.~J.}\ \bibnamefont {Dwayne~Miller}},\ }\href {\doibase
  10.1126/science.1162697} {\bibfield  {journal} {\bibinfo  {journal}
  {Science}\ }\textbf {\bibinfo {volume} {323}},\ \bibinfo {pages} {1033}
  (\bibinfo {year} {2009})}\BibitemShut {NoStop}%
\bibitem [{\citenamefont {Giret}, \citenamefont {Gell{\'e}},\ and\
  \citenamefont {Arnaud}(2011)}]{Giret_2011}%
  \BibitemOpen
  \bibfield  {author} {\bibinfo {author} {\bibfnamefont {Y.}~\bibnamefont
  {Giret}}, \bibinfo {author} {\bibfnamefont {A.}~\bibnamefont {Gell{\'e}}}, \
  and\ \bibinfo {author} {\bibfnamefont {B.}~\bibnamefont {Arnaud}},\ }\href
  {\doibase 10.1103/PhysRevLett.106.155503} {\bibfield  {journal} {\bibinfo
  {journal} {Phys. Rev. Lett.}\ }\textbf {\bibinfo {volume} {106}},\ \bibinfo
  {pages} {155503} (\bibinfo {year} {2011})}\BibitemShut {NoStop}%
\bibitem [{\citenamefont {Recoules}\ \emph {et~al.}(2006)\citenamefont
  {Recoules}, \citenamefont {Cl{\'e}rouin}, \citenamefont {Z{\'e}rah},
  \citenamefont {Anglade},\ and\ \citenamefont {Mazevet}}]{Recoules_2006}%
  \BibitemOpen
  \bibfield  {author} {\bibinfo {author} {\bibfnamefont {V.}~\bibnamefont
  {Recoules}}, \bibinfo {author} {\bibfnamefont {J.}~\bibnamefont
  {Cl{\'e}rouin}}, \bibinfo {author} {\bibfnamefont {G.}~\bibnamefont
  {Z{\'e}rah}}, \bibinfo {author} {\bibfnamefont {P.~M.}\ \bibnamefont
  {Anglade}}, \ and\ \bibinfo {author} {\bibfnamefont {S.}~\bibnamefont
  {Mazevet}},\ }\href {\doibase 10.1103/PhysRevLett.96.055503} {\bibfield
  {journal} {\bibinfo  {journal} {Phys. Rev. Lett.}\ }\textbf {\bibinfo
  {volume} {96}},\ \bibinfo {pages} {055503} (\bibinfo {year}
  {2006})}\BibitemShut {NoStop}%
\bibitem [{\citenamefont {Inogamov}\ \emph {et~al.}(2012)\citenamefont
  {Inogamov}, \citenamefont {Petrov}, \citenamefont {Zhakhovsky}, \citenamefont
  {Khokhlov}, \citenamefont {Demaske}, \citenamefont {Ashitkov}, \citenamefont
  {Khishchenko}, \citenamefont {Migdal}, \citenamefont {Agranat}, \citenamefont
  {Anisimov}, \citenamefont {Fortov},\ and\ \citenamefont
  {Oleynik}}]{Inogamov_2012}%
  \BibitemOpen
  \bibfield  {author} {\bibinfo {author} {\bibfnamefont {N.~A.}\ \bibnamefont
  {Inogamov}}, \bibinfo {author} {\bibfnamefont {Y.~V.}\ \bibnamefont
  {Petrov}}, \bibinfo {author} {\bibfnamefont {V.~V.}\ \bibnamefont
  {Zhakhovsky}}, \bibinfo {author} {\bibfnamefont {V.~A.}\ \bibnamefont
  {Khokhlov}}, \bibinfo {author} {\bibfnamefont {B.~J.}\ \bibnamefont
  {Demaske}}, \bibinfo {author} {\bibfnamefont {S.~I.}\ \bibnamefont
  {Ashitkov}}, \bibinfo {author} {\bibfnamefont {K.}~\bibnamefont
  {Khishchenko}}, \bibinfo {author} {\bibfnamefont {K.~P.}\ \bibnamefont
  {Migdal}}, \bibinfo {author} {\bibfnamefont {M.~B.}\ \bibnamefont {Agranat}},
  \bibinfo {author} {\bibfnamefont {S.~I.}\ \bibnamefont {Anisimov}}, \bibinfo
  {author} {\bibfnamefont {V.~E.}\ \bibnamefont {Fortov}}, \ and\ \bibinfo
  {author} {\bibfnamefont {I.~I.}\ \bibnamefont {Oleynik}},\ }\href@noop {}
  {\bibfield  {journal} {\bibinfo  {journal} {AIP Conf. Proc.}\ }\textbf
  {\bibinfo {volume} {1464}},\ \bibinfo {pages} {593} (\bibinfo {year}
  {2012})}\BibitemShut {NoStop}%
\bibitem [{\citenamefont {Wang}\ \emph {et~al.}(2017)\citenamefont {Wang},
  \citenamefont {Kuchmizhak}, \citenamefont {Li}, \citenamefont {Juodkazis},
  \citenamefont {Vitrik}, \citenamefont {Kulchin}, \citenamefont {Zhakhovsky},
  \citenamefont {Danilov}, \citenamefont {Ionin}, \citenamefont {Kudryashov},
  \citenamefont {Rudenko},\ and\ \citenamefont {Inogamov}}]{Wang_2018}%
  \BibitemOpen
  \bibfield  {author} {\bibinfo {author} {\bibfnamefont {X.~W.}\ \bibnamefont
  {Wang}}, \bibinfo {author} {\bibfnamefont {A.~A.}\ \bibnamefont
  {Kuchmizhak}}, \bibinfo {author} {\bibfnamefont {X.}~\bibnamefont {Li}},
  \bibinfo {author} {\bibfnamefont {S.}~\bibnamefont {Juodkazis}}, \bibinfo
  {author} {\bibfnamefont {O.~B.}\ \bibnamefont {Vitrik}}, \bibinfo {author}
  {\bibfnamefont {Y.~N.}\ \bibnamefont {Kulchin}}, \bibinfo {author}
  {\bibfnamefont {V.~V.}\ \bibnamefont {Zhakhovsky}}, \bibinfo {author}
  {\bibfnamefont {P.~A.}\ \bibnamefont {Danilov}}, \bibinfo {author}
  {\bibfnamefont {A.~A.}\ \bibnamefont {Ionin}}, \bibinfo {author}
  {\bibfnamefont {S.~I.}\ \bibnamefont {Kudryashov}}, \bibinfo {author}
  {\bibfnamefont {A.~A.}\ \bibnamefont {Rudenko}}, \ and\ \bibinfo {author}
  {\bibfnamefont {N.~A.}\ \bibnamefont {Inogamov}},\ }\href@noop {} {\bibfield
  {journal} {\bibinfo  {journal} {Phys. Rev. Appl.}\ }\textbf {\bibinfo
  {volume} {8}},\ \bibinfo {pages} {044016} (\bibinfo {year}
  {2017})}\BibitemShut {NoStop}%
\bibitem [{\citenamefont {Mueller}\ and\ \citenamefont
  {Rethfeld}(2013)}]{Mueller_2013}%
  \BibitemOpen
  \bibfield  {author} {\bibinfo {author} {\bibfnamefont {B.~Y.}\ \bibnamefont
  {Mueller}}\ and\ \bibinfo {author} {\bibfnamefont {B.}~\bibnamefont
  {Rethfeld}},\ }\href {\doibase 10.1103/PhysRevB.87.035139} {\bibfield
  {journal} {\bibinfo  {journal} {Phys. Rev. B}\ }\textbf {\bibinfo {volume}
  {87}},\ \bibinfo {pages} {035139} (\bibinfo {year} {2013})}\BibitemShut
  {NoStop}%
\bibitem [{\citenamefont {Brown}\ \emph {et~al.}(2016)\citenamefont {Brown},
  \citenamefont {Sundararaman}, \citenamefont {Narang}, \citenamefont
  {Goddard},\ and\ \citenamefont {Atwater}}]{Brown_2016_2}%
  \BibitemOpen
  \bibfield  {author} {\bibinfo {author} {\bibfnamefont {A.~M.}\ \bibnamefont
  {Brown}}, \bibinfo {author} {\bibfnamefont {R.}~\bibnamefont {Sundararaman}},
  \bibinfo {author} {\bibfnamefont {P.}~\bibnamefont {Narang}}, \bibinfo
  {author} {\bibfnamefont {W.~A.}\ \bibnamefont {Goddard}}, \ and\ \bibinfo
  {author} {\bibfnamefont {H.~A.}\ \bibnamefont {Atwater}},\ }\href {\doibase
  10.1021/acsnano.5b06199} {\bibfield  {journal} {\bibinfo  {journal} {ACS
  Nano}\ }\textbf {\bibinfo {volume} {10}},\ \bibinfo {pages} {957} (\bibinfo
  {year} {2016})}\BibitemShut {NoStop}%
\bibitem [{\citenamefont {Schoenlein}\ \emph {et~al.}(1987)\citenamefont
  {Schoenlein}, \citenamefont {Lin}, \citenamefont {Fujimoto},\ and\
  \citenamefont {Eesley}}]{Schoenlein_1987}%
  \BibitemOpen
  \bibfield  {author} {\bibinfo {author} {\bibfnamefont {R.~W.}\ \bibnamefont
  {Schoenlein}}, \bibinfo {author} {\bibfnamefont {W.~Z.}\ \bibnamefont {Lin}},
  \bibinfo {author} {\bibfnamefont {J.~G.}\ \bibnamefont {Fujimoto}}, \ and\
  \bibinfo {author} {\bibfnamefont {G.~L.}\ \bibnamefont {Eesley}},\ }\href
  {\doibase 10.1103/PhysRevLett.58.1680} {\bibfield  {journal} {\bibinfo
  {journal} {Phys. Rev. Lett.}\ }\textbf {\bibinfo {volume} {58}},\ \bibinfo
  {pages} {1680} (\bibinfo {year} {1987})}\BibitemShut {NoStop}%
\bibitem [{\citenamefont {Elsayed-Ali}\ \emph {et~al.}(1987)\citenamefont
  {Elsayed-Ali}, \citenamefont {Norris}, \citenamefont {Pessot},\ and\
  \citenamefont {Mourou}}]{Elsayed-Ali_1987}%
  \BibitemOpen
  \bibfield  {author} {\bibinfo {author} {\bibfnamefont {H.~E.}\ \bibnamefont
  {Elsayed-Ali}}, \bibinfo {author} {\bibfnamefont {T.~B.}\ \bibnamefont
  {Norris}}, \bibinfo {author} {\bibfnamefont {M.~A.}\ \bibnamefont {Pessot}},
  \ and\ \bibinfo {author} {\bibfnamefont {G.~A.}\ \bibnamefont {Mourou}},\
  }\href {\doibase 10.1103/PhysRevLett.58.1212} {\bibfield  {journal} {\bibinfo
   {journal} {Phys. Rev. Lett.}\ }\textbf {\bibinfo {volume} {58}},\ \bibinfo
  {pages} {1212} (\bibinfo {year} {1987})}\BibitemShut {NoStop}%
\bibitem [{\citenamefont {Elsayed-Ali}\ \emph {et~al.}(1991)\citenamefont
  {Elsayed-Ali}, \citenamefont {Juhasz}, \citenamefont {Smith},\ and\
  \citenamefont {Bron}}]{Elsayed-Ali_1991}%
  \BibitemOpen
  \bibfield  {author} {\bibinfo {author} {\bibfnamefont {H.~E.}\ \bibnamefont
  {Elsayed-Ali}}, \bibinfo {author} {\bibfnamefont {T.}~\bibnamefont {Juhasz}},
  \bibinfo {author} {\bibfnamefont {G.~O.}\ \bibnamefont {Smith}}, \ and\
  \bibinfo {author} {\bibfnamefont {W.~E.}\ \bibnamefont {Bron}},\ }\href
  {\doibase 10.1103/PhysRevB.43.4488} {\bibfield  {journal} {\bibinfo
  {journal} {Phys. Rev. B}\ }\textbf {\bibinfo {volume} {43}},\ \bibinfo
  {pages} {4488} (\bibinfo {year} {1991})}\BibitemShut {NoStop}%
\bibitem [{\citenamefont {Hohlfeld}\ \emph {et~al.}(2000)\citenamefont
  {Hohlfeld}, \citenamefont {Wellershoff}, \citenamefont {G{\"u}dde},
  \citenamefont {Conrad}, \citenamefont {J{\"a}hnke},\ and\ \citenamefont
  {Matthias}}]{Hohlfeld_2000}%
  \BibitemOpen
  \bibfield  {author} {\bibinfo {author} {\bibfnamefont {J.}~\bibnamefont
  {Hohlfeld}}, \bibinfo {author} {\bibfnamefont {S.-S.}\ \bibnamefont
  {Wellershoff}}, \bibinfo {author} {\bibfnamefont {J.}~\bibnamefont
  {G{\"u}dde}}, \bibinfo {author} {\bibfnamefont {U.}~\bibnamefont {Conrad}},
  \bibinfo {author} {\bibfnamefont {V.}~\bibnamefont {J{\"a}hnke}}, \ and\
  \bibinfo {author} {\bibfnamefont {E.}~\bibnamefont {Matthias}},\ }\href
  {\doibase 10.1016/S0301-0104(99)00330-4} {\bibfield  {journal} {\bibinfo
  {journal} {Chem. Phys.}\ }\textbf {\bibinfo {volume} {251}},\ \bibinfo
  {pages} {237} (\bibinfo {year} {2000})}\BibitemShut {NoStop}%
\bibitem [{\citenamefont {Tanaka}\ and\ \citenamefont
  {Tsuneyuki}(2022)}]{Tanaka_2021}%
  \BibitemOpen
  \bibfield  {author} {\bibinfo {author} {\bibfnamefont {Y.}~\bibnamefont
  {Tanaka}}\ and\ \bibinfo {author} {\bibfnamefont {S.}~\bibnamefont
  {Tsuneyuki}},\ }\href@noop {} {\bibfield  {journal} {\bibinfo  {journal} {J.
  Phys.: Condens. Matter}\ }\textbf {\bibinfo {volume} {34}},\ \bibinfo {pages}
  {165901} (\bibinfo {year} {2022})}\BibitemShut {NoStop}%
\bibitem [{\citenamefont {Lin}, \citenamefont {Zhigilei},\ and\ \citenamefont
  {Celli}(2008)}]{Lin_2008}%
  \BibitemOpen
  \bibfield  {author} {\bibinfo {author} {\bibfnamefont {Z.}~\bibnamefont
  {Lin}}, \bibinfo {author} {\bibfnamefont {L.~V.}\ \bibnamefont {Zhigilei}}, \
  and\ \bibinfo {author} {\bibfnamefont {V.}~\bibnamefont {Celli}},\ }\href
  {\doibase 10.1103/PhysRevB.77.075133} {\bibfield  {journal} {\bibinfo
  {journal} {Phys. Rev. B}\ }\textbf {\bibinfo {volume} {77}},\ \bibinfo
  {pages} {075133} (\bibinfo {year} {2008})}\BibitemShut {NoStop}%
\bibitem [{\citenamefont {Migdal}\ \emph {et~al.}(2016)\citenamefont {Migdal},
  \citenamefont {Petrov}, \citenamefont {Il`nitsky}, \citenamefont
  {Zhakhovsky}, \citenamefont {Inogamov}, \citenamefont {Khishchenko},
  \citenamefont {Knyazev},\ and\ \citenamefont {Levashov}}]{Migdal_2016}%
  \BibitemOpen
  \bibfield  {author} {\bibinfo {author} {\bibfnamefont {K.~P.}\ \bibnamefont
  {Migdal}}, \bibinfo {author} {\bibfnamefont {Y.~V.}\ \bibnamefont {Petrov}},
  \bibinfo {author} {\bibfnamefont {D.~K.}\ \bibnamefont {Il`nitsky}}, \bibinfo
  {author} {\bibfnamefont {V.~V.}\ \bibnamefont {Zhakhovsky}}, \bibinfo
  {author} {\bibfnamefont {N.~A.}\ \bibnamefont {Inogamov}}, \bibinfo {author}
  {\bibfnamefont {K.~V.}\ \bibnamefont {Khishchenko}}, \bibinfo {author}
  {\bibfnamefont {D.~V.}\ \bibnamefont {Knyazev}}, \ and\ \bibinfo {author}
  {\bibfnamefont {P.~R.}\ \bibnamefont {Levashov}},\ }\href {\doibase
  10.1007/s00339-016-9757-8} {\bibfield  {journal} {\bibinfo  {journal} {Appl.
  Phys. A}\ }\textbf {\bibinfo {volume} {122}},\ \bibinfo {pages} {408}
  (\bibinfo {year} {2016})}\BibitemShut {NoStop}%
\bibitem [{\citenamefont {Petrov}, \citenamefont {Inogamov},\ and\
  \citenamefont {Migdal}(2013)}]{Petrov_2013}%
  \BibitemOpen
  \bibfield  {author} {\bibinfo {author} {\bibfnamefont {Y.}~\bibnamefont
  {Petrov}}, \bibinfo {author} {\bibfnamefont {N.}~\bibnamefont {Inogamov}}, \
  and\ \bibinfo {author} {\bibfnamefont {K.}~\bibnamefont {Migdal}},\ }\href
  {\doibase 10.1134/S0021364013010098} {\bibfield  {journal} {\bibinfo
  {journal} {JETP. Lett.}\ }\textbf {\bibinfo {volume} {97}},\ \bibinfo {pages}
  {20} (\bibinfo {year} {2013})}\BibitemShut {NoStop}%
\bibitem [{\citenamefont {Migdal}\ \emph {et~al.}(2015)\citenamefont {Migdal},
  \citenamefont {Il`nitsky}, \citenamefont {Petrov},\ and\ \citenamefont
  {Inogamov}}]{Migdal_2015}%
  \BibitemOpen
  \bibfield  {author} {\bibinfo {author} {\bibfnamefont {K.~P.}\ \bibnamefont
  {Migdal}}, \bibinfo {author} {\bibfnamefont {V.~V.}\ \bibnamefont
  {Il`nitsky}}, \bibinfo {author} {\bibfnamefont {Y.~V.}\ \bibnamefont
  {Petrov}}, \ and\ \bibinfo {author} {\bibfnamefont {K.~V.}\ \bibnamefont
  {Inogamov}},\ }\href@noop {} {\bibfield  {journal} {\bibinfo  {journal} {J.
  Phys. Conf. Ser.}\ }\textbf {\bibinfo {volume} {653}},\ \bibinfo {pages}
  {012086} (\bibinfo {year} {2015})}\BibitemShut {NoStop}%
\bibitem [{\citenamefont {Hoover}(1985)}]{Hoover_1985}%
  \BibitemOpen
  \bibfield  {author} {\bibinfo {author} {\bibfnamefont {W.~G.}\ \bibnamefont
  {Hoover}},\ }\href {\doibase 10.1103/PhysRevA.31.1695} {\bibfield  {journal}
  {\bibinfo  {journal} {Phys. Rev. A}\ }\textbf {\bibinfo {volume} {31}},\
  \bibinfo {pages} {1695} (\bibinfo {year} {1985})}\BibitemShut {NoStop}%
\bibitem [{\citenamefont {Stukowski}(2010)}]{OVITO}%
  \BibitemOpen
  \bibfield  {author} {\bibinfo {author} {\bibfnamefont {A.}~\bibnamefont
  {Stukowski}},\ }\href@noop {} {\bibfield  {journal} {\bibinfo  {journal}
  {Model. Simul. Mater. Sci. Eng.}\ }\textbf {\bibinfo {volume} {18}},\
  \bibinfo {pages} {015012} (\bibinfo {year} {2010})}\BibitemShut {NoStop}%
\bibitem [{\citenamefont {Basinski}, \citenamefont {Duesbery},\ and\
  \citenamefont {Taylor}(1971)}]{Basinski_1971}%
  \BibitemOpen
  \bibfield  {author} {\bibinfo {author} {\bibfnamefont {Z.~S.}\ \bibnamefont
  {Basinski}}, \bibinfo {author} {\bibfnamefont {M.~S.}\ \bibnamefont
  {Duesbery}}, \ and\ \bibinfo {author} {\bibfnamefont {R.}~\bibnamefont
  {Taylor}},\ }\href@noop {} {\bibfield  {journal} {\bibinfo  {journal} {Can.
  J. Phys.}\ }\textbf {\bibinfo {volume} {49}},\ \bibinfo {pages} {2160}
  (\bibinfo {year} {1971})}\BibitemShut {NoStop}%
\bibitem [{\citenamefont {Linde}(2004)}]{Linde_2003}%
  \BibitemOpen
  \bibfield  {author} {\bibinfo {author} {\bibfnamefont {D.~R.}\ \bibnamefont
  {Linde}},\ }\href@noop {} {\emph {\bibinfo {title} {CRC Handbook of Chemistry
  and Physics}}},\ \bibinfo {edition} {84th}\ ed.\ (\bibinfo  {publisher} {SCRC
  Press},\ \bibinfo {address} {Florida},\ \bibinfo {year}
  {2003-2004})\BibitemShut {NoStop}%
\end{thebibliography}%
\end{document}